%% file: main.tex
\documentclass[12pt,twoside,openright,a4paper]{book}

\usepackage{graphicx}
\usepackage{array}
\usepackage{booktabs}
\usepackage{multirow}
\usepackage{multicol}
\usepackage{float}
\usepackage[justification=centering]{caption}
\usepackage{hyperref}
\usepackage{amsmath}
\usepackage{amsfonts}
\usepackage[dvipsnames]{xcolor}
\usepackage{color, colortbl}
\usepackage[backend=biber, style=ieee-alphabetic, natbib=true]{biblatex}
\renewcommand{\baselinestretch}{1.2}	

\newcommand{\norm}[1]{\left\lVert#1\right\rVert}
\DeclareMathOperator{\Tr}{Tr}

\definecolor{LightRed}{rgb}{1,0.88,0.88}
\definecolor{LightGreen}{rgb}{0.88,1,0.88}
\definecolor{LightYellow}{rgb}{1,1,0.88}

\begin{document}

\frontmatter
\begin{titlepage}
\renewcommand{\baselinestretch}{1.0}
\begin{center}
\includegraphics{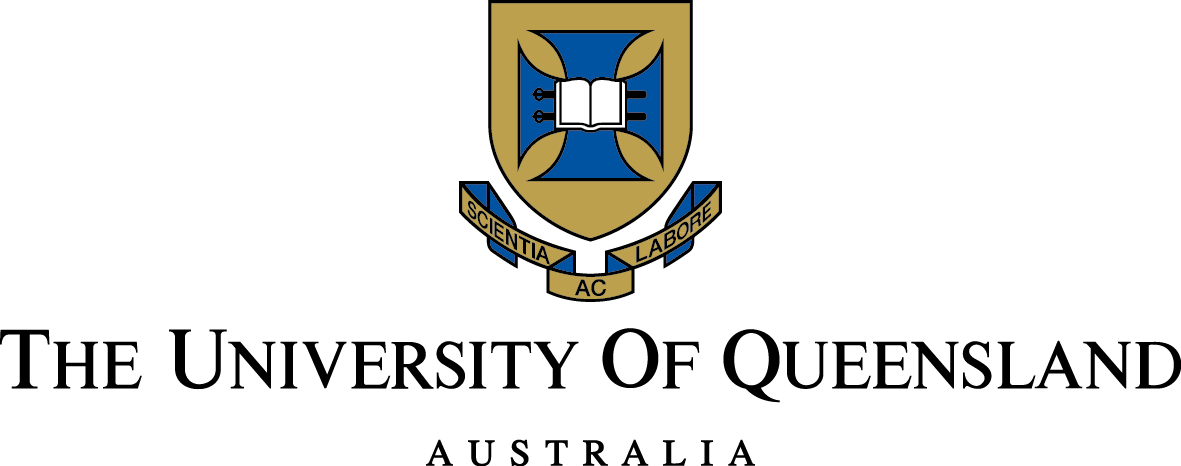}\\
\vfill
\Huge\bf
		SEGMENTATION AND GENERATION OF MAGNETIC RESONANCE IMAGES BY DEEP NEURAL NETWORKS\\
\vfill
\large\sl
		by\\
		\href{mailto:a.delplace@uq.net.au}{ANTOINE DELPLACE}
		\medskip\\
\rm
		School of Information Technology and Electrical Engineering,\\
		The University of Queensland.\\
\vfill
		Submitted for the degree of\\
		Master of Engineering Science
		\smallskip\\
\normalsize
		in the field of Software Engineering
		\medskip\\
\large
		OCTOBER 2019
		\vfill
\end{center}
\end{titlepage}
\cleardoublepage

\vspace*{70mm}
\begin{center}
\renewcommand{\baselinestretch}{1.0}
\sl
	To my parents,\\
	who always support me.
\end{center}

\chapter{Acknowledgments}
I would like to thank Dr.~Shekhar Chandra for his supervision throughout the whole year. His advice and insight were helpful in choosing challenging projects to help Medical Research and in the production of my Conference Paper and final Thesis.

I would also like to thank Mr.~Douglas Kosovic, for his help and his time when monitoring Goliath, the cluster of the University of Queensland. Finally, I thank Dr.~Tero Karras for answering my questions about the implementation of an optimized two-in-one upscale and convolution.
\cleardoublepage

\chapter{Abstract}
Magnetic Resonance Images (MRIs) are extremely used in the medical field to detect and better understand diseases. In order to fasten automatic processing of scans and enhance medical research, this project focuses on automatically segmenting targeted parts of MRIs and generating new MRI datasets from random noise. More specifically, a Deep Neural Network architecture called U-net is used to segment bones and cartilages of Knee MRIs, and several Generative Adversarial Networks (GANs) are compared and tuned to create new realistic and high quality brain MRIs that can be used as training set for more advanced models. Three main architectures are described: Deep Convolution GAN (DCGAN), Super Resolution Residual GAN (SRResGAN) and Progressive GAN (ProGAN), and five loss functions are tested: the Original loss, LSGAN, WGAN, WGAN\_GP and DRAGAN. Moreover, a quantitative benchmark is carried out thanks to evaluation measures using Principal Component Analysis.

The results show that U-net can achieve state-of-the-art performance in segmenting bones and cartilages in Knee MRIs (Accuracy of more than 99.5\%). Moreover, the three GAN architectures can successfully generate realistic brain MRIs even if some models have difficulties to converge. The main insights to stabilize the networks are using one-sided smoothing labels, regularization with gradient penalty in the loss function (like in WGAN\_GP or DRAGAN), adding a minibatch similarity layer in the Discriminator and a long training time.\\

All source code files and training animations are available on Github at the following links:

\href{https://github.com/antoinedelplace/MRI-Segmentation}{https://github.com/antoinedelplace/MRI-Segmentation}

\href{https://github.com/antoinedelplace/MRI-Generation}{https://github.com/antoinedelplace/MRI-Generation}

\tableofcontents

\listoffigures
\addcontentsline{toc}{chapter}{List of Figures}

\listoftables
\addcontentsline{toc}{chapter}{List of Tables}
\cleardoublepage

\mainmatter
\include{introduction}
\include{literature}
\include{theory}
\include{methodology}
\include{experiments}
\include{results}
\include{discussion}
\include{conclusion}

\appendix
\newpage
\addcontentsline{toc}{part}{Appendices}
\mbox{}
\newpage

\include{segmentation_results}
\include{generation_results}

\cleardoublepage
\nocite{*}
\printbibliography[heading=bibintoc]

\end{document}

%% file: introduction.tex
\chapter{Introduction}
For the past decade, Medical Research has been evolving to meet the requirements of an aging population and the new challenges that come with it. In that respect, Magnetic Resonance Imaging (MRI) is often used to detect diseases, to make a diagnosis or to monitor a treatment because it is a non-invasive image technology \cite{mri_def}. As more and more scans need to be analyzed, automatic techniques are created to assist practitioners and increase accuracy and efficiency of treatments. For example, new algorithms are now able to detect breast cancer from screening mammography automatically with high accuracy and precision \cite{breast_cancer_dnn}.

\begin{figure}
\centering
	\includegraphics{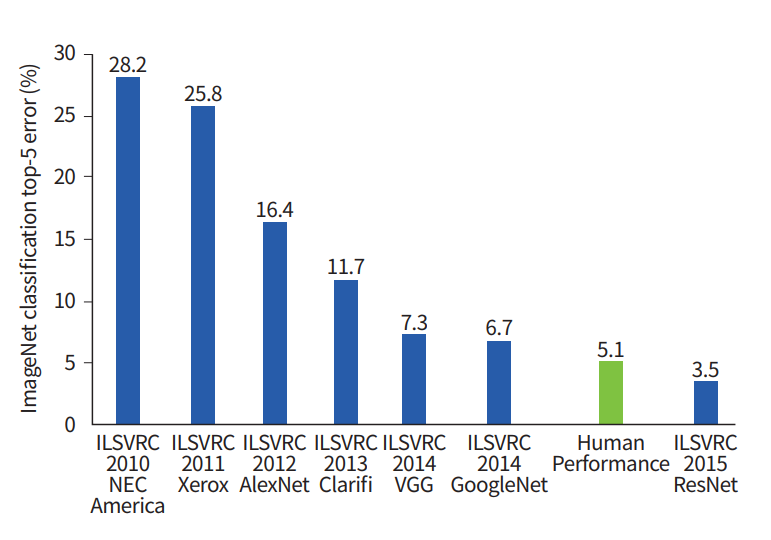}
    \caption{ImageNet Large Scale Visual Recognition Challenge\\(ILSVRC) Performance \cite{ai_intelligent}}
    \label{ilsvc_performance}
\end{figure}

The majority of these new methods use emerging techniques from machine learning and especially deep neural networks \cite{machine_learning_medical}. This is due to a huge increase in image recognition and classification performance, that has even outperformed human abilities in some domains (Figure \ref{ilsvc_performance}). However, training models such as neural networks requires computing power (often provided by GPU clusters) and a huge amount of data, that are not easily available in the medical field because of privacy concerns.

In that context, my work focuses on two main aspects: MRI Segmentation and MRI Generation. First, in order to facilitate the detection of abnormal shapes of bones in MRIs, I implemented a Deep Neural Network that segments cartilages and bones of knee scans. The method can also be used as a preprocessing part for a more complex task like image generation, classification or denoising. The main challenge is to find the relevant hyperparameters and architecture that suit the dataset and enable fast convergence. It takes inspiration from the U-net architecture by \citet{unet} and its performance at the International Symposium on Biomedical Imaging (ISBI) in 2015 but transfers it from Electron Microscope cells to Magnetic Resonance Images. Second, in order to increase the amount of data available for models to train, I made a comparison of three different Generative Adversarial Network (GAN) architectures that produce realistic brain MRIs from a random input latent space. The challenge here is to generate high quality and realistic images with few training images and limited computation power. In addition, GANs are very sensitive to initialization, architectures, hyperparameters and training datasets so a huge exploration process is needed to transfer what have been done with human faces \cite{progan} into Magnetic Resonance Images. This work can then be used to generate training datasets or give some guidance to stabilize and improve GAN convergence.

More specifically, the contributions of the thesis correspond to the following: a detailed performance comparison between different hyperparameters of a U-net trained to segment bones and cartilage of MusculoSkeletal Knee MRIs; and a benchmark of three main GAN architectures to generate $256 \times 256$ brain MRIs from latent noise, along with qualitative results for some variations in the models (hyperparameters and architecture).

We begin the thesis presentation by a review of the founding papers related to segmentation techniques and image generation by deep neural networks in Chapter~\ref{literature}. Then we give some theoretical background about Magnetic Resonance Imaging and Deep Neural Network architectures in Chapter~\ref{theory}. In Chapter~\ref{methodology}, we describe the methodology and the approach taken to tackle the encountered challenges and quantify the segmentation and generation results. Then, Chapter~\ref{experiments} presents the experiment framework: the training datasets, the chosen architectures to test and the different hyperparameters to tune. Finally, Chapter~\ref{results} gives the results of the experiments followed by a discussion on their relevance, the main outcomes and the possible improvements for future work. To conclude, Chapter~\ref{conclusion} gives a summary of the achievements of the thesis along with a critical review of performance.

%% file: literature.tex
\chapter{Background material and Related work}
\label{literature}
In this Chapter, we introduce reviews of important papers on which the thesis is based and an analysis of the main outcomes. Technical details are left for Chapter~\ref{theory}, which explains the theoretical background.

\section{Segmentation before Machine Learning: Shape Models}
\begin{figure}
\centering
	\includegraphics[width=1.0\textwidth]{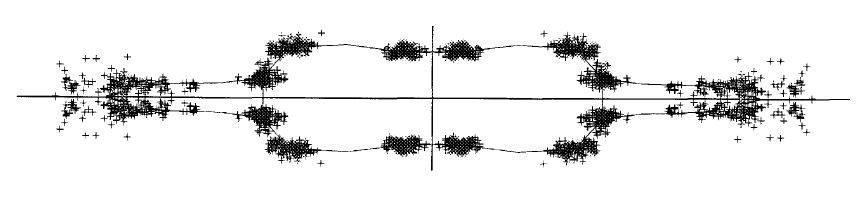}
    \caption{Scatter of training landmark points\\
    from the shape model of several resistors \cite{active_shape_models}}
    \label{active_shape_models_landmarks}
\end{figure}

Image segmentation has been studied for a long time and many methods that came before the age of Neural Networks are based on Shape Models. In \citeyear{active_shape_models}, \citet{active_shape_models} explained how they use an orthogonal representation of the variations of an object (Principal Component Analysis, see Section~\ref{pca}) to capture different shapes in images. This method enables to segment items with small specific variations and is robust to unlikely deformed objects. The shape mask is then aligned with the object in the image thanks to a boundary model based on edge detection. In their paper, \citet{active_shape_models} present successful segmentations of resistors, heart ventricles and hands even in noisy images. However, the method is a Point Distribution Model (PDM) and relies on landmarks that are manually positioned or adjusted to the edge of the object in the training dataset (see Figure \ref{active_shape_models_landmarks}). Thus, it requires a lot of preprocessing time and the results highly depend on the quality of the labelling.

\newpage
\section{Neural Network Revolution}
Since \citeyear{alex}, new methods have highly increased the quality of segmentation and produced new ways of generating samples from a complex manifold thanks to Neural Networks.

\subsection{Original papers}
Neural networks are not a brand new idea but the increasing performance of CPU, GPU and availability of RAM have taken them to the next level. In \citeyear{lecun_nn}, \citet{lecun_nn} use gradient descent and backpropagation (see Section~\ref{nn} for theory) to classify digits from the MNIST (Modified National Institute of Standards and Technology) dataset \cite{mnist}. It uses a Convolutional Neural Network to extract meaningful features from 2D images and succeeds in detecting ASCII characters from real-life documents.

However, the popularization of Neural Networks begins with \citet{alex} in \citeyear{alex} when their model outperformed other algorithms in the ImageNet Large Scale Visual Recognition Competition (ILSVRC). The paper introduces a Neural Network with 5 convolution layers, 3 dense layers and some max-pooling layers. Above all, it uses a new technique called Dropout to regulate the model and avoid overfitting.

Finally, \citet{vgg} open the way to Deep Neural Network by increasing the architecture depth and breaking a new record in the ILSVRC 2014. Their model, called VGG (for Visual Geometry Group), has 16 to 19 weight layers.

\subsection{Residual Network}
Nevertheless, training very deep Neural Networks is more difficult because it requires more computation time, more data and is more sensitive to weight initialization. To cope with this problem, \citet{resnet} introduce Residual Networks, a special architecture that uses Residual blocks that only learn the difference (or residues) with the identity function. This technique allows the network to be even deeper with up to 152 layers for the model introduced in the paper. This architecture reached a new threshold in the ILSVRC in 2015 by outperforming human abilities in image recognition (see Figure \ref{ilsvc_performance}).

\subsection{Auto-encoders}
\begin{figure}
\centering
	\includegraphics[width=0.7\textwidth]{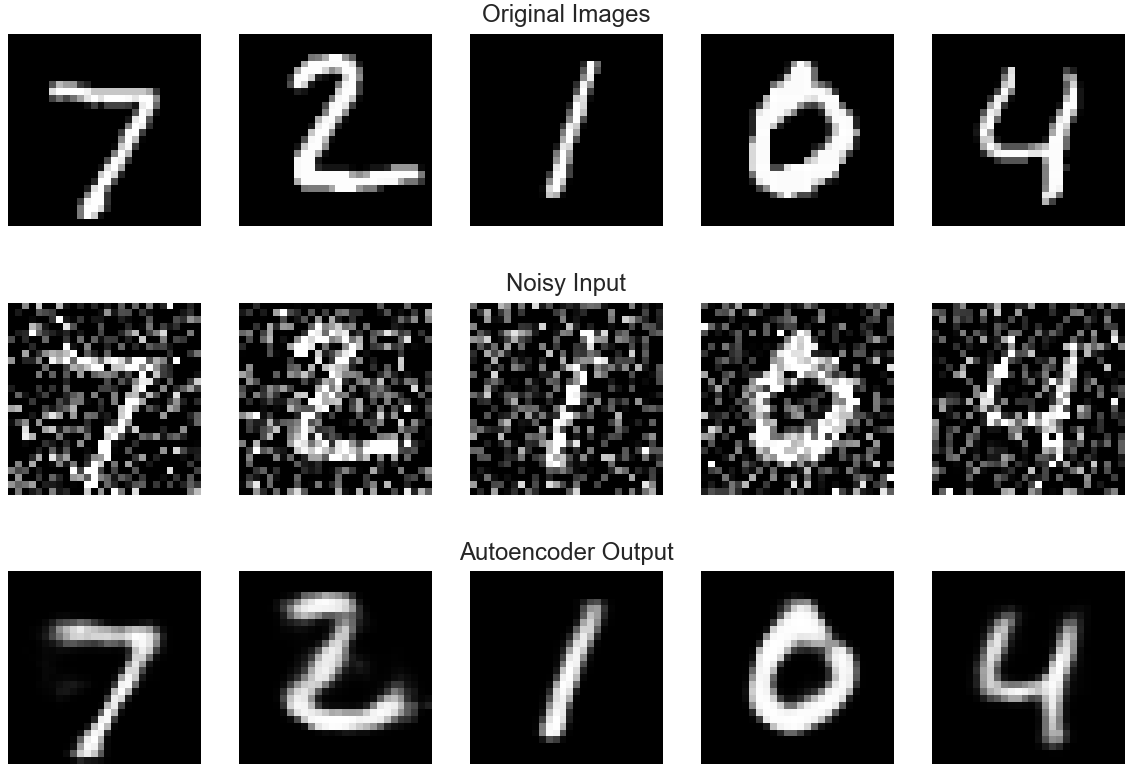}
    \caption{Denoising digits with Auto-encoders \cite{autoencoder}}
    \label{autoencoder_denoising}
\end{figure}

Most of the architectures presented so far focus on image classification tasks but image-to-image networks have also been used for image denoising (see Figure~\ref{autoencoder_denoising}) or compression. The mainly used architecture is called Auto-Encoder and has been known for a few decades \cite[Chapter~14]{deep_learning_book}. It is composed of two parts: an encoder that reduces the input image to a smaller latent space, and a decoder that tries to reproduce the input image from the learnt latent space. This idea is the fundamental principle that has lead to more complex architectures for segmentation and generation.

\section{Segmentation with Deep Neural Networks}
\begin{figure}
\centering
	\includegraphics[width=1.1\textwidth]{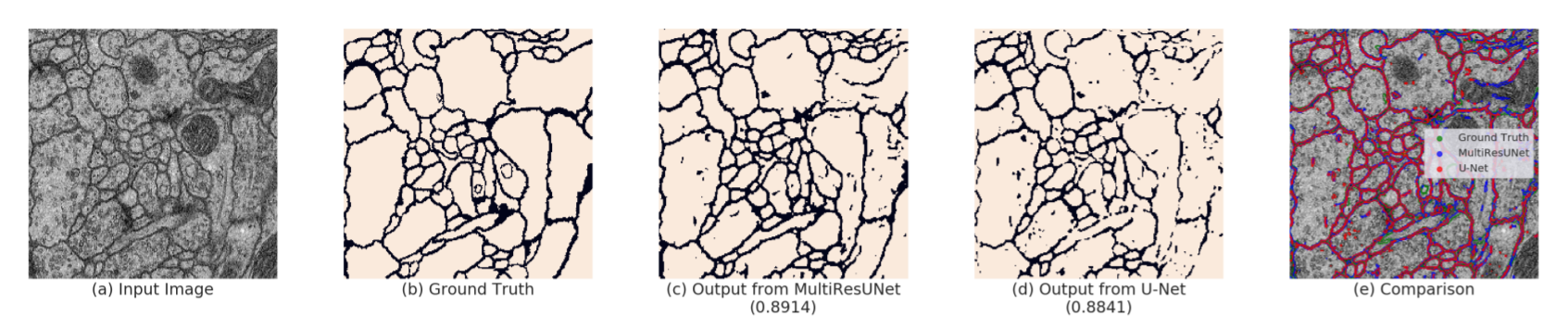}
    \caption{U-net Segmentation of cells \cite{multiresunet}}
    \label{unet_results}
\end{figure}

In \citeyear{unet}, \citet{unet} came with a new architecture that performs fast and accurate segmentation called U-net. It is based on the Auto-encoder architecture but adds skip connections (see Section \ref{unet}) between the downsampling path (encoder part) and the upsampling path (decoder part). It won the ISBI (International Symposium on Biomedical Imaging) Challenge by accurately segmenting neuronal structures in electron microscopic scans (see Figure~\ref{unet_results}). Moreover, it uses intensive data augmentation techniques to reduce the number of input training data needed. Recently, new architectures try to assemble ResNet and U-net like MultiResUNet \cite{multiresunet}.

\section{Image Generation: Generative Adversarial Networks}
Data augmentation techniques can be very useful and adapted for some training datasets (like cells that can be deformed easily), but may be inappropriate for other inputs (with more rigid shapes). That is why new networks are trying to generate data by learning and sampling complex manifolds.

\subsection{Original papers and applications}
\begin{figure}
\centering
	\includegraphics[width=0.7\textwidth]{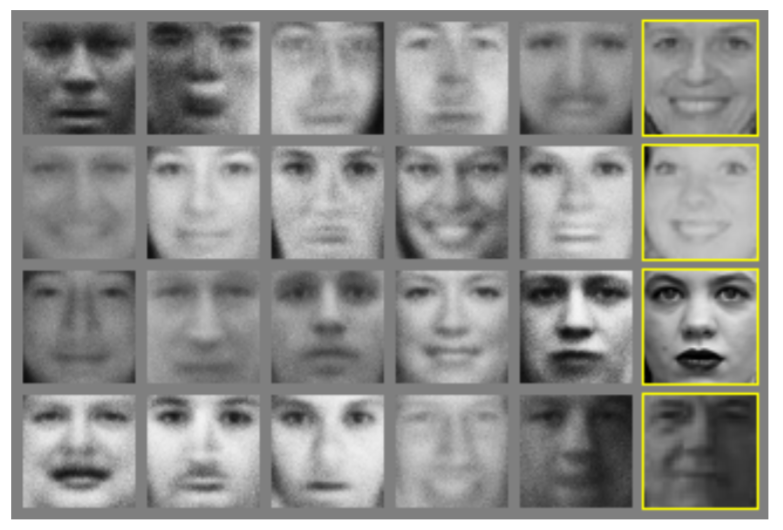}
    \caption[Human face Generation with GANs \cite{gan}]{Human face Generation with GANs \cite{gan}\\
    {\small (the rightmost column shows the nearest training example of the neighboring sample)}}
    \label{gan_tfd}
\end{figure}

In \citeyear{gan}, \citet{gan} introduce a new generative model, based on an adversarial process, called Generative Adversarial Network (GAN). The idea is to create two networks (a Generator and a Discriminator) that compete against each other: one to produce realistic fake images, and the other to detect if an input image is real or fake (see details in Section \ref{gan}). The paper presents theoretical results and successfully produces generated images with MNIST, TFD (Toronto Face Database, see Figure~\ref{gan_tfd}) and CIFAR-10 as training sets. It uses fully connected networks and convolution networks to sample new data.

\begin{figure}
\centering
	\includegraphics[width=0.7\textwidth]{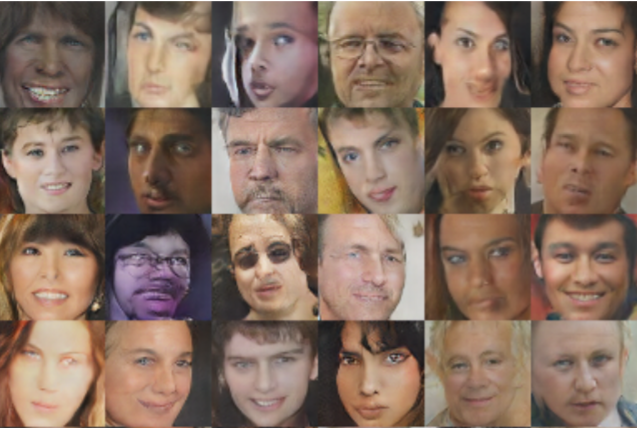}
    \caption{Human face Generation with DCGANs \cite{dcgan}}
    \label{dcgan_results}
\end{figure}

After that, \citet{dcgan} developed the concept to increase the performance with Deep Convolution Generative Adversarial Networks (DCGANs). They add some constraints to the network to stabilize convergence and tested their model on the LSUN (Large-scale Scene Understanding) dataset and on a human face dataset. Generating images is more challenging here because there are multiple face postures and colors that have to be learnt (see Figure~\ref{dcgan_results}).

Since then, a lot of applications have emerged, among which an image-to-image translator Pix2Pix by \citet{pix2pix} that can transfer aerial scenes to maps, labels to facades or edges to photos. More related to my work, \citet{mri_gan} have generated MRIs with a simple DCGAN at low resolutions.

\subsection{Loss function Investigations}
Because training GANs is very difficult, some papers have tried to adjust the loss functions used by the networks to stabilize the convergence.

\citet{lsgan} introduce LSGAN (Least Squares GAN) to cope with the vanishing gradient problem. It changes the original Jensen-Shannon divergence into the Pearson $\chi^2$ divergence.

Then, \citet{wgan} propose WGAN (Wasserstein GAN) that uses the Earth-Mover (EM) or Wasserstein distance instead of Jensen-Shannon divergence. This enables the use of smooth gradients to stabilize the convergence.

To improve WGAN, \citet{wgan_gp} remove the weight clipping by adding a gradient penalty that stabilizes training. This technique has allowed the training of deeper networks (up to a 101-layers Residual Network in the paper).

Finally, \citet{dragan} introduce DRAGAN (Deep Regret Analytic GAN) that adds an additional gradient penalty to the original GAN loss in order to reduce mode collapsing.

All these loss functions perform well on specific datasets and a benchmark is carried out in this paper to know which one suits the best with each architecture and the MRI dataset.

\subsection{Super-Resolution GAN}
\begin{figure}
\centering
	\includegraphics[width=0.9\textwidth]{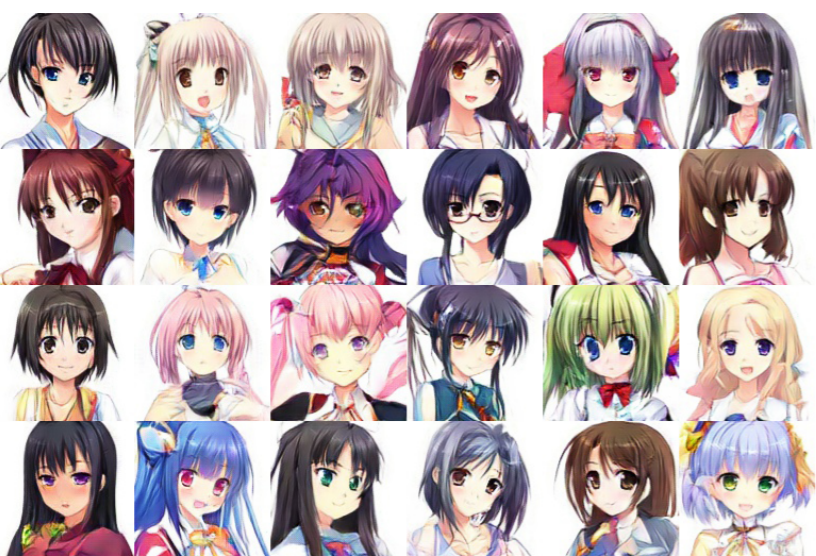}
    \caption{Anime character Generation with SRResGAN \cite{anime_characters_gan}}
    \label{anime_characters}
\end{figure}

In order to generate fine texture details in a synthetic image, \citet{resgan} present SRResGAN (Super Resolution Residual GAN), a model that can reproduce high frequency patterns when upscaling an image by 4. The input corresponds to the downscaled image and the paper succeeds in generating realistic details thanks to several residual blocks in the Generator and Discriminator. The general architecture is reused in \citet{anime_characters_gan} to generate anime characters from noise (see Figure~\ref{anime_characters}).

\subsection{Progressive GAN}
\begin{figure}
\centering
	\includegraphics[width=0.9\textwidth]{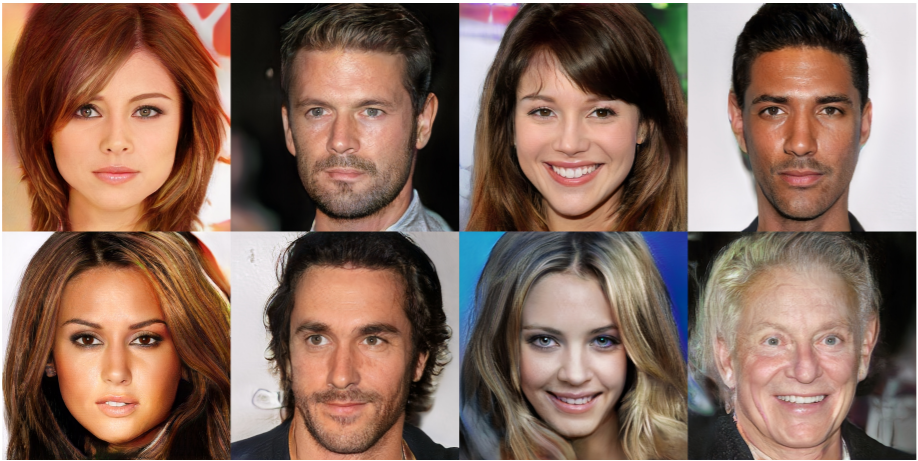}
    \caption{High Resolution Human Face Generation with ProGAN \cite{progan}}
    \label{progan_results}
\end{figure}

In the previous papers, GANs are successful in generating relatively small images but struggle to increase the resolution because of a sensitive stability. \citet{progan} propose a new method to train GANs progressively, by adding more layers in the Generator and in the Discriminator as the image resolution grows. This technique enables the generation of high quality images (see Figure~\ref{progan_results}) up to a resolution of $1024 \times 1024$ with fast and stabilized convergence. The training inputs correspond to a High Quality version of the {\sc CelebA} dataset. Moreover, the paper adds a mini-batch similarity layer in the Discriminator in order to increase the global variation of the generated images. Finally, some modifications in the architecture and implementation intend to avoid unhealthy competitions between the Generator and the Discriminator, and reduce instabilities in the convergence.

%% file: theory.tex
\chapter{Theory}
\label{theory}
In this Chapter, we briefly explain any theoretical result which is necessary for the understanding of later contents. More specifically, we introduce the concept of Magnetic Resonance Imaging (Section~\ref{mri}), Principal Component Analysis (Section~\ref{pca}), Neural Networks and Convolutions (Section~\ref{nn}), and we present two classic architectures: the U-net (Section~\ref{unet}) and the GAN (Section~\ref{gan}).

\section{Magnetic Resonance Imaging}
\label{mri}
Magnetic Resonance Imaging (MRI) is a non-invasive medical technique that can produce 3D images of bones, tissues and cartilages thanks to a strong magnetic field. The different levels of gray give information about the nature of the scanned tissues.

\begin{figure}
\centering
	\includegraphics{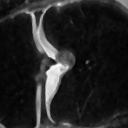}%
	\hspace{1pt}%
    \includegraphics{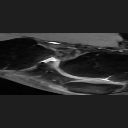}\\
	\includegraphics{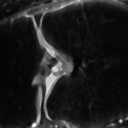}%
	\hspace{1pt}%
	\includegraphics{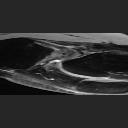}
    \caption[MusculoSkeletal Knee MRIs \cite{knee_dataset}]{MusculoSkeletal Knee MRIs \cite{knee_dataset}\\
    {\small (coronal plane on the left, sagittal plane on the right)}}
    \label{dataset_knee}
\end{figure}

\begin{figure}
\centering
	\includegraphics[width=0.4\textwidth]{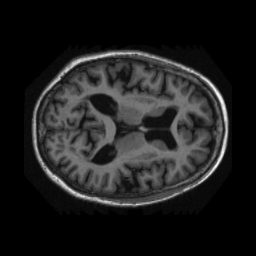}%
	\hspace{1pt}%
    \includegraphics[width=0.4\textwidth]{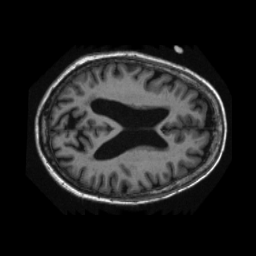}
	\includegraphics[width=0.4\textwidth]{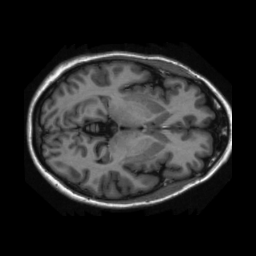}%
	\hspace{1pt}%
	\includegraphics[width=0.4\textwidth]{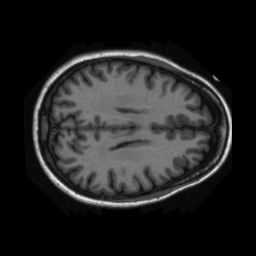}
    \caption{Brain MRIs from the OASIS dataset \cite{oasis}}
    \label{dataset_brain}
\end{figure}

There are different types of MRIs according to the sequences (frequency and pulse of the signal) used during the signal acquisition. The datasets of this project correspond to 2D slices of knee and brain MRIs with T1 weighted sequences (see Figure~\ref{dataset_knee} and \ref{dataset_brain}).

\begin{figure}
\centering
	\includegraphics[width=1.0\textwidth]{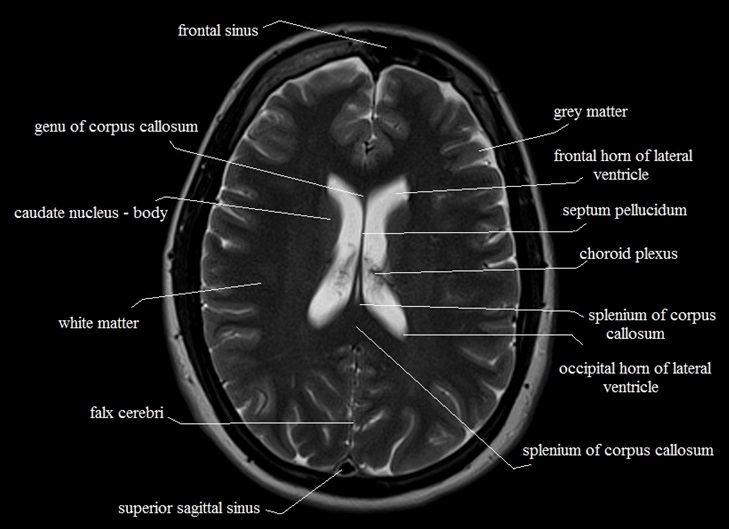}
    \caption{Brain anatomy \cite{brain_anatomy}}
    \label{brain_anatonmy}
\end{figure}

An annotated brain MRI is presented in Figure~\ref{brain_anatonmy}.

\section{Principal Component Analysis}
\label{pca}
Principal Component Analysis (PCA) is a method used to decompose the variations of a multi-dimensional set of observations into an orthogonal basis. The process consists in finding the eigenvalues and eigenvectors of the covariance matrix: the higher the eigenvalue, the more significant the associated eigenvector is to describe the set variance.

\begin{figure}
\centering
	\includegraphics{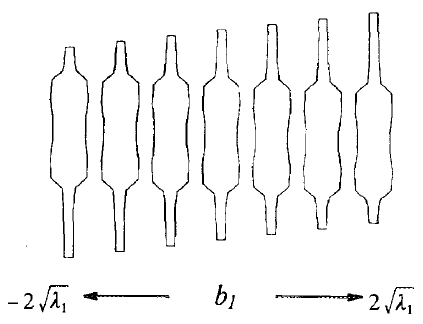}
    \caption{Variation of the resistor shape model with respect to\\
    the major component of PCA \cite{active_shape_models}}
    \label{active_shape_models_pca}
\end{figure}

The method is largely used to detect the main variations of a manifold (see for example Figure~\ref{active_shape_models_pca}), to decorrelate several variations (thanks to the orthogonality of the eigenvectors) or to reduce the dimension of a set (by projecting the observations on the most significant eigenvectors). In this project, it is used to evaluate the quality of the generative models (see Section~\ref{evaluation_measures_gan}).

\section{Neural Networks and Convolutions}
\label{nn}
\begin{figure}
\centering
	\includegraphics[width=0.7\textwidth]{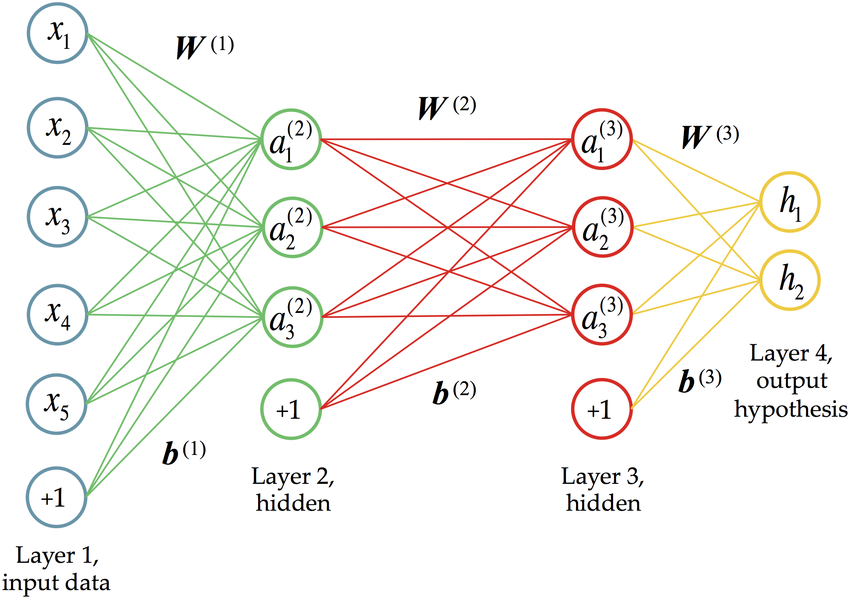}
    \caption{Example of a fully-connected neural network for classification \cite{neural_network_figure}}
    \label{neural_network}
\end{figure}

Neural Networks are a technique to approximate any function thanks to a range of neurons organized in layers. Each neuron processes input signals from some neurons of the previous layer and triggers a new signal to some neurons of the following layer (see Figure~\ref{neural_network}). Equation~\ref{nn_equation} illustrates how an input layer $\mathbf{X}$ is processed into an output signal $\mathbf{Y}$ thanks to a weight matrix $\mathbf{W}$, a bias vector $\mathbf{b}$ and an activation function $\alpha$.

\begin{equation}
    \mathbf{Y} = \alpha (\mathbf{W} \cdot \mathbf{X} + \mathbf{b})
    \label{nn_equation}
\end{equation}

The matrices $\mathbf{W}$ and $\mathbf{b}$ of all layers need to be learnt during the training process, whereas the activation functions, the number of layers and the number of neurons for each layer are hyperparameters that are chosen beforehand. This leads to a huge variety of architectures designed for different purposes (like classification, denoising, sampling, ...). Moreover, training uses a technique called backpropagation where the differences between the network output and the ground truth are spread across all the layers (with the chain rule) and needs lots of input data as a consequence.

\begin{figure}
\centering
	\includegraphics[width=1.0\textwidth]{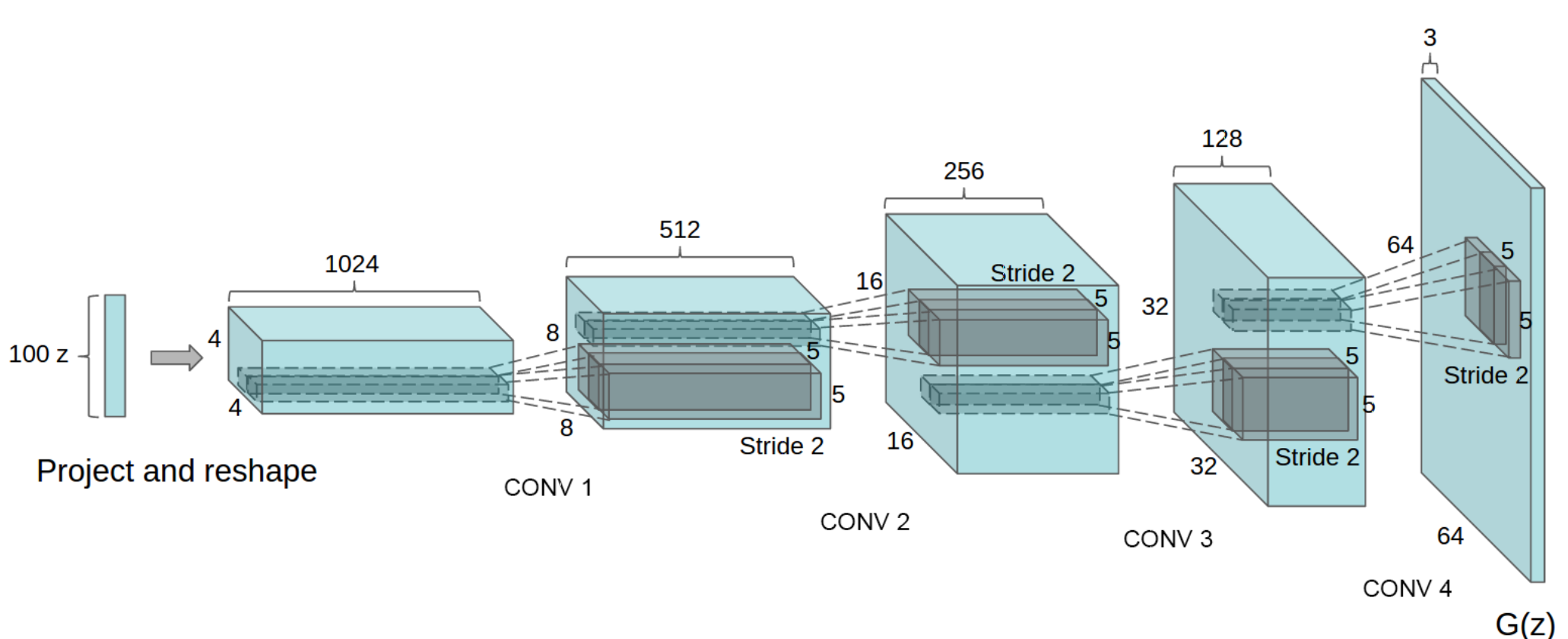}
    \caption{Upsampling network composed of several convolution layers \cite{dcgan}}
    \label{upsampling_network_convolution}
\end{figure}

When it comes to image processing, Convolutional Neural Networks (CNNs) are used to take into account spatial relationships. It consists in learning several kernel filters that are convolved with the image to extract meaningful features. An example of such architecture is presented in Figure~\ref{upsampling_network_convolution}: $5 \times 5$ kernels with stride 2 are used and the number of filters (on top of each block) decreases while the model upsamples the output.

\begin{figure}
\centering
	\includegraphics[width=0.5\textwidth]{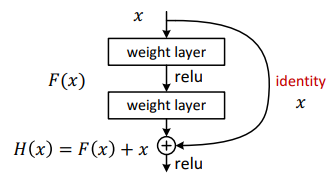}
    \caption{Residual block used in ResNet \cite{resnet}}
    \label{res_blocks}
\end{figure}

Because Deep Neural Networks are difficult to train, some models adapt their architecture to learn residuals (ie the difference with the identity function). It is the case of ResNet \cite{resnet}, which is made of several residual blocks (see Figure~\ref{res_blocks}).

\section{U-net}
\label{unet}
\begin{figure}
\centering
	\includegraphics[width=1.0\textwidth]{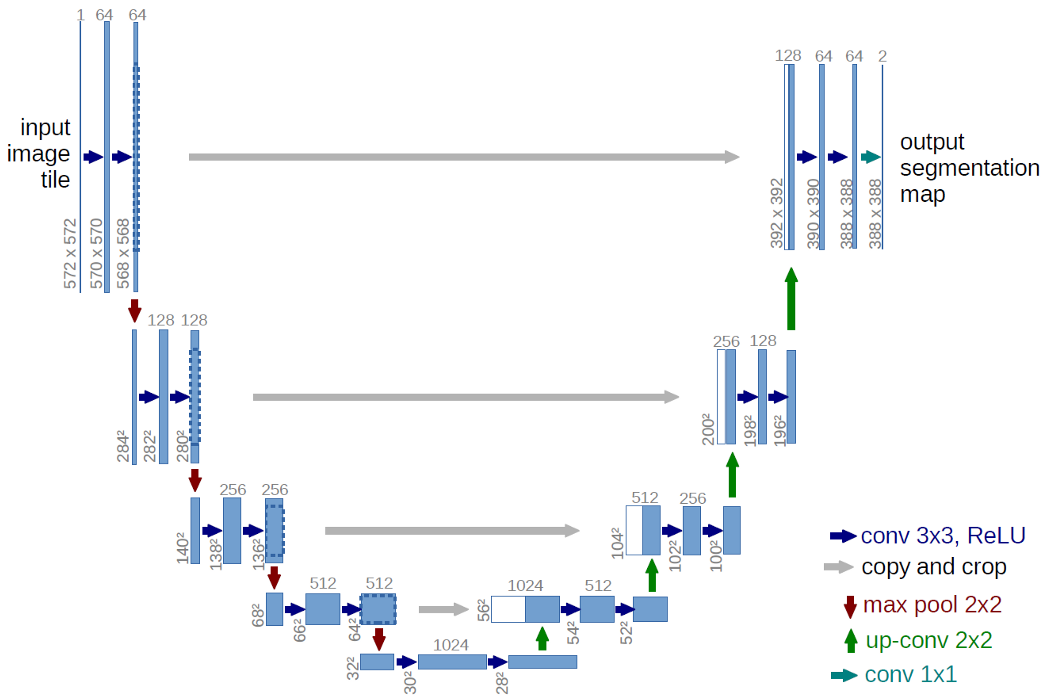}
    \caption{U-net architecture \cite{unet}}
    \label{unet_architecture}
\end{figure}

The U-net architecture, introduced by \citet{unet}, is a model that performs well on segmentation tasks. It is composed of an encoder and a decoder linked by skip connections (see Figure~\ref{unet_architecture}).

The encoder is made of convolution layers with an increasing number of filters and separated by max pooling layers. It extracts meaningful features from the input image. On the other hand, the decoder expands the image from the learnt latent space thanks to convolutions with decreasing number of filters and upsampling convolutions. The skip connections transfer less abstract features from the encoder to the decoder.

\section{Generative Adversarial Networks}
\label{gan}
\begin{figure}
\centering
	\includegraphics[width=1.0\textwidth]{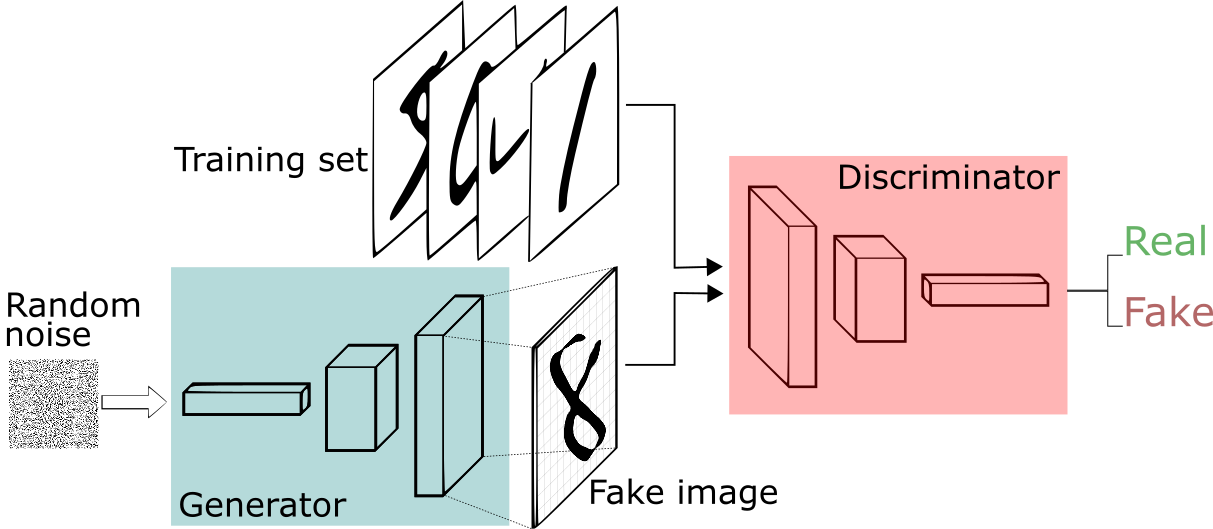}
    \caption{GAN Principle \cite{gan_principle}}
    \label{gan_principle}
\end{figure}

Generative Adversarial Networks (GANs) are a generative model that differs from classical methods by relying on another network to evaluate the ``distance'' (or accuracy) between the output and the original distribution. The first network, called Generator, is responsible for producing new unseen samples (or fake images) from a latent space made of random noise. The second network, called Discriminator, is in charge of detecting if an image has been generated by the Generator or comes from the real training set (see Figure~\ref{gan_principle}).

This principle has some advantages and some drawbacks. First, no explicit distance between the output and the original manifold is necessary so it expands the possibilities of sampling complex manifolds like image coloring (a simple distance would give a mean color to reduce the loss) or realistic images (a simple distance would get blurry edges to minimize the loss). Moreover, GANs can generate images that are not close to any training input but are still relevant in the whole manifold.

\begin{figure}
\centering
	\includegraphics{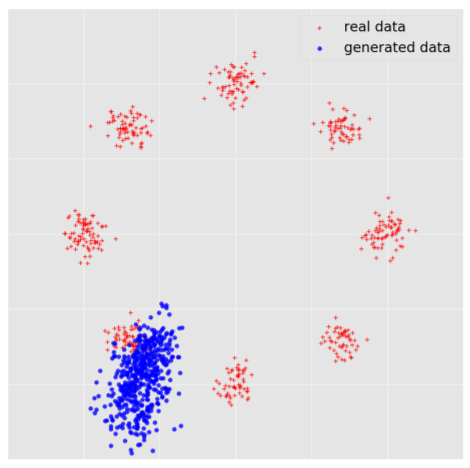}
    \caption{GAN with Mode collapse \cite{mgan}}
    \label{mode_collapse}
\end{figure}

However, some drawbacks arise. The convergence is not very stable because it is necessary that the Generator and the Discriminator learn at the same rate. If the Discriminator is too strong, the Generator does not succeed in generating realistic fake images anymore. Moreover, some architectures experience vanishing gradients and backpropagation becomes inefficient after a while. Finally, some GANs learn only parts of the manifold and discard relevant areas that need to be learnt: this problem is called Mode collapsing (see Figure~\ref{mode_collapse}).

New architectures and loss functions try to cope with these problems and some of them are  reviewed in this project (see Section~\ref{mri_generation_experiment}).

%% file: methodology.tex
\chapter{Methodology and Approach}
\label{methodology}
In this Chapter, are presented a detailed explanation of the approach, the major decisions taken and the evaluation measures used to quantify the performance of the models.

\section{MRI Segmentation with U-net}
\subsection{Motivation}
The aim of the first part of my thesis is to adapt the U-net architecture developed for neuron cells in \citet{unet} to bones and cartilages of Knee MRIs. It is the first step before advanced work on 3D segmentation can be processed.

The main challenges are: limited possibilities for data augmentation (for bones compared to cells) and the hyperparameter tuning of the Neural Network. Moreover, fast convergence and stability are required in order to pave the way for 3D segmentation.

\subsection{Metrics}
To evaluate the performance of my model, I will use the Dice-Sørensen Coefficient (DSC) along with the mean accuracy (See Equations~\ref{dsc_equation} and \ref{acc_equation}).

\begin{equation}
\label{dsc_equation}
    DSC = \frac{2 \cdot |X \cap Y|}{|X|+|Y|} = \frac{2 \cdot TP}{(TP + FN) + (TP + FP)}
\end{equation}

\begin{equation}
\label{acc_equation}
    acc = \frac{|X \cap Y| + |\overline{X} \cap \overline{Y}|}{|X|+|\overline{X}|} = \frac{TP + TN}{(TP + FN) + (TN + FP)}
\end{equation}
with:
\vspace{-12pt}
\begin{itemize}
    \itemsep0em
    \item $X$: the ground truth segmentation (mask of 1 and 0)
    \item $Y$: the output segmentation of the U-net
    \item $TP$: the number of True Positive
    \item $TN$: the number of True Negative
    \item $FP$: the number of False Positive
    \item $FN$: the number of False Negative
\end{itemize}

Efficiency will be assessed through the computation time needed to train the network and the number of trainable parameters.

\section{MRI Generation with GANs}
\subsection{Purpose}
The second part of the project is dedicated to generating synthetic brain MRIs so that they can be used as a training dataset for more advanced models. It can also give insights about a latent representation of brain MRIs that could be used in other medical problems. As GANs have proven to perform very well with human faces (see \cite{gan}, \cite{dcgan} and \cite{progan}), the challenge is to transpose the architectures to generate MRIs.

Because they are a lot of different GAN models that try to cope with instabilities and slow convergence, I decided to compare three main architectures: a simple Deep Convolutional GAN (DCGAN), a GAN using residual blocks called Super Resolution Residual GAN (SRResGAN) and a Progessive GAN (ProGAN). On top of that, a benchmark of five different loss functions will be presented. Furthermore, my project goes beyond the work of \citet{mri_gan} by increasing the resolution to $256 \times 256$.

The main difficulties would be to achieve realistic images with high quality (high resolution) and high variation (by avoiding mode collapsing). Intensive hyperparameter tuning is necessary to obtain stable convergence. Moreover, training must be relatively fast as I am limited in time and in computation power (I cannot train a group of GANs for 2 weeks on very efficient GPUs like \cite{progan}).

\subsection{Evalutation measures}
\label{evaluation_measures_gan}
Evaluating the performance of GANs is also a challenging task because the loss function does not represent the accuracy or quality of generated images. Many papers, like \cite{anime_characters_gan}, use the Fréchet Inception Distance (FID), but this measure requires a pre-trained CNN (Convolutional Neural Network) to extract relevant features. This network comes often from another project dedicated to classify images from the same dataset. Because I did not have access to such network, I decided to use PCA to evaluate the quality of GANs.

\subsubsection{Realism}
In order to evaluate the realism of generated images, I performed a Principal Component Analysis over the training data\footnote{The training data is composed of $N=11328$ images (see Section~\ref{brain_mri_dataset})} to extract 16 orthonormal eigenvectors $\mathbf{E}_i$ that represent the most the variations of the input distribution (each vector represents more than 1\% of the total variation and all vectors describe 55\% of the total variation).

Then, the realism measure $\rho$ is calculated by projecting $N=11328$ normalized generated images $\mathbf{G}$ onto the vector space induced by the selected covariance matrix eigenvectors, and retrieve the mean of the vector norms (see Equation~\ref{rho}).

\begin{equation}%
\label{rho}
\rho = \frac{1}{N}\sum_{\mathbf{G}} \sqrt{\sum_{i=1}^{16} (\mathbf{G}\cdot\mathbf{E}_i)^2 }
\end{equation}%

With this measure, my idea is to evaluate the necessity for generated images to be represented with eigenvectors orthogonal to the manifold composed of the main variations.

\subsubsection{Variation}
Along with realism, it is necessary to evaluate the amount of variation in the generated manifold to see if the model is not always generating the same images. To do so, a comparison between the total variance $\sigma$ of the input and output distributions is performed.

\begin{equation}%
\sigma = \Tr(XX^T)
\end{equation}%
with:
\vspace{-12pt}
\begin{itemize}
    \itemsep0em
    \item $X$: the matrix of observations
    \item $XX^T$: the covariance matrix
    \item $\Tr$: the trace of the matrix
\end{itemize}

\subsubsection{Diversity}
On the top of the global variation, it is necessary to detect whether parts of the variations are missing in the generated manifold (mode collapse problem see Figure~\ref{mode_collapse}). To do so, a comparison of the number of eigenvectors that represent more than 1\% of the total variation is performed.

\begin{equation}%
\delta = |\big\{\lambda_i, XX^T\mathbf{E_i} = \lambda_i\mathbf{E_i}\text{  and  }\lambda_i>\frac{\sigma}{100}\big\}|
\end{equation}%
with:
\vspace{-12pt}
\begin{itemize}
    \itemsep0em
    \item $XX^T$: the covariance matrix
    \item $\mathbf{E_i}$: the i\textsuperscript{th} eigenvector of the covariance matrix
    \item $\lambda_i$: the eigenvalue associated with $\mathbf{E_i}$
    \item $\sigma$: the total variation of the manifold
\end{itemize}

\subsubsection{Overfitting estimation}
Many machine learning models tend to overfit so a check is needed to verify that GANs are not just reproducing some input images but are really learning the underlying manifold. That is why a visual analysis of generated images from the interpolation of two random latent spaces is performed. If these images all look realistic, it means the model is not overfitting and is able to sample images that are not represented in the input observations.

\subsubsection{Computational efficiency}
Computational efficiency is also an important part of the evaluation to compare the different architectures. It is simply performed by comparing the time needed to train the models.

%% file: experiments.tex
\chapter{Experiments}
\label{experiments}
In this Chapter, we present the framework of the different experiments: the used datasets, the tested architectures and the tuned hyperparameters.

\section{MRI Segmentation}
\subsection{Training datasets and Data augmentation}
\begin{figure}
\centering
	\includegraphics{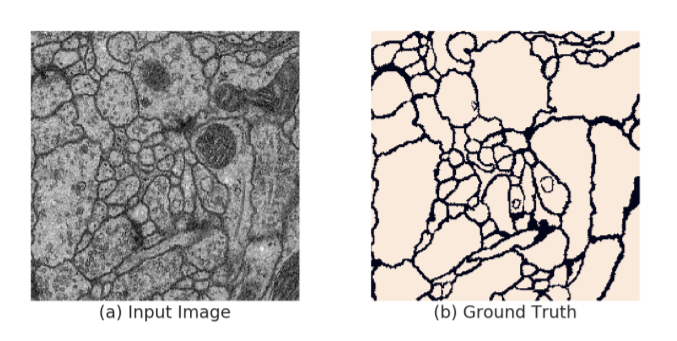}
    \caption{ISBI 2D EM segmentation challenge dataset \cite{isbi}}
    \label{isbi_dataset}
\end{figure}

In order to achieve high accuracy segmentation on MRIs, a pre-tuning process is carried out to reach state-of-the-art performance on the ISBI 2D EM segmentation challenge dataset (See Figure~\ref{isbi_dataset}). This dataset is composed of 30 grayscaled images of size $512 \times 512$ for training (with a black and white mask as ground truth) and 30 images for testing.

Then, the model is applied to 2D Knee MRIs from \citet{knee_dataset} (See Figure~\ref{dataset_knee}). These correspond to $128 \times 128$ grayscaled slices of 3D Knee MRIs from 30 patients, with 2 types of MRI sequences for each subject. The first experiment is carried out on 300 coronal planes\footnote{\label{note_split_train_test}2/3 is used for training, 1/3 is used for testing} (5 slices by sequence and by patient) in order to segment bones and cartilages of the tibia and the femur. The second experiment is performed on 660 sagittal planes (11 slices by sequence and by patient) to segment bones and cartilages of the tibia, the femur and the patella.

Data augmentation is an important cornerstone to achieve good segmentation performance, especially when very few data is available. For the ISBI dataset, many things are possible, as neuron cells do not have a particular shape: we will use rotations (from 0 to 180$^\circ$), horizontal flips, shifts (up to 10\%) and zooms (up to 20\%). However, there are more constraints on Knee MRIs because the shape of bones and cartilages matters: we will use rotations (from 0 to 5$^\circ$) and shifts (up to 5\%) only.

\subsection{Architecture and Hyperparameter tuning}
The architecture used for segmentation is a U-net as shown in Table~\ref{architecture_unet}. Batch Normalization (BN) and ReLU activation functions are used after each Convolution layer. Moreover, the downsampling layers in the encoder correspond to maxpooling layers.

The tuning part focuses on the use of Dropout, the use of Batch Normalization, the number of filters, the batch size, the number of epochs, etc (See Section~\ref{segmentation_pretuning}).

\begin{table}
    \centering
    \caption{U-net Architecture}
    \label{architecture_unet}
    \scriptsize
    \setlength\extrarowheight{2pt}
    \begin{tabular}[t]{|c@{ $\leftarrow$ }c|c|c@{ $\times$ }c@{ $\times$ }c|}%
        \hline
        \multicolumn{2}{|c|}{\textbf{Encoder}} & Act. & \multicolumn{3}{c|}{Output shape} \\
        \hline
        \multicolumn{2}{|c|}{Input image} & - & 1 & 128 & 128 \\
        \hline
        \multicolumn{2}{|c|}{Conv $3 \times 3$} & BN+ReLU & 32 & 128 & 128 \\
        \textit{$l_1$} & Conv $3 \times 3$ & BN+ReLU & 32 & 128 & 128 \\
        \multicolumn{2}{|c|}{Downsample} & - & 32 & 64 & 64 \\
        \hline
        \multicolumn{2}{|c|}{Conv $3 \times 3$} & BN+ReLU & 64 & 64 & 64 \\
        \textit{$l_2$} & Conv $3 \times 3$ & BN+ReLU & 64 & 64 & 64 \\
        \multicolumn{2}{|c|}{Downsample} & - & 64 & 32 & 32 \\
        \hline
        \multicolumn{2}{|c|}{Conv $3 \times 3$} & BN+ReLU & 128 & 32 & 32 \\
        \textit{$l_3$} & Conv $3 \times 3$ & BN+ReLU & 128 & 32 & 32 \\
        \multicolumn{2}{|c|}{Downsample} & - & 128 & 16 & 16 \\
        \hline
        \multicolumn{2}{|c|}{Conv $3 \times 3$} & BN+ReLU & 256 & 16 & 16 \\
        \textit{$l_4$} & Conv $3 \times 3$ & BN+ReLU & 256 & 16 & 16 \\
        \multicolumn{2}{|c|}{Downsample} & - & 256 & 8 & 8 \\
        \hline
        \multicolumn{2}{|c|}{Conv $3 \times 3$} & BN+ReLU & 512 & 8 & 8 \\
        \multicolumn{2}{|c|}{Conv $3 \times 3$} & BN+ReLU & 512 & 8 & 8 \\
        \hline
    \end{tabular}%
    \hspace{3pt}%
    \begin{tabular}[t]{|c|c|c@{ $\times$ }c@{ $\times$ }c|}%
        \hline
        \textbf{Decoder} & Act. & \multicolumn{3}{c|}{Output shape} \\
        \hline
        Conv Trans $3 \times 3$ & - & 256 & 16 & 16 \\
        Concatenate \textit{$l_4$} & - & 512 & 16 & 16 \\
        Conv $3 \times 3$ & BN+ReLU & 256 & 16 & 16 \\
        Conv $3 \times 3$ & BN+ReLU & 256 & 16 & 16 \\
        \hline
        Conv Trans $3 \times 3$ & - & 128 & 32 & 32 \\
        Concatenate \textit{$l_3$} & - & 256 & 32 & 32 \\
        Conv $3 \times 3$ & BN+ReLU & 128 & 32 & 32 \\
        Conv $3 \times 3$ & BN+ReLU & 128 & 32 & 32 \\
        \hline
        Conv Trans $3 \times 3$ & - & 64 & 64 & 64 \\
        Concatenate \textit{$l_2$} & - & 128 & 64 & 64 \\
        Conv $3 \times 3$ & BN+ReLU & 64 & 64 & 64 \\
        Conv $3 \times 3$ & BN+ReLU & 64 & 64 & 64 \\
        \hline
        Conv Trans $3 \times 3$ & - & 32 & 128 & 128 \\
        Concatenate \textit{$l_1$} & - & 64 & 128 & 128 \\
        Conv $3 \times 3$ & BN+ReLU & 32 & 128 & 128 \\
        Conv $3 \times 3$ & BN+ReLU & 32 & 128 & 128 \\
        \hline
        Conv $1 \times 1$ & Sigmoid & 1 & 128 & 128 \\
        \hline
    \end{tabular}%
\end{table}

\section{MRI Generation}
\label{mri_generation_experiment}
\subsection{Training dataset}
\label{brain_mri_dataset}
The dataset used to train the generative models is composed of brain MRIs from the OASIS (Open Access
Series of Imaging Studies) dataset \cite{oasis}. It is composed of 11328 grayscaled images of resolution $256 \times 256$ that are rescaled in the range $[-1, 1]$ (see Figure~\ref{dataset_brain}). These 2D images come from 3D MRIs of 354 different patients where 32 slices have been extracted.

\subsection{Architectures}
We consider three main GAN architectures as basis for fine tuning and experimentation: a Deep Convolution GAN (DCGAN), a Super Resolution Residual GAN (SRResGAN) and a Progressive GAN (ProGAN). These stable models have been chosen after a pre-tuning process (See Section~\ref{gan_pretuning}).

\subsubsection{DCGAN}
The DCGAN architecture is inspired by the work of \citet{dcgan}. The Generator begins with a dense layer that reshapes the 256-long input latent vector, drawn from a uniform distribution $\mathcal{U}(-1, 1)$, to a $256 \times 8 \times 8$ tensor. Then, 9 transpose convolution layers with $5 \times 5$ kernels upscale the image until it reaches the size $1 \times 256 \times 256$ (See Table~\ref{architecture_dcgan}). Batch normalization (BN) and ReLU activation functions are used after each convolution layer. On the other hand, the Discriminator is composed of 5 downscaling convolutions with $5 \times 5$ kernels, followed by 2 dense layers to get a unique scalar as output. Batch normalization (except for the first convolution and the dense layers) and Leaky ReLU activation functions with a slope of 0.2 are used.

\begin{table}
    \centering
    \caption{DCGAN Architecture}
    \label{architecture_dcgan}
    \scriptsize
    \setlength\extrarowheight{2pt}
    \begin{tabular}[t]{|c|c|c@{ $\times$ }c@{ $\times$ }c|}%
        \hline
        \textbf{Generator} & Act. & \multicolumn{3}{c|}{Output shape} \\
        \hline
        Latent vector & - & 256 & 1 & 1 \\
        Dense & BN+ReLU & 256 & 8 & 8 \\
        \hline
        Conv Trans $5 \times 5$ & BN+ReLU & 256 & 16 & 16 \\
        Conv Trans $5 \times 5$ & BN+ReLU & 256 & 16 & 16 \\
        \hline
        Conv Trans $5 \times 5$ & BN+ReLU & 256 & 32 & 32 \\
        Conv Trans $5 \times 5$ & BN+ReLU & 256 & 32 & 32 \\
        \hline
        Conv Trans $5 \times 5$ & BN+ReLU & 256 & 64 & 64 \\
        Conv Trans $5 \times 5$ & BN+ReLU & 256 & 64 & 64 \\
        \hline
        Conv Trans $5 \times 5$ & BN+ReLU & 128 & 128 & 128 \\
        Conv Trans $5 \times 5$ & BN+ReLU & 64 & 256 & 256 \\
        Conv Trans $5 \times 5$ & Tanh & 1 & 256 & 256 \\
        \hline
    \end{tabular}%
    \hspace{3pt}%
    \begin{tabular}[t]{|c|c|c@{ $\times$ }c@{ $\times$ }c|}%
        \hline
        \textbf{Discriminator} & Act. & \multicolumn{3}{c|}{Output shape} \\
        \hline
        Input image & - & 1 & 256 & 256 \\
        \hline
        Conv $5 \times 5$ & LReLU & 64 & 128 & 128 \\
        Conv $5 \times 5$ & BN+LReLU & 128 & 64 & 64 \\
        Conv $5 \times 5$ & BN+LReLU & 256 & 32 & 32 \\
        Conv $5 \times 5$ & BN+LReLU & 512 & 16 & 16 \\
        Conv $5 \times 5$ & BN+LReLU & 1024 & 8 & 8 \\
        \hline
        Dense & LReLU & 1024 & 1 & 1 \\
        Dense & Sigmoid & 1 & 1 & 1 \\
        \hline
    \end{tabular}%
\end{table}

\subsubsection{SRResGAN}
The SRResGAN architecture uses residual blocks to increase the depth of the network as presented by \citet{resgan} or \citet{anime_characters_gan}. Compared to the DCGAN with 10 layers, the SRResGAN Generator has 38 layers (see Table~\ref{architecture_srresgan}). The first layer of the Generator is a simple dense layer that upscales the 256-long input latent vector, drawn from a normal distribution $\mathcal{N}(0, 1)$ but normalized afterwards to belong to the unit hypersphere. Then, 16 residual blocks, composed of 2 convolution layers with $3 \times 3$ kernels, aim at generating relevant features to form the output image. 4 upscaling blocks, made of a convolution layer ($3 \times 3$ kernels) and a PixelShuffle layer, progressively increase the size of the output to reach the final convolution layer ($9 \times 9$ kernels) that produces the generated MRI. The PixelShuffle layer is a simple layer that transfers features from depth (channel dimension) into space blocks (width and height dimensions).

The Discriminator is also deeper with a convolution layer ($4 \times 4$ kernels) followed by 2 residual blocks repeated 6 times to reach a final convolution layer ($3 \times 3$ kernels) and a dense layer to output a scalar. As in the DCGAN, Batch Normalization and ReLU activation functions are used in the Generator. However, Batch Normalization has been removed from the Discriminator to avoid correlations within a generated batch. Leaky ReLU activation functions with a slope of 0.2 are still used in the Discriminator.

\begin{table}
    \centering
    \caption{SRResGAN Architecture}
    \label{architecture_srresgan}
    \scriptsize
    \setlength\extrarowheight{2pt}
    \begin{tabular}[t]{|c|c|c|c@{ $\times$ }c@{ $\times$ }c|}%
        \hline
        \multicolumn{2}{|c|}{\textbf{Generator}} & Act. & \multicolumn{3}{c|}{Output shape} \\
        \hline
        \multicolumn{2}{|c|}{Latent vector} & - & 256 & 1 & 1 \\
        \multicolumn{2}{|c|}{Dense} & BN+ReLU & 64 & 16 & 16 \\
        \hline
        \multirow{3}{*}{\textit{$\times$16}} & Conv $3 \times 3$ & BN+ReLU & 64 & 16 & 16 \\
        & Conv $3 \times 3$ & BN & 64 & 16 & 16 \\
        & Add & - & 64 & 16 & 16 \\
        \hline
        \multicolumn{2}{|c|}{-} & BN+ReLU & 64 & 16 & 16 \\
        \multicolumn{2}{|c|}{Add} & - & 64 & 16 & 16 \\
        \hline
        \multicolumn{2}{|c|}{Conv $3 \times 3$} & - & 256 & 16 & 16 \\
        \multicolumn{2}{|c|}{PixelShuffle} & BN+ReLU & 64 & 32 & 32 \\
        \hline
         \multicolumn{2}{|c|}{Conv $3 \times 3$} & - & 256 & 32 & 32 \\
        \multicolumn{2}{|c|}{PixelShuffle} & BN+ReLU & 64 & 64 & 64 \\
        \hline
         \multicolumn{2}{|c|}{Conv $3 \times 3$} & - & 256 & 64 & 64 \\
        \multicolumn{2}{|c|}{PixelShuffle} & BN+ReLU & 64 & 128 & 128 \\
        \hline
         \multicolumn{2}{|c|}{Conv $3 \times 3$} & - & 256 & 128 & 128 \\
        \multicolumn{2}{|c|}{PixelShuffle} & BN+ReLU & 64 & 256 & 256 \\
        \hline
        \multicolumn{2}{|c|}{Conv $9 \times 9$} & Tanh & 1 & 256 & 256 \\
        \hline
    \end{tabular}%
    \hspace{3pt}%
    \begin{tabular}[t]{|c|c|c|c@{ $\times$ }c@{ $\times$ }c|}%
        \hline
        \multicolumn{2}{|c|}{\textbf{Discriminator}} & Act. & \multicolumn{3}{c|}{Output shape} \\
        \hline
        \multicolumn{2}{|c|}{Input image} & - & 1 & 256 & 256 \\
        \hline
        \multicolumn{2}{|c|}{Conv $4 \times 4$} & LReLU & 32 & 128 & 128 \\
        \hline
        \multirow{3}{*}{\textit{$\times$2}} & Conv $3 \times 3$ & LReLU & 32 & 128 & 128 \\
        & Conv $3 \times 3$ & - & 32 & 128 & 128 \\
        & Add & LReLU & 32 & 128 & 128 \\
        \hline
        \multicolumn{2}{|c|}{Conv $4 \times 4$} & LReLU & 64 & 64 & 64 \\
        \hline
        \multirow{3}{*}{\textit{$\times$2}} & Conv $3 \times 3$ & LReLU & 64 & 64 & 64 \\
        & Conv $3 \times 3$ & - & 64 & 64 & 64 \\
        & Add & LReLU & 64 & 64 & 64 \\
        \hline
        \multicolumn{2}{|c|}{Conv $4 \times 4$} & LReLU & 128 & 32 & 32 \\
        \hline
        \multirow{3}{*}{\textit{$\times$2}} & Conv $3 \times 3$ & LReLU & 128 & 32 & 32 \\
        & Conv $3 \times 3$ & - & 128 & 32 & 32 \\
        & Add & LReLU & 128 & 32 & 32 \\
        \hline
        \multicolumn{2}{|c|}{Conv $4 \times 4$} & LReLU & 256 & 16 & 16 \\
        \hline
        \multirow{3}{*}{\textit{$\times$2}} & Conv $3 \times 3$ & LReLU & 256 & 16 & 16 \\
        & Conv $3 \times 3$ & - & 256 & 16 & 16 \\
        & Add & LReLU & 256 & 16 & 16 \\
        \hline
        \multicolumn{2}{|c|}{Conv $4 \times 4$} & LReLU & 512 & 8 & 8 \\
        \hline
        \multirow{3}{*}{\textit{$\times$2}} & Conv $3 \times 3$ & LReLU & 512 & 8 & 8 \\
        & Conv $3 \times 3$ & - & 512 & 8 & 8 \\
        & Add & LReLU & 512 & 8 & 8 \\
        \hline
        \multicolumn{2}{|c|}{Conv $4 \times 4$} & LReLU & 1024 & 4 & 4 \\
        \hline
        \multirow{3}{*}{\textit{$\times$2}} & Conv $3 \times 3$ & LReLU & 1024 & 4 & 4 \\
        & Conv $3 \times 3$ & - & 1024 & 4 & 4 \\
        & Add & LReLU & 1024 & 4 & 4 \\
        \hline
        \multicolumn{2}{|c|}{Conv $3 \times 3$} & LReLU & 2048 & 2 & 2 \\
        \multicolumn{2}{|c|}{Dense} & Sigmoid & 1 & 1 & 1 \\
        \hline
    \end{tabular}%
\end{table}

\subsubsection{ProGAN}
\begin{figure}
\centering
	\includegraphics{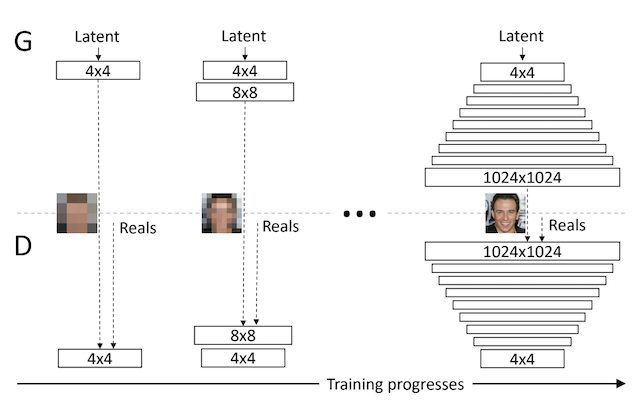}
    \caption{Progressive increase in resolution during ProGAN training \cite{progan}}
    \label{progan_increase_res}
\end{figure}

Finally, the last tested architecture is a Progressive GAN, introduced by \citet{progan}. The idea is to gradually deepen both the Generator and the Discriminator while increasing the input image resolution (see Figure~\ref{progan_increase_res}). In order to take advantage of what have been learnt from the previous step, a period of transition enables a smooth and linear progression (in the input images and in the network layers) to the next resolution.

\begin{table}
    \centering
    \caption{Progressive GAN Architecture}
    \label{architecture_proggan}
    \scriptsize
    \setlength\extrarowheight{2pt}
    \begin{tabular}[t]{|c|c|c@{ $\times$ }c@{ $\times$ }c|}
        \hline
        \textbf{Generator} & Act. & \multicolumn{3}{c|}{Output shape} \\
        \hline
        Latent vector & - & 512 & 1 & 1 \\
        Conv $4 \times 4$ & LReLU & 512 & 4 & 4 \\
        Conv $3 \times 3$ & PN+LReLU & 512 & 4 & 4 \\
        \hline
        Upsample & - & 512 & 8 & 8 \\
        Conv $5 \times 5$ & PN+LReLU & 512 & 8 & 8 \\
        Conv $5 \times 5$ & PN+LReLU & 512 & 8 & 8 \\
        \hline
        Upsample & - & 512 & 16 & 16 \\
        Conv $5 \times 5$ & PN+LReLU & 256 & 16 & 16 \\
        Conv $5 \times 5$ & PN+LReLU & 256 & 16 & 16 \\
        \hline
        Upsample & - & 256 & 32 & 32 \\
        Conv $5 \times 5$ & PN+LReLU & 128 & 32 & 32 \\
        Conv $5 \times 5$ & PN+LReLU & 128 & 32 & 32 \\
        \hline
        Upsample & - & 128 & 64 & 64 \\
        Conv $5 \times 5$ & PN+LReLU & 64 & 64 & 64 \\
        Conv $5 \times 5$ & PN+LReLU & 64 & 64 & 64 \\
        \hline
        Upsample & - & 64 & 128 & 128 \\
        Conv $5 \times 5$ & PN+LReLU & 32 & 128 & 128 \\
        Conv $5 \times 5$ & PN+LReLU & 32 & 128 & 128 \\
        \hline
        Upsample & - & 32 & 256 & 256 \\
        Conv $5 \times 5$ & PN+LReLU & 16 & 256 & 256 \\
        Conv $5 \times 5$ & PN+LReLU & 16 & 256 & 256 \\
        \hline
        Conv $1 \times 1$ & Tanh & 1 & 256 & 256 \\
        \hline
    \end{tabular}
    \begin{tabular}[t]{|c|c|c@{ $\times$ }c@{ $\times$ }c|}
        \hline
        \textbf{Discriminator} & Act. & \multicolumn{3}{c|}{Output shape} \\
        \hline
        Input image & - & 1 & 256 & 256 \\
        Conv $1 \times 1$ & LReLU & 16 & 256 & 256 \\
        \hline
        Conv $5 \times 5$ & LReLU & 16 & 256 & 256 \\
        Conv $5 \times 5$ & - & 32 & 256 & 256 \\
        Downsample & LReLU & 32 & 128 & 128 \\
        \hline
        Conv $5 \times 5$ & LReLU & 32 & 128 & 128 \\
        Conv $5 \times 5$ & - & 64 & 128 & 128 \\
        Downsample & LReLU & 64 & 64 & 64 \\
        \hline
        Conv $5 \times 5$ & LReLU & 64 & 64 & 64 \\
        Conv $5 \times 5$ & - & 128 & 64 & 64 \\
        Downsample & LReLU & 128 & 32 & 32 \\
        \hline
        Conv $5 \times 5$ & LReLU & 128 & 32 & 32 \\
        Conv $5 \times 5$ & - & 256 & 32 & 32 \\
        Downsample & LReLU & 256 & 16 & 16 \\
        \hline
        Conv $5 \times 5$ & LReLU & 256 & 16 & 16 \\
        Conv $5 \times 5$ & - & 512 & 16 & 16 \\
        Downsample & LReLU & 512 & 8 & 8 \\
        \hline
        Conv $5 \times 5$ & LReLU & 512 & 8 & 8 \\
        Conv $5 \times 5$ & - & 512 & 8 & 8 \\
        Downsample & LReLU & 512 & 4 & 4 \\
        \hline
        Minibatch std & - & 513 & 4 & 4 \\
        Conv $3 \times 3$ & LReLU & 512 & 4 & 4 \\
        Conv $4 \times 4$ & LReLU & 512 & 1 & 1 \\
        Dense & Sigmoid & 1 & 1 & 1 \\
        \hline
    \end{tabular}
\end{table}

The architecture is illustrated in Table~\ref{architecture_proggan}. The Generator is made of upsampling layers followed by 2 convolution layers with $5 \times 5$ kernels (except for the first 2 convolutions which have $4 \times 4$ and $3 \times 3$ kernels). This pattern is repeated 6 times before reaching the final convolution layer ($1 \times 1$ kernels).Leaky ReLU activation functions with a slope of 0.2 are used but Batch Normalization is replaced by Pixel Normalization (see Equation~\ref{pixel_normalization}) in order to avoid unhealthy competition with the Discriminator \cite{progan}.

\begin{equation}%
\label{pixel_normalization}%
    b_{x, y} = \frac{a_{x, y}}{\sqrt{\frac{1}{N}\sum_{j=0}^{N-1}(a_{x, y}^j)^2+\epsilon}}
\end{equation}%
with:
\vspace{-12pt}
\begin{itemize}
    \itemsep0em
    \item $b_{x, y}$: the output pixel at position $(x, y)$
    \item $a_{x, y}$: the input pixel at position $(x, y)$
    \item $N$: the number of feature maps (channels)
    \item $a_{x, y}^j$: the j\textsuperscript{th} feature value of the input pixel at position $(x, y)$
    \item $\epsilon = 10^{-8}$ (to avoid divison by zero)
\end{itemize}

The Discriminator is symmetrical to the Generator with 2 convolution layers with $5 \times 5$ kernels followed by a downsampling layer (average pooling), repeated 6 times before reaching the final 2 convolution layers (with $3 \times 3$ and $4 \times 4$ kernels) and the dense layer. No normalization is used but Leaky ReLU activation functions with a slope of 0.2 are still present. Moreover, a minibatch similarity layer is added before the final two convolutions in the Discriminator to increase diversity in the generated batch. This layer corresponds to a constant feature map representing the average over all feature maps of the standard deviation among all the generated spatial locations of the minibatch (part of the batch). In addition, the input noise vector is a 512-long array drawn from a normal distribution $\mathcal{N}(0, 1)$ and normalized afterwards to belong to the unit hypersphere.

\subsection{Loss functions}
In order to stabilize GAN convergence, we compare 5 different loss functions: the Original loss, LSGAN, WGAN, WGAN\_GP and DRAGAN.\\

The following notations are used thereafter:
\begin{itemize}
    \item the input noise (from latent space): $\mathbf{z} \sim p_\mathbf{z}(\mathbf{z})$ (uniform or normalized normal)
    \item the Generator output $G(\mathbf{z}) \sim p_g$
    \item the input data $\mathbf{x} \sim p_\mathrm{data}(\mathbf{x})$
    \item the ``probability'' $D(\mathbf{y})$, computed by the Discriminator, that $\mathbf{y}$ comes from $p_\mathrm{data}$ rather than $p_g$
\end{itemize}

\subsubsection{Original loss}
The Original loss, introduced by \citet{gan}, is inspired by the binary cross entropy loss but the Generator loss is modified to cope with vanishing gradients (see Equation~\ref{loss_gan}).

\begin{equation}%
\label{loss_gan}%
\begin{split}%
    L_G^\mathrm{GAN} &= -~\mathbb{E} \bigg[ \log \big( D(G(\mathbf{z})) \big) \bigg]\\
    L_D^\mathrm{GAN} &= -~\mathbb{E} \bigg[ \log \big( D(\mathbf{x}) \big) \bigg] - \mathbb{E} \bigg[ \log \big( 1-D(G(\mathbf{z})) \big) \bigg]
\end{split}%
\end{equation}%

\subsubsection{LSGAN}
The LSGAN loss uses Least Squares instead of logarithms to try to cope with vanishing gradients (See Equation~\ref{loss_lsgan}).

\begin{equation}%
\label{loss_lsgan}%
\begin{split}%
    L_G^\mathrm{LSGAN} &= \mathbb{E} \bigg[ \big( D(G(\mathbf{z})) - 1 \big)^2 \bigg]\\
    L_D^\mathrm{LSGAN} &= \mathbb{E} \bigg[ \big( D(\mathbf{x}) - 1 \big)^2 \bigg] + \mathbb{E} \bigg[ D(G(\mathbf{z}))^2 \bigg]
\end{split}%
\end{equation}%

\subsubsection{WGAN}
\begin{figure}
\centering
	\includegraphics[width=0.7\textwidth]{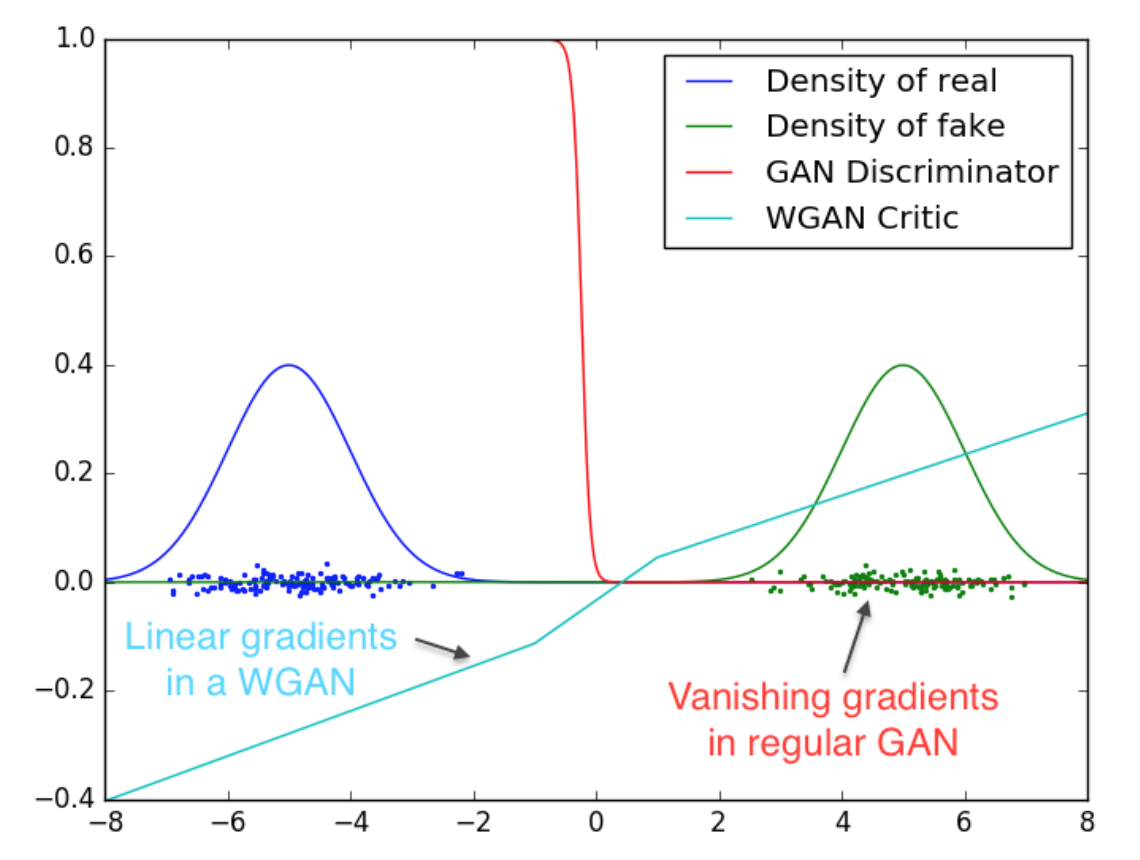}
    \caption{Gradient smoothing with WGAN loss \cite{wgan}}
    \label{wgan_smooth}
\end{figure}

WGAN gets also rid of logarithms (see Equation~\ref{loss_wgan}) and intends to achieve better convergence by replacing a discontinuous gradient by a smooth one (see Figure~\ref{wgan_smooth}). This helps the learning of networks through backpropagation. The method also involves clipping tensor weights to avoid exploding gradients and divergence.

\begin{equation}%
\label{loss_wgan}%
\begin{split}%
    L_G^\mathrm{WGAN} &= -~\mathbb{E} \big[ D(G(\mathbf{z})) \big]\\
    L_D^\mathrm{WGAN} &= -~\mathbb{E} \big[ D(\mathbf{x}) \big] + \mathbb{E} \big[ D(G(\mathbf{z})) \big]
\end{split}%
\end{equation}%

\subsubsection{WGAN\_GP}
A suggested improvement to the previous loss function is to remove weights clipping and add a regularization term to penalize huge gradients (See Equation~\ref{loss_wgan_gp}). This is especially important when dealing with deep networks according to \citet{wgan_gp}.

\begin{equation}%
\label{loss_wgan_gp}%
\begin{split}%
    L_G^\mathrm{WGAN\_GP} &= L_G^\mathrm{WGAN}\\
    L_D^\mathrm{WGAN\_GP} &= L_D^\mathrm{WGAN} + \lambda \mathbb{E} \bigg[ \big( \norm{\nabla D(\mathbf{x_m})} - 1 \big)^2 \bigg]\\
    \mathbf{x_m} &= \alpha \cdot \mathbf{x} + (1-\alpha) \cdot  G(\mathbf{z})
\end{split}%
\end{equation}%
with:
\vspace{-12pt}
\begin{itemize}
    \itemsep0em
    \item $\lambda$: a hyperparameter to balance the gradient penalty
    \item $\alpha \sim \mathcal{U}(0, 1)$: a random parameter that combines the real and fake images
\end{itemize}

\subsubsection{DRAGAN}
Finally, DRAGAN is another method that regulates the gradient but uses the Original loss as baseline and interpolates with random noise rather than a generated image (see Equation~\ref{loss_dragan}).

\begin{equation}%
\label{loss_dragan}%
\begin{split}%
    L_G^\mathrm{DRAGAN} &= L_G^\mathrm{GAN}\\
    L_D^\mathrm{DRAGAN} &= L_D^\mathrm{GAN} + \lambda \mathbb{E} \bigg[ \big( \norm{\nabla D(\mathbf{x_m'})} - 1 \big)^2 \bigg]\\
    \mathbf{x_m'} &= \alpha \cdot \mathbf{x} + (1-\alpha) \cdot  \mathbf{x_p}
\end{split}%
\end{equation}%
with:
\vspace{-12pt}
\begin{itemize}
    \itemsep0em
    \item $\lambda$: a hyperparameter to balance the gradient penalty
    \item $\alpha \sim \mathcal{U}(0, 1)$: a random parameter that combines the real images and random noise
    \item $\mathbf{x_p} \sim \mathcal{U}(0, \frac{1}{2}\sigma_\mathbf{x})$, a pixel-scaled random noise
    \item $\sigma_\mathbf{x}$, the standard deviation of the input image
\end{itemize}

%% file: results.tex
\chapter{Results}
\label{results}
This Chapter presents the results of pre-tuning hyperparameters and the final evaluations of the best selected architectures.

\section{MRI Segmentation}
\subsection{Pre-tuning}
\label{segmentation_pretuning}
In order to reach state-of-the-art performance, pre-tuning is performed on the ISBI dataset. The different experiments and their results are summarized in Table~\ref{hyperparameter_tuning_segmentation} in chronological order (from left to right). A Tesla K80 GPU was used for the experiments.

Several things can be noticed from the hyperparameter tuning. Adding dropout after each block does not improve accuracy. The Adam Optimizer preforms better than Stochastic Gradient Descent and using Dice coefficient as loss function instead of binary cross entropy improves the performance. Moreover, Batch Normalization increases the accuracy and the higher the input resolution, the higher the final score. Finally, one can notice that reducing the initial number of filters from 64 to 32 does not penalize the accuracy much but highly increases time efficiency (because the number of parameters to train is 4 times smaller).

\begin{table}
    \centering
    \caption{U-net Hyperparameter tuning}
    \label{hyperparameter_tuning_segmentation}
    \scriptsize
    \setlength\extrarowheight{2pt}
    \begin{tabular}{|c|c|c|c|c|c|c|c|}
        \hline
        \multirow{10}{*}{\rotatebox[origin=c]{90}{Inputs}} & Input size & 256 & 256 & 256 & 256 & 256 & 256 \\
        \cline{2-8}
        & Filters & 64 & 64 & 64 & 64 & \textcolor{red}{32} & 64 \\
        \cline{2-8}
        & Dropout & 0 & 0 & \textcolor{red}{0.5} & 0 & 0 & 0 \\
        \cline{2-8}
        & Batch normalization & 0 & \textcolor{red}{1} & 1 & 1 & 1 & 1 \\
        \cline{2-8}
        & Batch size & 2 & 2 & 2 & 2 & 2 & \textcolor{red}{4} \\
        \cline{2-8}
        & Steps per epoch & 1000 & 1000 & 1000 & 1000 & 1000 & 1000 \\
        \cline{2-8}
        & Epochs & 10 & 10 & 10 & 10 & 10 & 10 \\
        \cline{2-8}
        & Learning rate & 1e-4 & 1e-4 & 1e-4 & \textcolor{red}{1e-3} & 1e-3 & 1e-3 \\
        \cline{2-8}
        & Loss function & \texttt{BCE} & \texttt{BCE} & \texttt{BCE} & \texttt{BCE} & \texttt{BCE} & \texttt{BCE} \\
        \cline{2-8}
        & Optimizer & \texttt{A} & \texttt{A} & \texttt{A} & \texttt{A} & \texttt{A} & \texttt{A} \\
        \hline
        \hline
        \multirow{7}{*}{\rotatebox[origin=c]{90}{Outputs}} & Total params & 34,512,193 & 34,535,745 & 34,535,745 & 34,535,745 & 8,641,697 & 34,535,745 \\
        \cline{2-8}
        & Trainable params & 34,512,193 & 34,523,969 & 34,523,969 & 34,523,969 & 8,635,809 & 34,523,969 \\
        \cline{2-8}
        & Training Time (s) & 2289.4 & 2588.7 & 2987.2 & 2615.9 & \textcolor{OliveGreen}{1110.8} & 4395.5 \\
        \cline{2-8}
        & Final training loss & 0.1986 & 0.2053 & 0.2029 & 0.1852 & 0.1838 & 0.1839 \\
        \cline{2-8}
        & Final training accuracy & 0.8182 & 0.8170 & 0.8168 & 0.8230 & 0.8234 & 0.8235 \\
        \cline{2-8}
        & Final validation loss & 0.2607 & 0.2320 & 0.2535 & 0.2233 & 0.2253 & 0.2259 \\
        \cline{2-8}
        & Final validation accuracy & 0.7893 & \textcolor{OliveGreen}{0.7908} & 0.7896 & \textcolor{OliveGreen}{0.7933} & 0.7920 & 0.7918 \\
        \hline
    \end{tabular}\\
    \vspace{10pt}
    \begin{tabular}{|c|c|c|c|c|c|c|c|}
        \hline
        \multirow{10}{*}{\rotatebox[origin=c]{90}{Inputs}} & Input size & 256 & \textcolor{red}{512} & \textcolor{red}{512} & 512 & 512 & 512 \\
        \cline{2-8}
        & Filters & 64 & 64 & 64 & 64 & 64 & 64 \\
        \cline{2-8}
        & Dropout & 0 & 0 & 0 & 0 & 0 & 0 \\
        \cline{2-8}
        & Batch normalization & 1 & 1 & 1 & 1 & 1 & 1 \\
        \cline{2-8}
        & Batch size & \textcolor{red}{4} & 4 & \textcolor{red}{2} & 2 & 2 & 2 \\
        \cline{2-8}
        & Steps per epoch & \textcolor{red}{500} & 500 & \textcolor{red}{1000} & 1000 & 1000 & 1000 \\
        \cline{2-8}
        & Epochs & 10 & 10 & 10 & 10 & 10 & \textcolor{red}{20} \\
        \cline{2-8}
        & Learning rate & 1e-3 & 1e-3 & 1e-3 & 1e-3 & 1e-3 & 1e-3 \\
        \cline{2-8}
        & Loss function & \texttt{BCE} & \texttt{BCE} & \texttt{BCE} & \texttt{BCE} & \textcolor{red}{\texttt{DSC}} & \texttt{DSC} \\
        \cline{2-8}
        & Optimizer & \texttt{A} & \texttt{A} & \texttt{A} & \textcolor{red}{\texttt{SGD++}} & \texttt{A} & \texttt{A} \\
        \hline
        \hline
        \multirow{7}{*}{\rotatebox[origin=c]{90}{Outputs}} & Total params & 34,535,745 & \textsc{Ram}\_Error & 34,535,745 & 34,535,745 & 34,535,745 & 34,535,745 \\
        \cline{2-8}
        & Trainable params & 34,523,969 & \textsc{Ram}\_Error & 34,523,969 & 34,523,969 & 34,523,969 & 34,523,969 \\
        \cline{2-8}
        & Training Time (s) & 2223.5 & \textsc{Ram}\_Error & 9006.8 & 8993.2 & 9116.5 & 18301.2 \\
        \cline{2-8}
        & Final training loss & 0.1899 & \textsc{Ram}\_Error & 0.1710 & 0.2034 & 0.0444 & 0.0428 \\
        \cline{2-8}
        & Final training accuracy & 0.8212 & \textsc{Ram}\_Error & 0.8794 & 0.8667 & 0.8851 & 0.8875 \\
        \cline{2-8}
        & Final validation loss & 0.2233 & \textsc{Ram}\_Error & 0.2041 & 0.2218 & 0.0565 & 0.0569 \\
        \cline{2-8}
        & Final validation accuracy & 0.7925 & \textsc{Ram}\_Error & \textcolor{OliveGreen}{0.8598} & 0.8506 & \textcolor{OliveGreen}{0.8611} & 0.8611 \\
        \hline
    \end{tabular}
    \begin{flushright}
    \texttt{BCE} = Binary Cross Entropy; \texttt{DSC} = Dice S{\o}rensen Coefficient; \texttt{A} = Adam\\
    \texttt{SGD++} = Stochastic Gradient Descent with Momentum=0.9 and Nesterov momentum
    \end{flushright}
\end{table}

\subsection{Final results}
The final hyperparameters chosen to be tested on the knee MRI dataset are summarized in Table~\ref{chosen_hyperparameters_segmentation}.

\begin{table}
    \centering
    \caption{Chosen Hyperparameters for MRI Segmentation}
    \label{chosen_hyperparameters_segmentation}
    \scriptsize
    \setlength\extrarowheight{2pt}
    \begin{tabular}{|c|c|c|}
        \hline
        \multirow{10}{*}{\rotatebox[origin=c]{90}{Inputs}} & Input size & 128 \\
        \cline{2-3}
        & Filters & 32 \\
        \cline{2-3}
        & Dropout & 0 \\
        \cline{2-3}
        & Batch normalization & 1 \\
        \cline{2-3}
        & Batch size & 8 \\
        \cline{2-3}
        & Steps per epoch & 250 \\
        \cline{2-3}
        & Epochs & 10 \\
        \cline{2-3}
        & Learning rate & 1e-3 \\
        \cline{2-3}
        & Loss function & \texttt{DSC} \\
        \cline{2-3}
        & Optimizer & \texttt{A} \\
        \hline
        \hline
        \multirow{3}{*}{\rotatebox[origin=c]{90}{Outputs}} & Total params & 8,641,697 \\
        \cline{2-3}
        & Trainable params & 8,635,809 \\
        \cline{2-3}
        & Training Time (s) & 273.7 \\
        \hline
    \end{tabular}
\end{table}

The performances on the different bones and cartilages are presented in Table~\ref{segmentation_results_coronal} for the coronal plane and Table~\ref{segmentation_results_sagittal} for the sagittal plane. Note that the patella is not displayed on the coronal plane and the tibia bone cannot be seen in the sagittal plane. Results show state-of-the-art performance.

\begin{table}
    \centering
    \caption{Segmentation Accuracy for Coronal plane}
    \label{segmentation_results_coronal}
    \small
    \setlength\extrarowheight{2pt}
    \begin{tabular}{|c|c|c|c|c|}
        \hline
        & \multicolumn{2}{c|}{\textbf{Cartilage}} & \multicolumn{2}{c|}{\textbf{Bone}} \\
        \hline
        Results & Femur & Tibia & Femur & Tibia \\
        \hline
        Final training loss & 0.1663 & 0.1790 & 0.0091 & 0.0147 \\
        \hline
        Final training accuracy & 0.9813 & 0.9861 & 0.9777 & 0.9840 \\
        \hline
        Final validation loss & 0.1358 & 0.1534 & 0.0141 & 0.0214 \\
        \hline
        Final validation accuracy & 0.9834 & 0.9852 & 0.9739 & 0.9816 \\
        \hline
    \end{tabular}
\end{table}

\begin{table}
    \centering
    \caption{Segmentation Accuracy for Sagittal plane}
    \label{segmentation_results_sagittal}
    \small
    \setlength\extrarowheight{2pt}
    \begin{tabular}{|c|c|c|c|c|c|}
        \hline
        & \multicolumn{2}{c|}{\textbf{Cartilage}} & \multicolumn{3}{c|}{\textbf{Bone}} \\
        \hline
        Results & Femur & Patella & Femur & Tibia & Patella \\
        \hline
        Final training loss & 0.0840 & 0.1150 & 0.0137 & 0.0127 & 0.0373 \\
        \hline
        Final training accuracy & 0.9983 & 0.9985 & 0.9975 & 0.9979 & 0.9975 \\
        \hline
        Final validation loss & 0.0917 & 0.0809 & 0.0211 & 0.0252 & 0.0542 \\
        \hline
        Final validation accuracy & 0.9981 & 0.9989 & 0.9963 & 0.9960 & 0.9961 \\
        \hline
    \end{tabular}
\end{table}

\begin{figure}
\centering
	\includegraphics[width=1.0\textwidth]{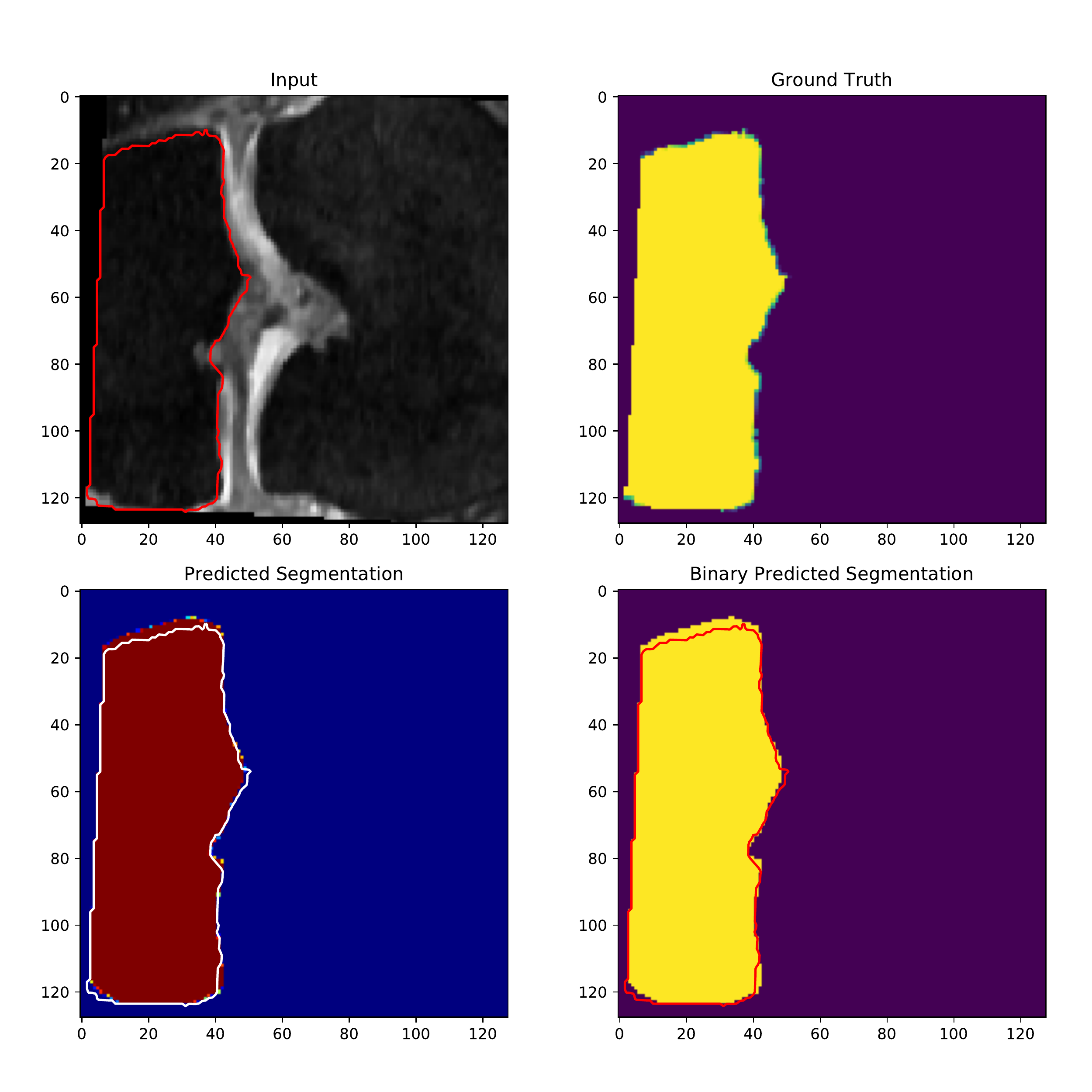}
    \caption{Segmentation of the tibia bone with U-net in the coronal plane}
    \label{plot_segmentation_mri}
\end{figure}

The visual segmentation output is illustrated in Figure~\ref{plot_segmentation_mri}. We can see that the ground truth (red line in the Figure) is not really accurate in some places. This is due to the fact that the segmentation is semi-automated on the 3D MRIs to fasten labelling. One can even notice that the U-net segmentation is more accurate than the ground truth in some parts of the images. More visual results are presented in Appendix~\ref{segmentation_results_appendix}.

\begin{figure}
\centering
	\includegraphics[width=0.9\textwidth]{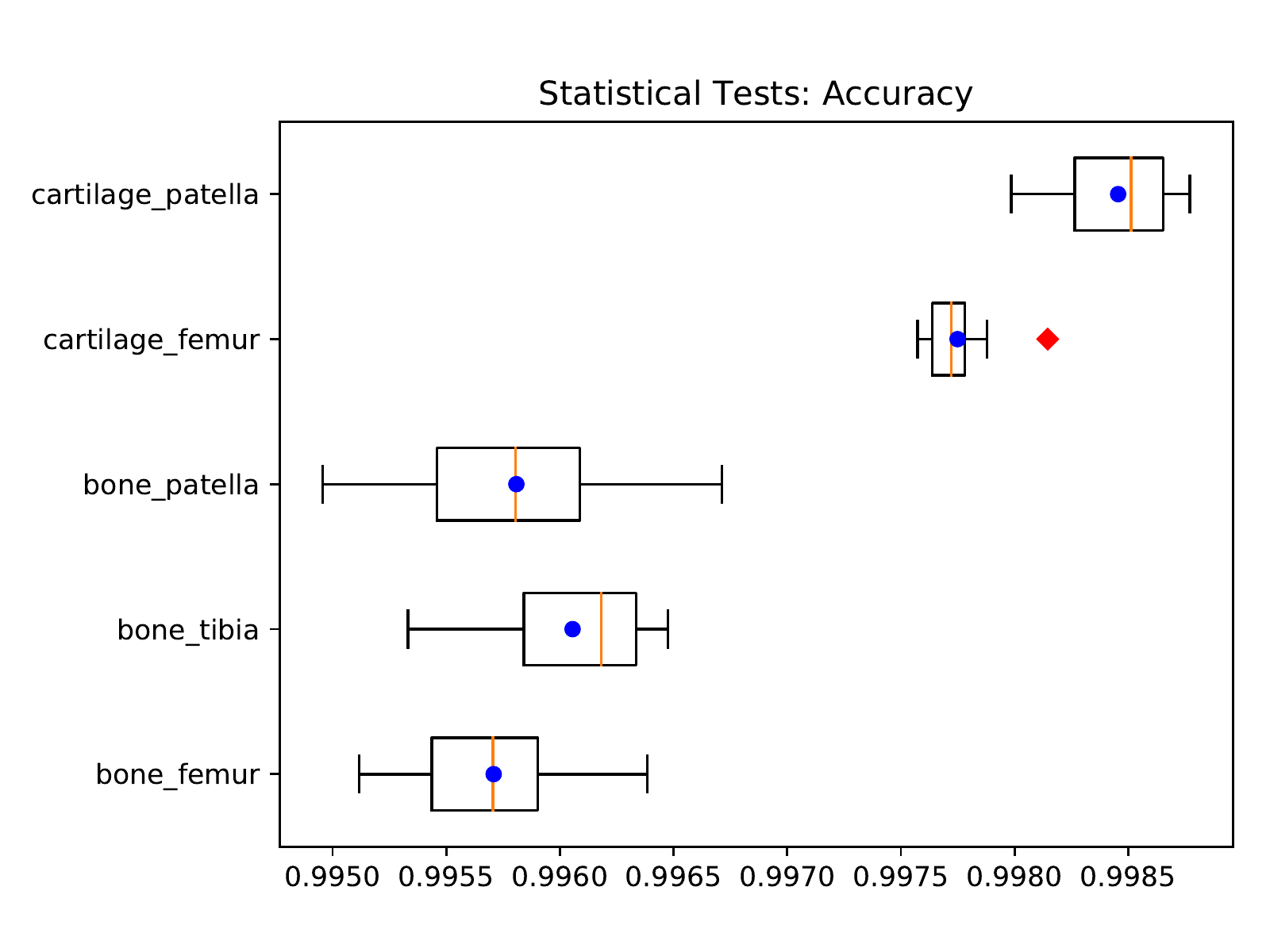}
    \caption{Accuracy boxplots for the sagittal plane}
    \label{boxplot_accuracy}
\end{figure}

\begin{figure}
\centering
	\includegraphics[width=0.9\textwidth]{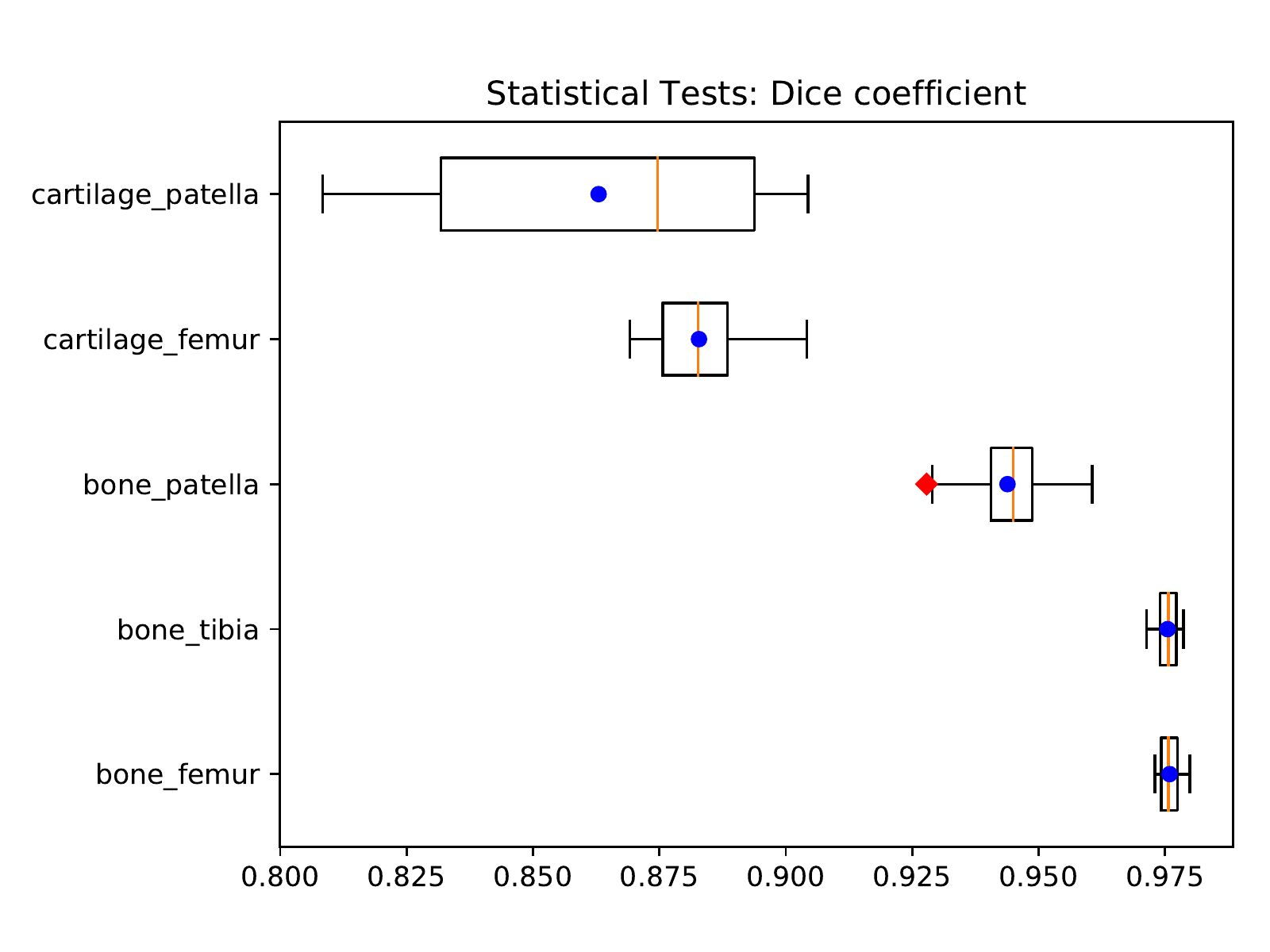}
    \caption{Dice coefficient boxplots for the sagittal plane}
    \label{boxplot_dice}
\end{figure}

Finally, in order to take into account the variations in performance between different batchs, boxplots are displayed in Figures~\ref{boxplot_accuracy} and \ref{boxplot_dice}. Red dots correspond to outliers and blue dots represent mean measurements. The statistical tests show high performance and very few variations.

\begin{figure}
\centering
	\includegraphics[width=0.9\textwidth]{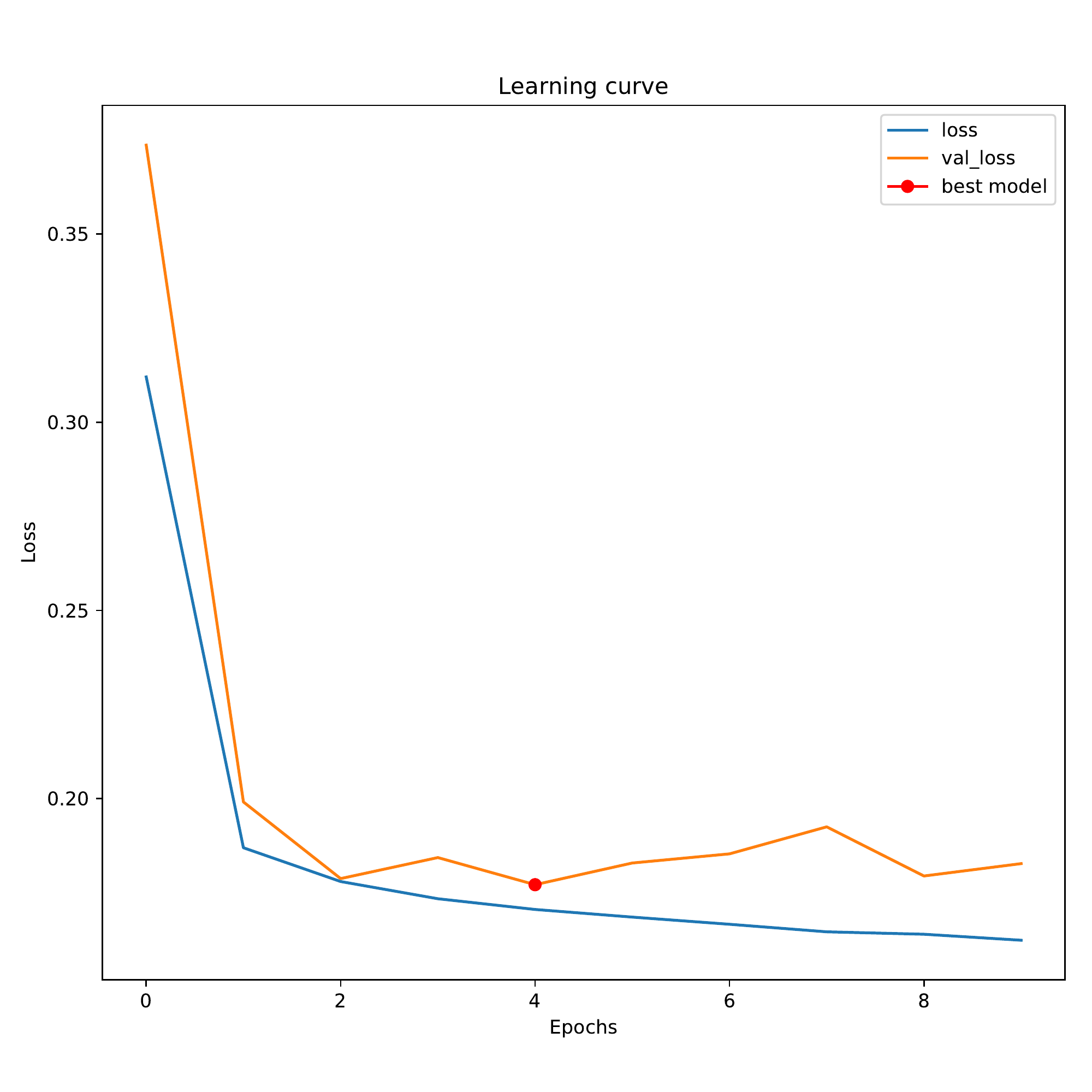}
    \caption{Training loss wrt epochs for the femur cartilage\\
    segmentation in the coronal plane}
    \label{learning_curve_segmentation}
\end{figure}

The Figure~\ref{learning_curve_segmentation} shows the decrease of the loss with respect to the epochs. In order to avoid overfitting, the chosen model is the one minimizing the validation loss.

\newpage
\section{MRI Generation}
\subsection{Qualitative Pre-tuning}
\label{gan_pretuning}
Before being able to compare quantitatively several networks, an important pre-tuning process is necessary. The aim is to reach satisfactory visual results in a relatively short training time.

\subsubsection{DCGAN}
The different modifications of DCGAN for pre-tuning are presented in Table~\ref{dcgan_pretuning}. All models are trained with 20 epochs and a Telsa K80 GPU.

DCGAN v1 corresponds to the architecture of Table~\ref{architecture_dcgan} with some modifications: images are scaled between 0 and 1, the loss function is LSGAN, the Generator/Discriminator training rate is 2:1 and the batch size is 64.

DCGAN v2 is the same as DCGAN v1 with images scaled between -1 and 1, the Original loss function, one-sided label smoothing and a Generator/Discriminator training rate of 3:1.

\begin{table}
    \centering
    \caption{DCGAN pre-tuning}
    \label{dcgan_pretuning}
    \scriptsize
    \setlength\extrarowheight{2pt}
    \newcolumntype{C}{>{\centering\let\newline\\\arraybackslash\hspace{0pt}}p{12cm}}
    \begin{tabular}{|c|C|c|}
        \hline
        \textbf{Baseline} & \textbf{Tested changes} & \textbf{Improv.} \\
        \specialrule{1pt}{0pt}{0pt}
        \multirow{10}{*}{DCGAN v1} & Ouput between -1 and 1 & Yes \\
        \cline{2-3}
        & Input Gaussian noise & No \\
        \cline{2-3}
        & \cellcolor{LightRed} Batch size of 32 & \cellcolor{LightRed} Worse \\
        \cline{2-3}
        & \cellcolor{LightRed} Rate of Generator training 1:1 & \cellcolor{LightRed} Worse \\
        \cline{2-3}
        & \cellcolor{LightGreen} Rate of Generator training 3:1 & \cellcolor{LightGreen} Better\\
        \cline{2-3}
        & Replace dense layers in Discriminator by convolutions & No \\
        \cline{2-3}
        & Replace dense layers in Discriminator by convolutions with Batch Normalization & No \\
        \cline{2-3}
        & Add one deconvolution layer in the generator & No \\
        \cline{2-3}
        & \cellcolor{LightGreen} Change loss function to original GAN loss & \cellcolor{LightGreen} Better \\
        \cline{2-3}
        & \cellcolor{LightGreen} Change loss function to original GAN loss + one-sided label smoothing & \cellcolor{LightGreen} Better \\
        \specialrule{1pt}{0pt}{0pt}
        \multirow{8}{*}{DCGAN v2} & Use Normalized Gaussian noise & No \\
        \cline{2-3}
        & Replace original loss by LSGAN & No \\
        \cline{2-3}
        & \cellcolor{LightYellow} Replace original loss by WGAN & \cellcolor{LightYellow} ? \\
        \cline{2-3}
        & \cellcolor{LightYellow} Replace original loss by WGAN\_GP & \cellcolor{LightYellow} ? \\
        \cline{2-3}
        & Replace original loss by DRAGAN & Yes \\
        \cline{2-3}
        & Increase Generator learning rate instead of iteration & No \\
        \cline{2-3}
        & Add noise to the real and generated images before feeding them into the Discriminator & No \\
        \cline{2-3}
        & \cellcolor{LightYellow} Add minibatch similarity layer & \cellcolor{LightYellow} ? \\
        \specialrule{1pt}{0pt}{0pt}
    \end{tabular}
\end{table}

\subsubsection{SRResGAN}
The different modifications of SRResGAN for pre-tuning are presented in Table~\ref{srresgan_pretuning}. All models are trained with 20 epochs and a Telsa K80 GPU.

SRResGAN v1 correponds to the architecture of Table~\ref{architecture_srresgan} with DRAGAN loss function, uniform input noise and Normal initialization for weight tensors.

SRResGAN v2 is the same as SRResGAN v1 with the Original loss function, additional Gaussian noise in the Discriminator, one-sided label smoothing and normalized normal input noise.

Finally, SRResGAN v3 is an improvement of SRResGAN v2 with a He Normal initialization of weight tensors.

\begin{table}
    \centering
    \caption{SRResGAN pre-tuning}
    \label{srresgan_pretuning}
    \scriptsize
    \setlength\extrarowheight{2pt}
    \newcolumntype{C}{>{\centering\let\newline\\\arraybackslash\hspace{0pt}}p{11cm}}
    \begin{tabular}{|c|C|c|}
        \hline
        \textbf{Baseline} & \textbf{Tested changes} & \textbf{Improv.} \\
        \specialrule{1pt}{0pt}{0pt}
        \multirow{9}{*}{SRResGAN v1} & \cellcolor{LightYellow} Increase size of residual blocks in the Generator & \cellcolor{LightYellow} ? \\
        \cline{2-3}
        & \cellcolor{LightYellow} Use Gaussian input noise instead of Uniform noise & \cellcolor{LightYellow} ? \\
        \cline{2-3}
        & \cellcolor{LightYellow} Replace DRAGAN loss by original GAN loss & \cellcolor{LightYellow} ? \\
        \cline{2-3}
        & \cellcolor{LightYellow} Decrease $\lambda_{gp}$ from 0.5 to 0.25 & \cellcolor{LightYellow} ? \\
        \cline{2-3}
        & \cellcolor{LightYellow} Increase $\lambda_{gp}$ from 0.5 to 0.75 & \cellcolor{LightYellow} ? \\
        \cline{2-3}
        & One-sided label smoothing & Yes \\
        \cline{2-3}
        & \cellcolor{LightYellow} Replace PixelShuffle layers by Transpose convolutions & \cellcolor{LightYellow} ? \\
        \cline{2-3}
        & \cellcolor{LightGreen} Change kernel size from 3 to 5 & \cellcolor{LightGreen} Better \\
        \cline{2-3}
        & \cellcolor{LightYellow} Use Batch Normalization in Discriminator & \cellcolor{LightYellow} ? \\
        \specialrule{1pt}{0pt}{0pt}
        \multirow{3}{*}{SRResGAN v2} & \cellcolor{LightYellow} Change kernel size from 3 to 5 + batch 32 & \cellcolor{LightYellow} ? \\
        \cline{2-3}
        & \cellcolor{LightGreen} ReLU and not LReLU in Discriminator residual blocks & \cellcolor{LightGreen} Better \\
        \cline{2-3}
        & \cellcolor{LightGreen} Use He Normal initializer instead of Normal & \cellcolor{LightGreen} Better \\
        \specialrule{1pt}{0pt}{0pt}
        \multirow{7}{*}{SRResGAN v3} & ReLU and not LReLU in discriminator residual blocks & No \\
        \cline{2-3}
        & Change kernel size from 3 to 5 + batch 32 & Yes \\
        \cline{2-3}
        & Use normalized Uniform noise input & No \\
        \cline{2-3}
        & Remove noise in the Discriminator & No \\
        \cline{2-3}
        & \cellcolor{LightRed} Replace original loss by LSGAN & \cellcolor{LightRed} Worse \\
        \cline{2-3}
        & \cellcolor{LightRed} Replace original loss by WGAN & \cellcolor{LightRed} Worse \\
        \cline{2-3}
        & \cellcolor{LightYellow} Increase size of residual blocks in the Generator & \cellcolor{LightYellow} ? \\
        \specialrule{1pt}{0pt}{0pt}
    \end{tabular}
\end{table}

\subsubsection{ProGAN}
The different modifications of ProGAN for pre-tuning are presented in Table~\ref{progan_pretuning}. All models are trained with 20 epochs (for the final resolution) and a Telsa K80 GPU.

ProGAN v1 has the same architecture as in Table~\ref{architecture_proggan} with a batch size of 16, WGAN\_GP loss function, the Adam Optimizer with a learning rate of 0.001 and beta equal to 0, one epoch for low resolution and transition steps, ReLU activation functions in the Generator and Discriminator and $3 \times 3$ kernel size.

ProGAN v2 improves ProGAN v1 by increasing the batch size to 64, giving 20 training epochs for low resolution and transition steps and using Leaky ReLU activation functions in the Generator and Discriminator.

\begin{table}
    \centering
    \caption{Progressive GAN pre-tuning}
    \label{progan_pretuning}
    \scriptsize
    \setlength\extrarowheight{2pt}
    \newcolumntype{C}{>{\centering\let\newline\\\arraybackslash\hspace{0pt}}p{12cm}}
    \begin{tabular}{|c|C|c|}
        \hline
        \textbf{Baseline} & \textbf{Tested changes} & \textbf{Improv.} \\
        \specialrule{1pt}{0pt}{0pt}
        \multirow{20}{*}{ProGAN v1} & \cellcolor{LightYellow} Change batch 16 into batch 32 & \cellcolor{LightYellow} ? \\
        \cline{2-3}
        & \cellcolor{LightYellow} Replace upscale and downscale layers by strides in convolutions & \cellcolor{LightYellow} ? \\
        \cline{2-3}
        & \cellcolor{LightYellow} Delete normalization of input noise & \cellcolor{LightYellow} ? \\
        \cline{2-3}
        & \cellcolor{LightYellow} Delete $\epsilon_{drift}$ & \cellcolor{LightYellow} ? \\
        \cline{2-3}
        & \cellcolor{LightYellow} Remove minibatch similarity layer & \cellcolor{LightYellow} ? \\
        \cline{2-3}
        & \cellcolor{LightYellow} True upscale and downscale & \cellcolor{LightYellow} ? \\
        \cline{2-3}
        & \cellcolor{LightYellow} Remove Pixel Normalization & \cellcolor{LightYellow} ? \\
        \cline{2-3}
        & \cellcolor{LightYellow} Increase epochs for intermediate layers from 1 to 2 & \cellcolor{LightYellow} ? \\
        \cline{2-3}
        & Decrease learning rate to 0.0002 and increase $\beta_1$ to 0.5 & No \\
        \cline{2-3}
        & \cellcolor{LightYellow} Increase epochs during transition from 1 to 20 & \cellcolor{LightYellow} ? \\
        \cline{2-3}
        & Increase Generator/Discriminator training rate from 1:1 to 3:1 & No \\
        \cline{2-3}
        & \cellcolor{LightRed} Replace WGAN\_GP loss by original loss & \cellcolor{LightRed} Worst \\
        \cline{2-3}
        & \cellcolor{LightRed} Replace WGAN\_GP loss by LSGAN loss & \cellcolor{LightRed} Worst \\
        \cline{2-3}
        & \cellcolor{LightRed} Replace WGAN\_GP loss by WGAN loss & \cellcolor{LightRed} Worst \\
        \cline{2-3}
        & \cellcolor{LightRed} Replace WGAN\_GP loss by DRAGAN loss & \cellcolor{LightRed} Worst \\
        \cline{2-3}
        & \cellcolor{LightRed} Use He Normal initializer instead dynamic normalization of weights & \cellcolor{LightRed} Worst \\
        \cline{2-3}
        & \cellcolor{LightGreen} Replace all ReLU by LReLU & \cellcolor{LightGreen} Better \\
        \cline{2-3}
        & \cellcolor{LightYellow} Replace Conv $3 \times 3$ by Conv $5 \times 5$ except for the first layer & \cellcolor{LightYellow} ? \\
        \cline{2-3}
        & \cellcolor{LightYellow} Increase the number of filters (not lower than 64) & \cellcolor{LightYellow} ? \\
        \cline{2-3}
        & \cellcolor{LightYellow} Increase batch from 16 to 64 & \cellcolor{LightYellow} ? \\
        \specialrule{1pt}{0pt}{0pt}
        \multirow{6}{*}{ProGAN v2} & Replace Conv $3 \times 3$ by Conv $5 \times 5$ except for the first layer & Yes \\
        \cline{2-3}
        & Increase the number of filters (not lower than 64) & Yes \\
        \cline{2-3}
        & \cellcolor{LightRed} Replace WGAN\_GP loss by original loss & \cellcolor{LightRed} Worst \\
        \cline{2-3}
        & \cellcolor{LightRed} Replace WGAN\_GP loss by LSGAN loss & \cellcolor{LightRed} Worst \\
        \cline{2-3}
        & \cellcolor{LightRed} Replace WGAN\_GP loss by WGAN loss & \cellcolor{LightRed} Worst \\
        \cline{2-3}
        & \cellcolor{LightRed} Replace WGAN\_GP loss by DRAGAN loss & \cellcolor{LightRed} Worst \\
        \specialrule{1pt}{0pt}{0pt}
    \end{tabular}
\end{table}

\subsubsection{Summary}
The qualitative pre-tuning process has led to 3 baseline architectures ready for further quantitative experiments (see Section~\ref{final_quantitative_comparison}).\\

From these tests, one can conclude that:
\begin{itemize}
    \item One sided label smoothing (ie weakening the Discriminator by labelling real images as 0.9 instead of 1) helps stabilizing the model
    \item Choosing Gaussian input noise or Uniform input noise does not have any impact on stability or quality of the results
    \item Regularization of the gradient norm seems important to stabilize and fasten convergence (like in WGAN\_GP or DRAGAN)
    \item The models are very sensitive to any small changes in hyperparameters
\end{itemize}

\newpage
\subsection{Final quantitative comparison}
\label{final_quantitative_comparison}
We introduce here the final architectures to benchmark thanks to the evaluation measures presented in Section~\ref{evaluation_measures_gan}.

\subsubsection{DCGAN}
The final baseline architecture (DCGAN 1) corresponds to Table~\ref{architecture_dcgan} with the following hyperparameters:
\vspace{-12pt}
\begin{multicols}{2}
\begin{itemize}
    \itemsep0em
    \footnotesize
    \item \textbf{Input noise size:} 256
    \item \textbf{Type input noise:} $\sim \mathcal{U}(-1, 1)$
    \item \textbf{Batch size:} 64
    \item \textbf{Input image range: } $[-1, 1]$
    \item \textbf{Optimizer:} Adam
    \item \textbf{Learning rate:} 0.0002
    \item \textbf{$\beta_1$:} 0.5
    \item \textbf{Epochs:} 20
    \item \textbf{Loss function:} DRAGAN
    \item \textbf{$\lambda_{adv}$:} 1
    \item \textbf{$\lambda_{gp}$:} 0.25
    \item \textbf{Rate (Generator/Discriminator):} 3
    \item \textbf{One-sided label smoothing:} Yes
    \item \textbf{Weight initialization:} $\sim \mathcal{N}(0, 0.02)$
    \item \textbf{Add noise in the discriminator:} No
    \item \textbf{Add minibatch similarity layer:} No
\end{itemize}
\end{multicols}

Two other architectures are compared with the baseline. The small modifications of the model are illustrated in Table~\ref{dcgan_final_tuning}.

\begin{table}
    \centering
    \caption{DCGAN architectures for quantitative comparison}
    \label{dcgan_final_tuning}
    \small
    \setlength\extrarowheight{2pt}
    \newcolumntype{C}{>{\centering\let\newline\\\arraybackslash\hspace{0pt}}p{12cm}}
    \begin{tabular}{|c|C|}
        \hline
        \textbf{Model name} & \textbf{Tested changes} \\
        \specialrule{1pt}{0pt}{0pt}
        DCGAN 2 & Replace DRAGAN by WGAN\_GP \\
        \hline
        DCGAN 3 & Add minibatch similarity layer \\
        \specialrule{1pt}{0pt}{0pt}
    \end{tabular}
\end{table}

\subsubsection{SRResGAN}
The final baseline architecture (SRResGAN 1) corresponds to Table~\ref{architecture_srresgan} with the following hyperparameters:
\vspace{-12pt}
\begin{multicols}{2}
\begin{itemize}
    \itemsep0em
    \footnotesize
    \item \textbf{Input noise size:} 256
    \item \textbf{Type input noise:} $\sim \mathcal{N}(0, 1)$ normalized
    \item \textbf{Batch size:} 64
    \item \textbf{Input image range: } $[-1, 1]$
    \item \textbf{Optimizer:} Adam
    \item \textbf{Learning rate:} 0.0002
    \item \textbf{$\beta_1$:} 0.5
    \item \textbf{Epochs:} 20
    \item \textbf{Loss function:} DRAGAN
    \item \textbf{$\lambda_{adv}$:} 1
    \item \textbf{$\lambda_{gp}$:} 0.25
    \item \textbf{Rate (Generator/Discriminator):} 2
    \item \textbf{One-sided label smoothing:} Yes
    \item \textbf{Weight initialization:} He Normal, \\
    ie $\displaystyle\sim \mathcal{N} \big(0, \sqrt{\frac{2}{256 \cdot 256 \cdot N_\mathrm{filters}}} \big)$
    \item \textbf{Add noise in the discriminator:} No
    \item \textbf{Add minibatch similarity layer:} No
\end{itemize}
\end{multicols}

Four other architectures are compared with the baseline. The small modifications of the model are illustrated in Table~\ref{srresgan_final_tuning}.

\begin{table}
    \centering
    \caption{SRResGAN architectures for quantitative comparison}
    \label{srresgan_final_tuning}
    \small
    \setlength\extrarowheight{2pt}
    \newcolumntype{C}{>{\centering\let\newline\\\arraybackslash\hspace{0pt}}p{12cm}}
    \begin{tabular}{|c|C|}
        \hline
        \textbf{Model name} & \textbf{Tested changes} \\
        \specialrule{1pt}{0pt}{0pt}
        SRResGAN 2 & Change kernel size from 3 to 5 + batch 32 \\
        \hline
        SRResGAN 3 & Increase size of residual blocks in the Generator \\
        \hline
        SRResGAN 4 & ReLU and not LReLU in Discriminator residual blocks \\
        \hline
        SRResGAN 5 & Add minibatch similarity layer \\
        \specialrule{1pt}{0pt}{0pt}
    \end{tabular}
\end{table}

\newpage
\subsubsection{ProGAN}
The final baseline architecture (ProGAN 1) corresponds to Table~\ref{architecture_proggan} with the following hyperparameters:
\vspace{-12pt}
\begin{multicols}{2}
\begin{itemize}
    \itemsep0em
    \footnotesize
    \item \textbf{Input noise size:} 512
    \item \textbf{Type input noise:} $\sim \mathcal{N}(0, 1)$ normalized
    \item \textbf{Batch size:} 64
    \item \textbf{Input image range: } $[-1, 1]$
    \item \textbf{Optimizer:} Adam
    \item \textbf{Learning rate:} 0.001
    \item \textbf{$\beta_1$:} 0.0
    \item \textbf{Epochs:} 20
    \item \textbf{Loss function:} WGAN\_GP
    \item \textbf{$\lambda_{adv}$:} 1
    \item \textbf{$\lambda_{gp}$:} 0.25
    \item \textbf{Rate (Generator/Discriminator):} 1
    \item \textbf{One-sided label smoothing:} $\emptyset$
    \item \textbf{Weight initialization:} $\sim \mathcal{N}(0, 1)$ but scaled dynamically with $\displaystyle \sqrt{\frac{2}{256 \cdot 256 \cdot N_\mathrm{filters}}}$
    \item \textbf{Add noise in the discriminator:} No
    \item \textbf{Add minibatch similarity layer:} Yes
\end{itemize}
\end{multicols}

Five other architectures are compared with the baseline. The small modifications of the model are illustrated in Table~\ref{progan_final_tuning}.

\begin{table}
    \centering
    \caption{ProGAN architectures for quantitative comparison}
    \label{progan_final_tuning}
    \small
    \setlength\extrarowheight{2pt}
    \newcolumntype{C}{>{\centering\let\newline\\\arraybackslash\hspace{0pt}}p{12cm}}
    \begin{tabular}{|c|C|}
        \hline
        \textbf{Model name} & \textbf{Tested changes} \\
        \specialrule{1pt}{0pt}{0pt}
        ProGAN 2 & Increase the number of filters (not lower than 64) + batch 32\\
        \hline
        ProGAN 3 & Remove minibatch similarity layer\\
        \hline
        ProGAN 4 & Delete $\epsilon_{drift}$\\
        \hline
        ProGAN 5 & Remove Pixel Normalization\\
        \hline
        ProGAN 6 & Define usual learning rate 0.0002 and usual $\beta_1$ 0.5\\
        \specialrule{1pt}{0pt}{0pt}
    \end{tabular}
\end{table}

\subsubsection{Results}
The evaluation of the introduced GAN architectures is presented in Table~\ref{evaluation}. Are displayed: $\rho$ the realism measure, $\sigma$ the total variation, $\delta$ the diversity measure, $t$ the training time (d-hh:mm), and the visual quality verification. All models have been trained with a Tesla K80 GPU for 20 epochs, except the \textit{long} models that have been trained for 60 epochs.

\begin{figure}
\centering
	\includegraphics[width=0.25\textwidth]{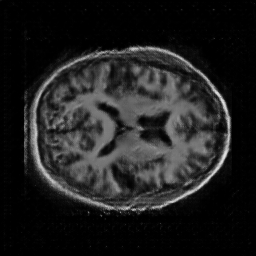}%
	\includegraphics[width=0.25\textwidth]{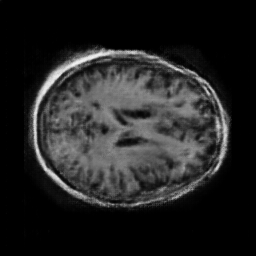}%
	\includegraphics[width=0.25\textwidth]{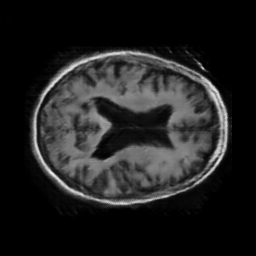}%
	\includegraphics[width=0.25\textwidth]{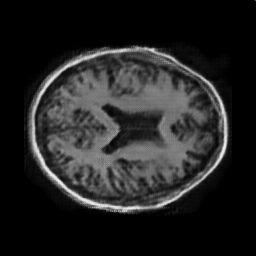}%
    \caption[DCGAN Visual results]{DCGAN Visual results\\
    {\small from left to right: DCGAN 1, DCGAN 2, DCGAN 3, DCGAN 3 long}}
    \label{dcgan_visual_results}
\end{figure}

\begin{figure}
\centering
	\includegraphics[width=1.0\textwidth]{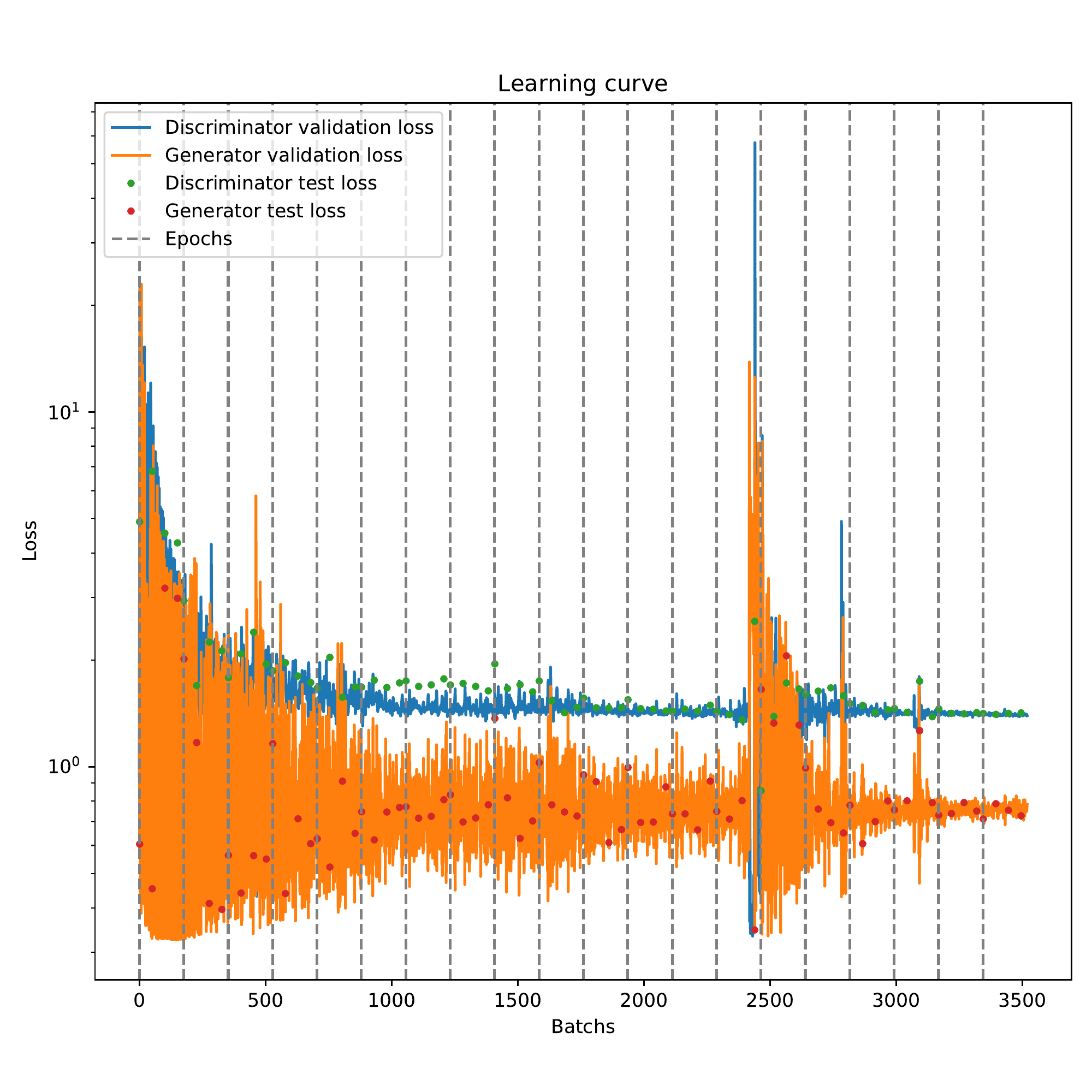}
    \caption{Learning curve of DCGAN 3}
    \label{dcgan3_learning_curve}
\end{figure}

Among the DCGAN architectures, the one using a minibatch similarity layer in the Discriminator (DCGAN 3) outperforms the others in realism and diversity. This performance can be improved by increasing the number of training epochs as shown by the scores of DCGAN 3 long. A visual comparison is done in Figure~\ref{dcgan_visual_results}. Also, Figure~\ref{dcgan3_learning_curve} shows the evolution of the different losses with respect to time: it can be noticed that convergence is difficult in some periods of time (at batch 2500 in the Figure) but a stabilized state is reached again soon after.

\begin{figure}
\centering
	\includegraphics[width=0.25\textwidth]{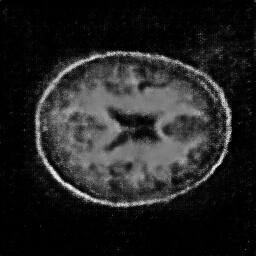}%
	\includegraphics[width=0.25\textwidth]{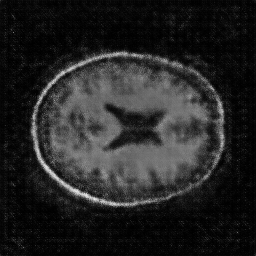}%
	\includegraphics[width=0.25\textwidth]{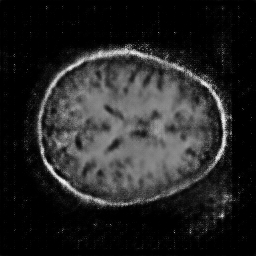}%
	\includegraphics[width=0.25\textwidth]{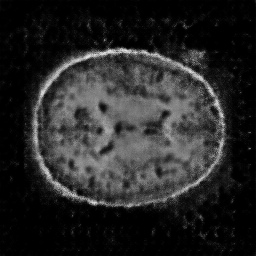}%
    \caption[SRResGAN Visual results]{SRResGAN Visual results\\
    {\small from left to right: SRResGAN 1, SRResGAN 3, SRResGAN 4, SRResGAN 5}}
    \label{srresgan_visual_results}
\end{figure}

Among the SRResGAN architecture, the baseline (SRResGAN 1) is the best model. However, some instabilities persist as shown by the divergence of SRResGAN 2 or SRResGAN 1 long. Also, the visual results seem a bit blurry as shown in Figure~\ref{srresgan_visual_results}.

\begin{figure}
\centering
	\includegraphics[width=0.25\textwidth]{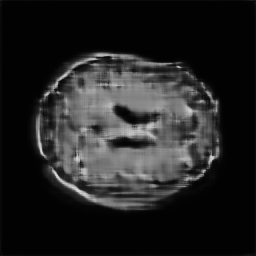}%
	\includegraphics[width=0.25\textwidth]{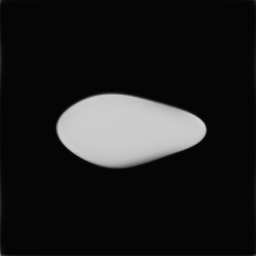}%
	\includegraphics[width=0.25\textwidth]{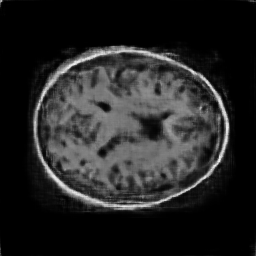}%
	\includegraphics[width=0.25\textwidth]{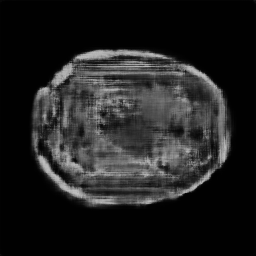}%
    \caption[ProGAN Visual results]{ProGAN Visual results\\
    {\small from left to right: ProGAN 1, ProGAN 4, ProGAN 5, ProGAN 6}}
    \label{progan_visual_results}
\end{figure}

\begin{figure}
\centering
	\includegraphics[width=1.0\textwidth]{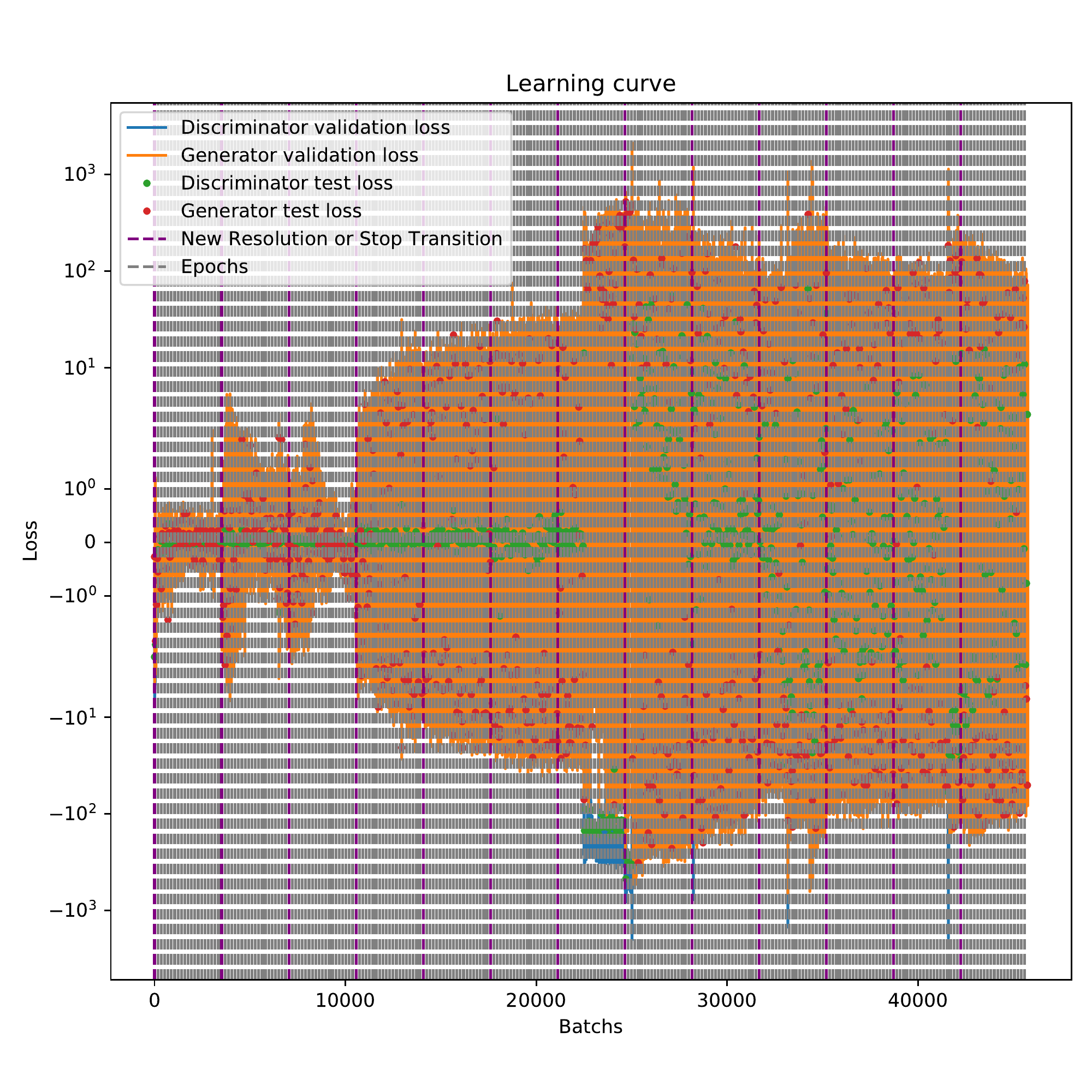}
    \caption{Learning curve of ProGAN 5}
    \label{progan5_learning_curve}
\end{figure}

\begin{figure}
\centering
	\includegraphics[width=1.0\textwidth]{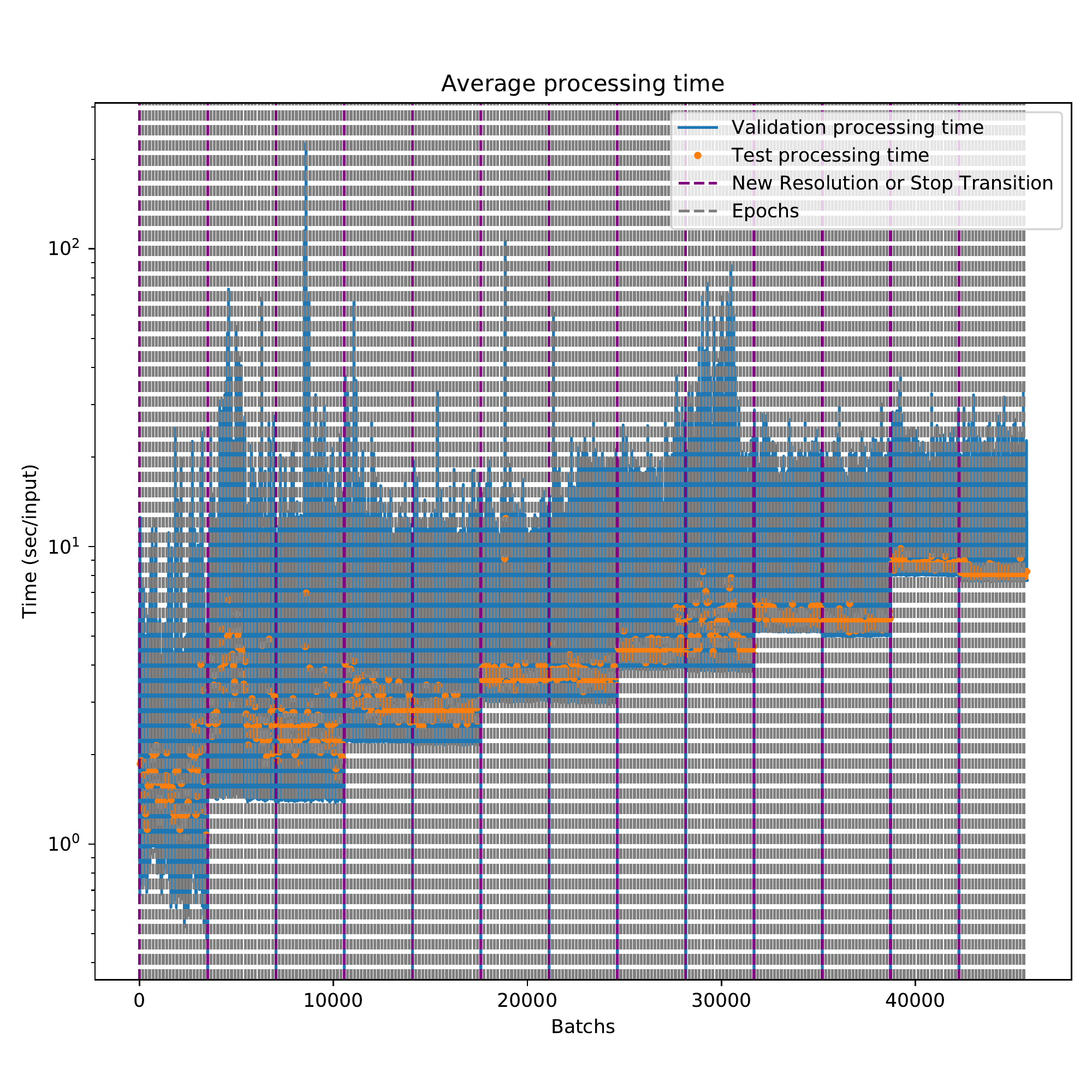}
    \caption{Processing time progression of ProGAN 5}
    \label{progan5_processing_time}
\end{figure}

ProGAN is surprisingly the architecture that is the least stable with 3 models out of 6 that have not converged, despite a huge training time. Figure~\ref{progan_visual_results} shows that the visual results are not satisfactory, except for ProGAN 5 (and potentially ProGAN 1). Moreover, Figure~\ref{progan5_learning_curve} displays the evolution of the different losses with respect to time and illustrates instabilities in the learning process (for example, generation is very difficult around batch 2300 because the Discriminator loss is very low). Nevertheless, Figure~\ref{progan5_processing_time} demonstrates that training the networks progressively helps fasten the process since computational time increases with additional layers and resolution expansion.

More visual results are available in Appendix~\ref{generation_results_appendix}.

\begin{table}
    \centering
    \small
    \begin{tabular}{cccccc}
        \toprule
        \textbf{Model} & $\rho$ & $\sigma$ & $\delta$ & $t$ & visual quality\\
        \midrule
        Training dataset & 0.739 & 409 & 15 & - & - \\
        \midrule
        DCGAN 1 & 0.690 & 243 & \color{red}{7} & 17:14 & Yes \\
        DCGAN 2 & 0.568 & \color{red}{51} & 3 & 14:54 & Yes \\
        DCGAN 3 & \textbf{0.709} & 315 & 12 & 15:57 & Yes \\
        \midrule
        SRResGAN 1 & \textbf{0.678} & 329 & 16 & 11:19 & Yes \\
        \color{red}{SRResGAN 2} & - & 0 & 0 & 20:18 & \color{red}{No} \\
        SRResGAN 3 & 0.640 & \color{red}{271} & 14 & 11:35 & Yes \\
        SRResGAN 4 & 0.658 & 327 & 15 & 11:42 & Yes \\
        SRResGAN 5 & 0.638 & 309 & 16 & 12:43 & Yes \\
        \midrule
        ProGAN 1 & \textbf{0.610} & 308 & 14 & 2\textendash{}13:36 & Yes \\
        \color{red}{ProGAN 2} & - & - & - & 3\textendash{}15:58 & \color{red}{No} \\
        \color{red}{ProGAN 3} & 0.273 & 0 & 6 & 2\textendash{}06:54 & \color{red}{No} \\
        \color{red}{ProGAN 4} & 0.610 & 340 & 2 & 2\textendash{}14:36 & \color{red}{No} \\
        ProGAN 5 & \textbf{0.601} & 331 & 13 & 2\textendash{}09:46 & Yes \\
        ProGAN 6 & 0.480 & 251 & \color{red}{5} & 2\textendash{}14:06 & Yes \\
        \midrule
        DCGAN 3 long & \textbf{0.718} & 352 & 15 & 1\textendash{}18:51 & Yes \\
        \color{red}{SRResGAN 1 long} & - & 0 & 0 & 1\textendash{}06:18 & \color{red}{No} \\
        \bottomrule
    \end{tabular}
    \caption{Benchmark of different GAN architectures}
    \label{evaluation}
\end{table}

\subsubsection{Overfitting estimation}
\begin{figure}
\centering
	\includegraphics[width=0.25\textwidth]{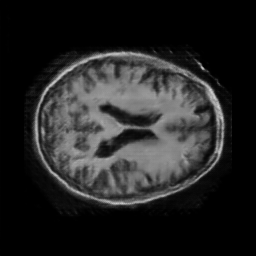}%
	\includegraphics[width=0.25\textwidth]{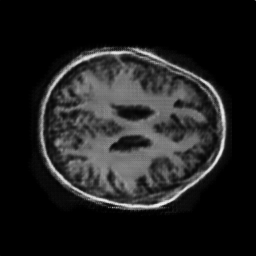}%
	\includegraphics[width=0.25\textwidth]{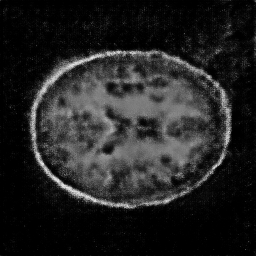}%
	\includegraphics[width=0.25\textwidth]{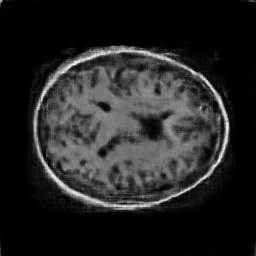}\\
	\vspace{-1pt}
    \includegraphics[width=0.25\textwidth]{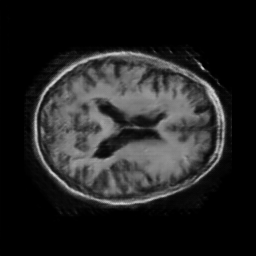}%
	\includegraphics[width=0.25\textwidth]{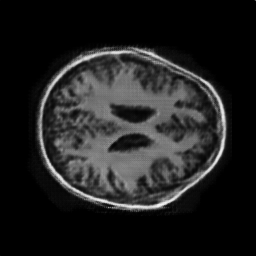}%
	\includegraphics[width=0.25\textwidth]{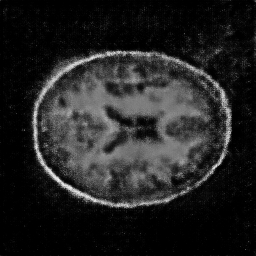}%
	\includegraphics[width=0.25\textwidth]{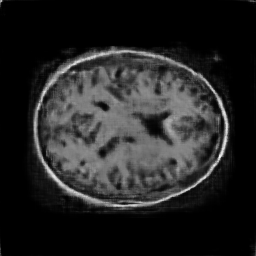}\\
	\vspace{-1pt}
    \includegraphics[width=0.25\textwidth]{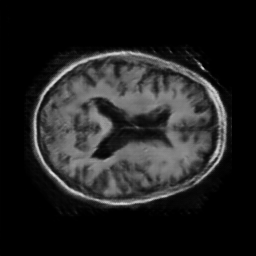}%
	\includegraphics[width=0.25\textwidth]{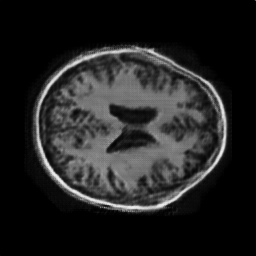}%
	\includegraphics[width=0.25\textwidth]{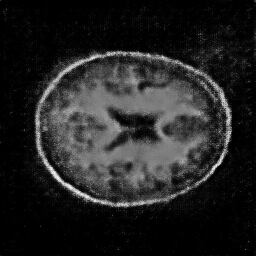}%
	\includegraphics[width=0.25\textwidth]{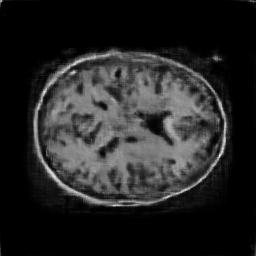}\\
	\vspace{-1pt}
    \includegraphics[width=0.25\textwidth]{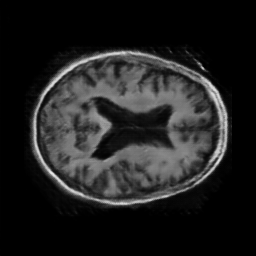}%
	\includegraphics[width=0.25\textwidth]{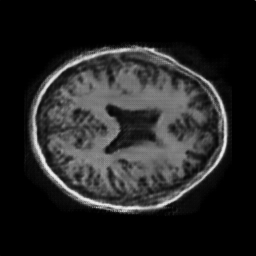}%
	\includegraphics[width=0.25\textwidth]{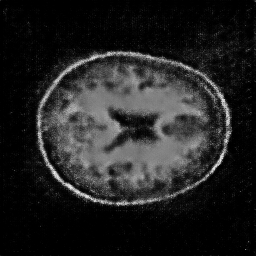}%
	\includegraphics[width=0.25\textwidth]{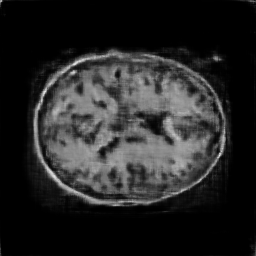}\\
	\vspace{-1pt}
    \includegraphics[width=0.25\textwidth]{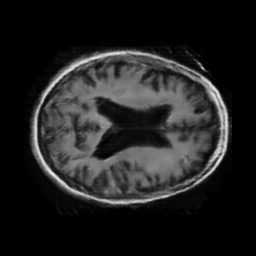}%
	\includegraphics[width=0.25\textwidth]{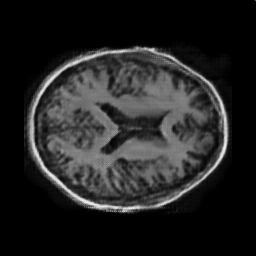}%
	\includegraphics[width=0.25\textwidth]{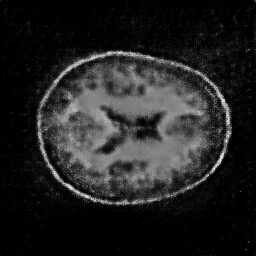}%
	\includegraphics[width=0.25\textwidth]{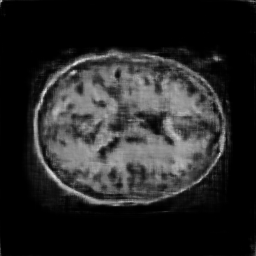}\\
	\vspace{-1pt}
    \includegraphics[width=0.25\textwidth]{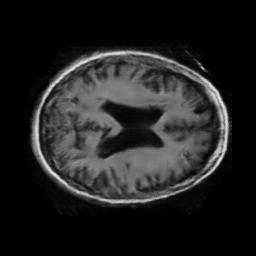}%
	\includegraphics[width=0.25\textwidth]{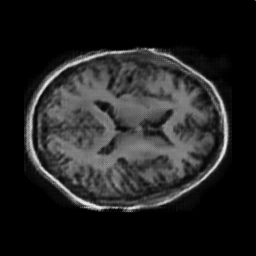}%
	\includegraphics[width=0.25\textwidth]{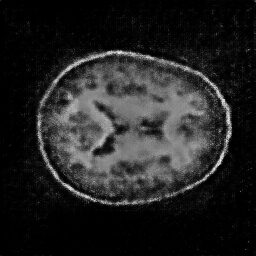}%
	\includegraphics[width=0.25\textwidth]{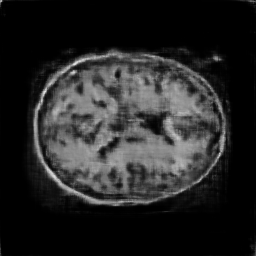}
    \caption[Visual interpolation between two random latent input vectors]{Visual interpolation between two random latent input vectors\\
    {\small in columns, from left to right: DCGAN 3, DCGAN 3 long, SRResGAN 1, ProGAN 5}}
    \label{visual_interpolation}
\end{figure}

In order to see if the generative models learn the underlying manifold and do not just reproduce some input data, images are generated using several interpolations of two random latent vectors (see Figure~\ref{visual_interpolation}).

It can be observed that the generation does not produce incoherent MRIs. We can then conclude that these GAN models are not overfitting.

\subsubsection{Summary}
To conclude, among all the tested GAN architectures, DCGAN is the most stable and so the one that performs the best.\\

Moreover, it has been shown that:
\begin{itemize}
    \item Adding a minibatch similarity layer in the Discriminator improves the diversity but also the quality of generated images
    \item Increasing the training time improves the realism and quality of the generation (if the model is stable)
    \item Using additional Gaussian noise in the Discriminator in order to slow down the learning and stabilize the model is not efficient (in our cases)
\end{itemize}

%% file: discussion.tex
\chapter{Discussion}
\label{discussion}
In this Chapter, segmentation and generation results are discussed to enlighten their strengths and weaknesses. It also gives insights on how the outcomes can be improved and enhanced.

\section{Relevance}
Segmentation performed with the U-net architecture reaches state-of-the-art results (more than 99.5\% of accuracy for all bones and cartilages in the sagittal plane). However, it can be noticed in the visual outputs that the ground truth labelling is not very accurate, especially on the small asperities or at the corners of the bone/cartilage. This is due to a pseudo-automatic method used to segment the 3D scans and give the ground truth for training. Thus, it is now difficult to increase the model performance, as the threshold where some U-net segmentations outperform the ground truth has been reached.

When it comes to MRI generation using GANs, the results seem very encouraging for a near use to better understand MRIs and train more advanced models. Nevertheless, one can notice that tuning the architectures to stabilize the model and evaluating the final performance are difficult tasks. In particular, the quantitative measures introduced in Section \ref{evaluation_measures_gan} have some drawbacks. The realism measure $\rho$ detects the necessity to use vectors orthogonal to the main variations of the input manifold to describe the generated image, but it does not take into account the distribution of the outputs inside the ``main'' vector space. That can explain why ProGAN 1 and ProGAN 5 have similar realism $\rho$ in Table~\ref{evaluation} but the latter visually outperforms the former (see Figure~\ref{progan_visual_results}). Moreover, the diversity $\delta$ gives insights on mode collapse but does not indicate if the main variations are the same as in the input manifold.

In addition, the training process is stochastic and very sensitive to tiny changes. It means that two launches of the same model can end up with very different outcomes (one giving good results, and the other diverging for example). To cope with this problem, a statistical analysis should be performed but requires a lot of time and computation power.

Furthermore, it is difficult to know when training needs to be stopped because the loss functions are not representative of the quality of the generated images. Only visual feedback or quantitative analysis (which needs the generation of a huge amount of images) can give insights on the performance of the model. For example, ProGAN 5 seems to perform poorly compared to DCGAN or SRResGAN but the model training was stopped in the middle of an abrupt and temporary loss in performance (as can be seen when looking over generated images while the training is running\footnote{See \href{https://github.com/antoinedelplace/MRI-Generation}{Github repository}}). One idea could be to save the weights at each iteration and then keep the best model but this is unrealistic as it is very time and memory consuming. Another idea could be to relaunch the training until the results are satisfactory.

Finally, the results can be improved by training the models for a longer period of time (usually, GANs are trained for several weeks on a GPU equivalent to a Telsa K80). This was not possible in the time constraint I had for this project.

\section{Possible future work}
Several avenues are possible to continue the work carried out so far.
With MRI segmentation, the next challenge is to develop a model that can process 3D scans. This is not an easy task because it requires a huge amount of memory to store all the model weights (hardwares available at the University of Queensland are currently not adapted to such project).

When it comes to MRI generation with GANs, a new architecture called StyleGAN by \citet{stylegan} could be tested. The architecture gives state-of-the-art performance on human face generation and is interesting in the fact that the input latent space becomes meaningful (each dimension in the latent space impacts one visual feature in the generated image).

Also, the idea of generating 3D MRIs with GANs is tempting but faces the same challenges with 3D segmentation: the memory consumption is too large.

Finally, the idea of combining 2D segmentation with GAN generation to produce 3D segmentation is a bad idea in my opinion, because it would not fit the input image (the loss function should not be adversarial\footnote{So GANs are not adapted}, it should be the Dice coefficient).

%% file: conclusion.tex
\chapter{Conclusion}
\label{conclusion}
To conclude, this project has achieved its two objectives: using Deep Neural Networks to achieve state-of-the-art performance in segmentation of bones and cartilages of Knee MRIs, and generate realistic brain MRIs from random noise.

Following the results of the U-net and GAN models, insights have been given to improve performance and efficiency by tuning some hyperparameters. Moreover, a benchmark has presented several GAN architectures to compare their efficiency at generating high resolution images with short training time.

In that sense, it goes beyond the work of \citet{unet} that only segments neural cells, \citet{progan} that focuses on human face generation and \citet{mri_gan} that only tried one GAN architecture with low resolution images.

Finally, ideas to improve and balance the results have been discussed in Chapter~\ref{discussion}.

One can remember that the U-net architecture is very efficient at segmenting 2D MRIs but requires too much memory to be applied as it is on 3D MRIs. Also, GANs are able to produce realistic high resolution MRIs but some architectures are difficult to stabilize. Using one-sided label smoothing, a minibatch similarity layer in the Discriminator and increasing the training time are good ways to improve stability and quality of generated MRIs.

With the work done in this thesis project, Medical research has methods to segment and generate Magnetic Resonance Images to develop more advanced techniques and process MRIs more easily.

%% file: segmentation_results.tex
\chapter{Visual results of Segmentation by the U-net model}
\label{segmentation_results_appendix}
\vspace{-100pt}
\begin{figure}
\centering
	\includegraphics[width=1.0\textwidth]{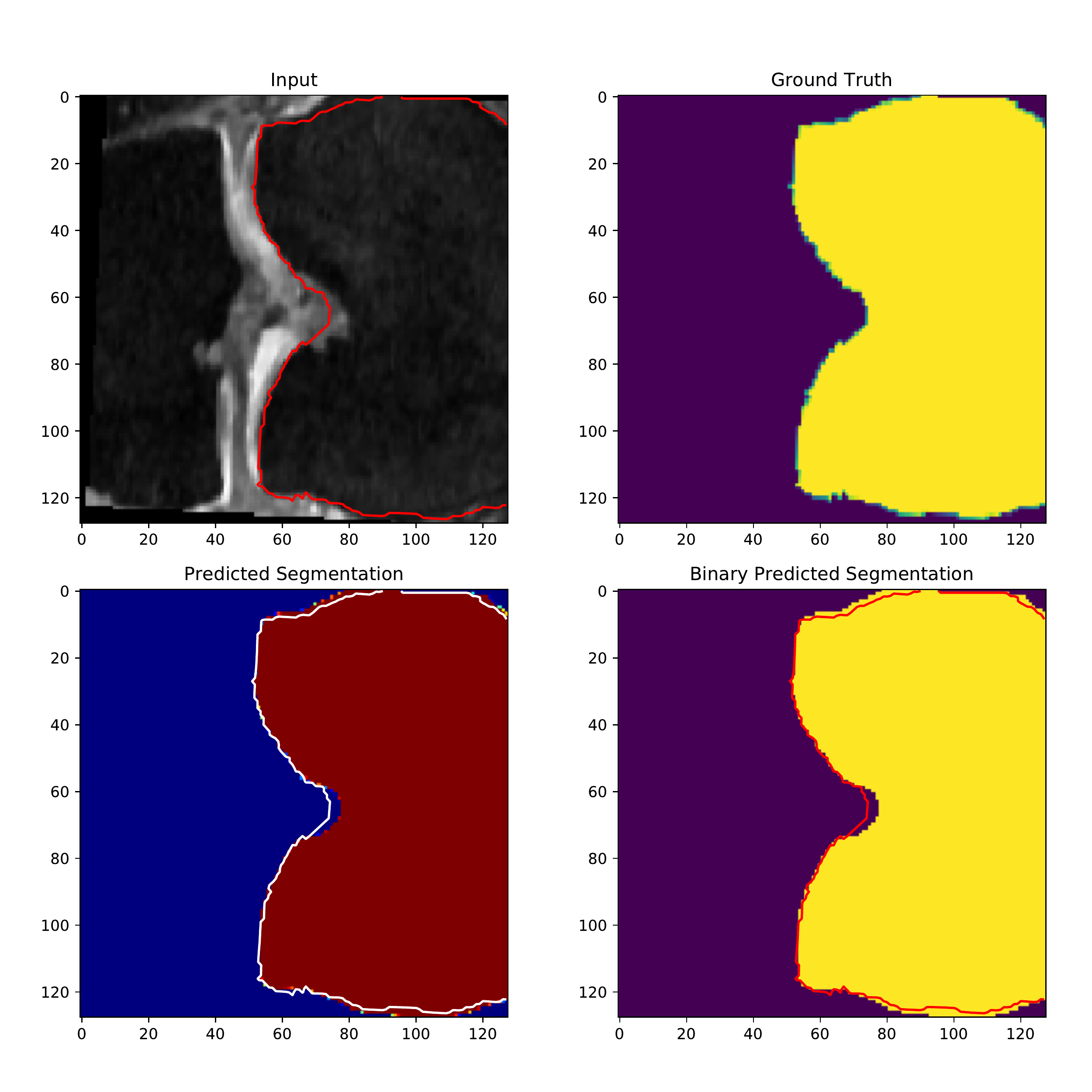}
    \caption{Segmentation of the femur bone with U-net in the coronal plane}
\end{figure}

\begin{figure}
\centering
	\includegraphics[width=1.0\textwidth]{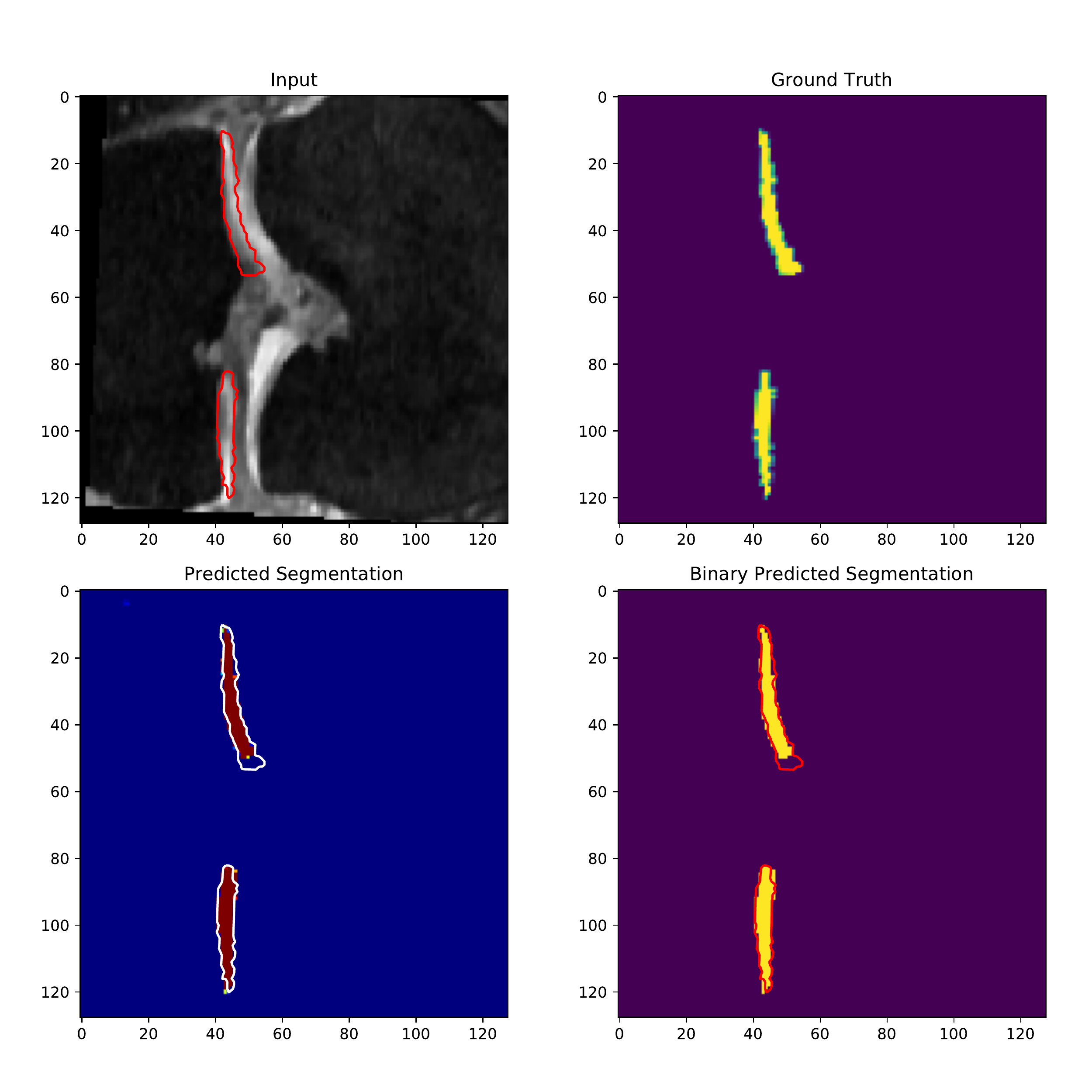}
    \caption{Segmentation of the tibia cartilage with U-net in the coronal plane}
\end{figure}

\begin{figure}
\centering
	\includegraphics[width=1.0\textwidth]{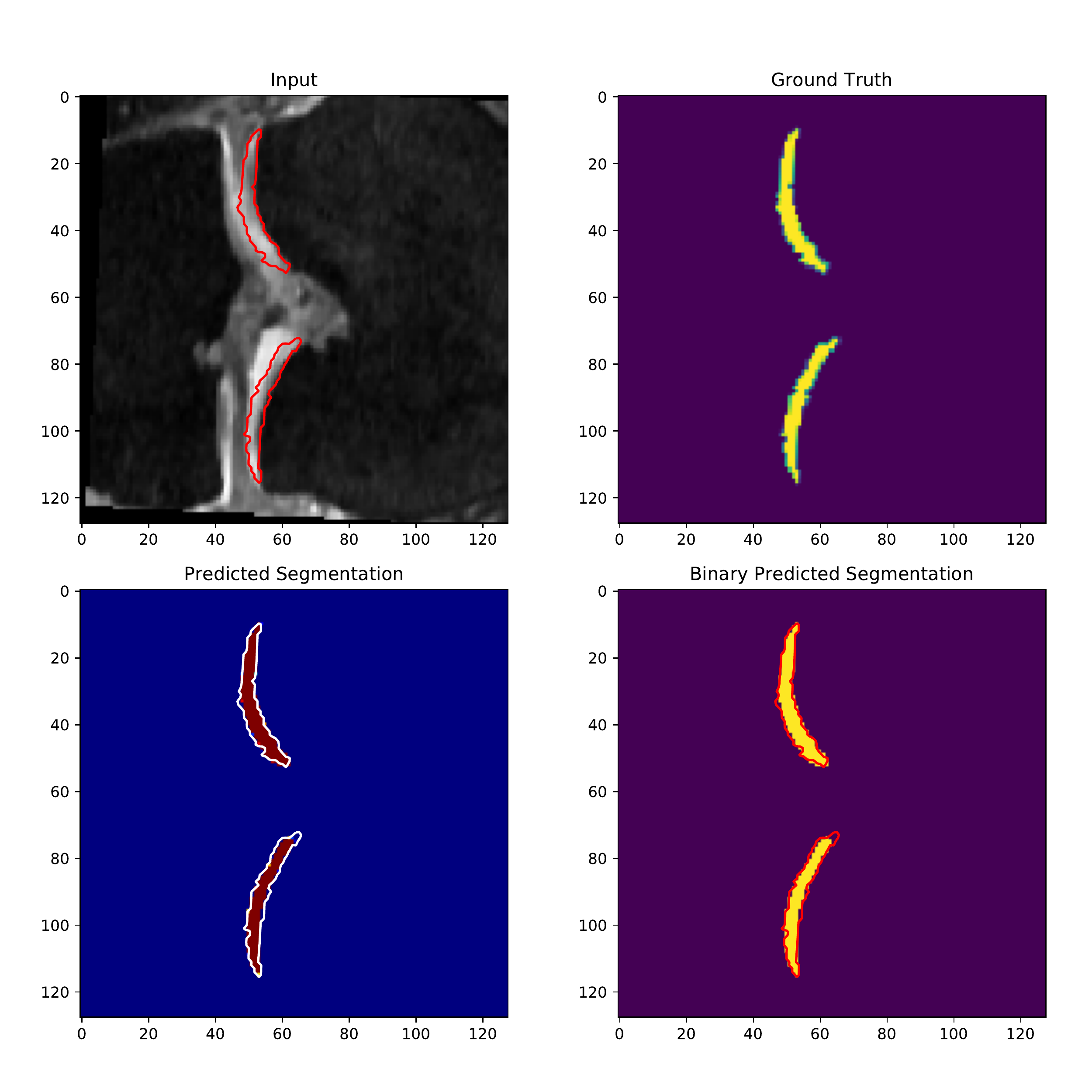}
    \caption{Segmentation of the femur cartilage with U-net in the coronal plane}
\end{figure}

\begin{figure}
\centering
	\includegraphics[width=1.0\textwidth]{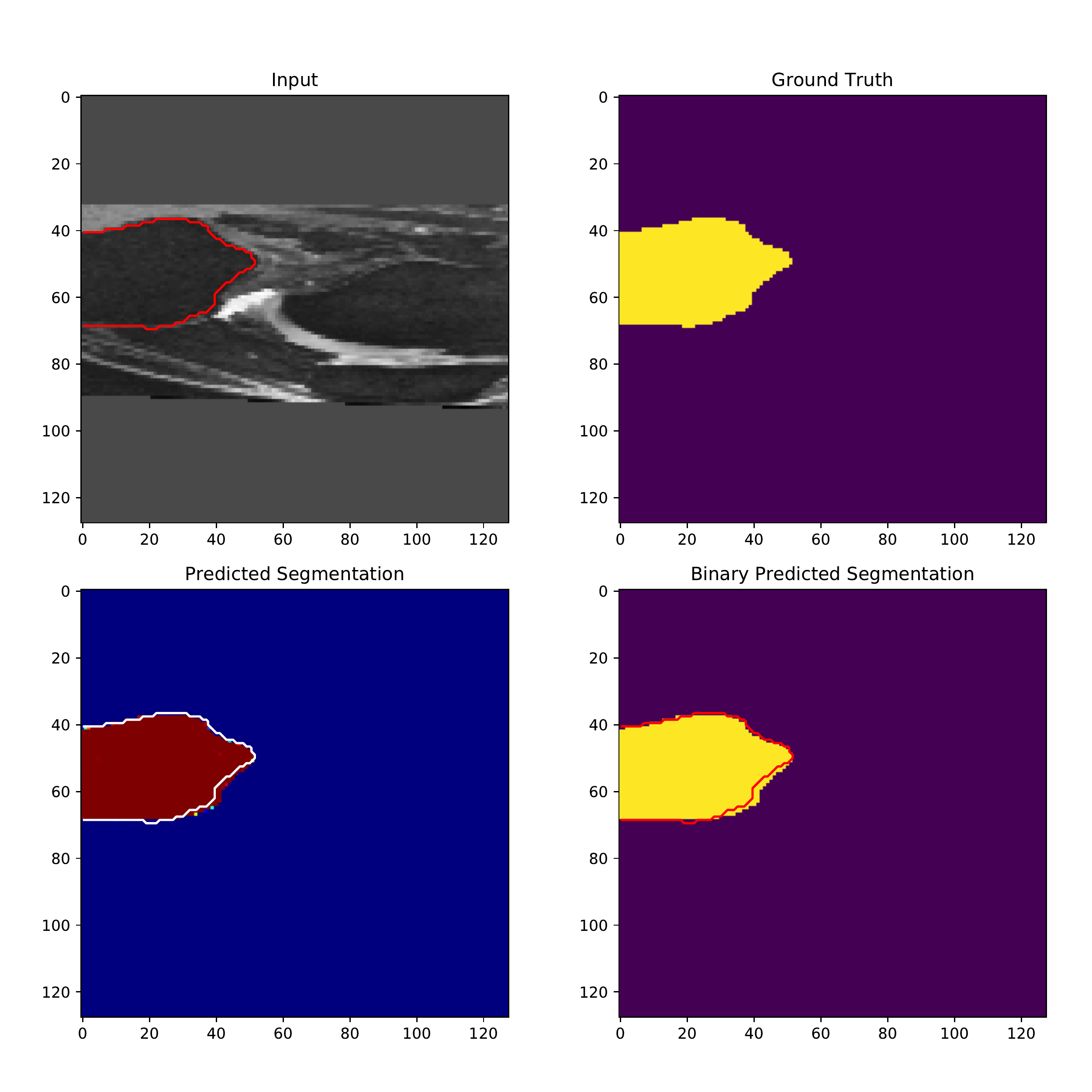}
    \caption{Segmentation of the tibia bone with U-net in the sagittal plane}
\end{figure}

\begin{figure}
\centering
	\includegraphics[width=1.0\textwidth]{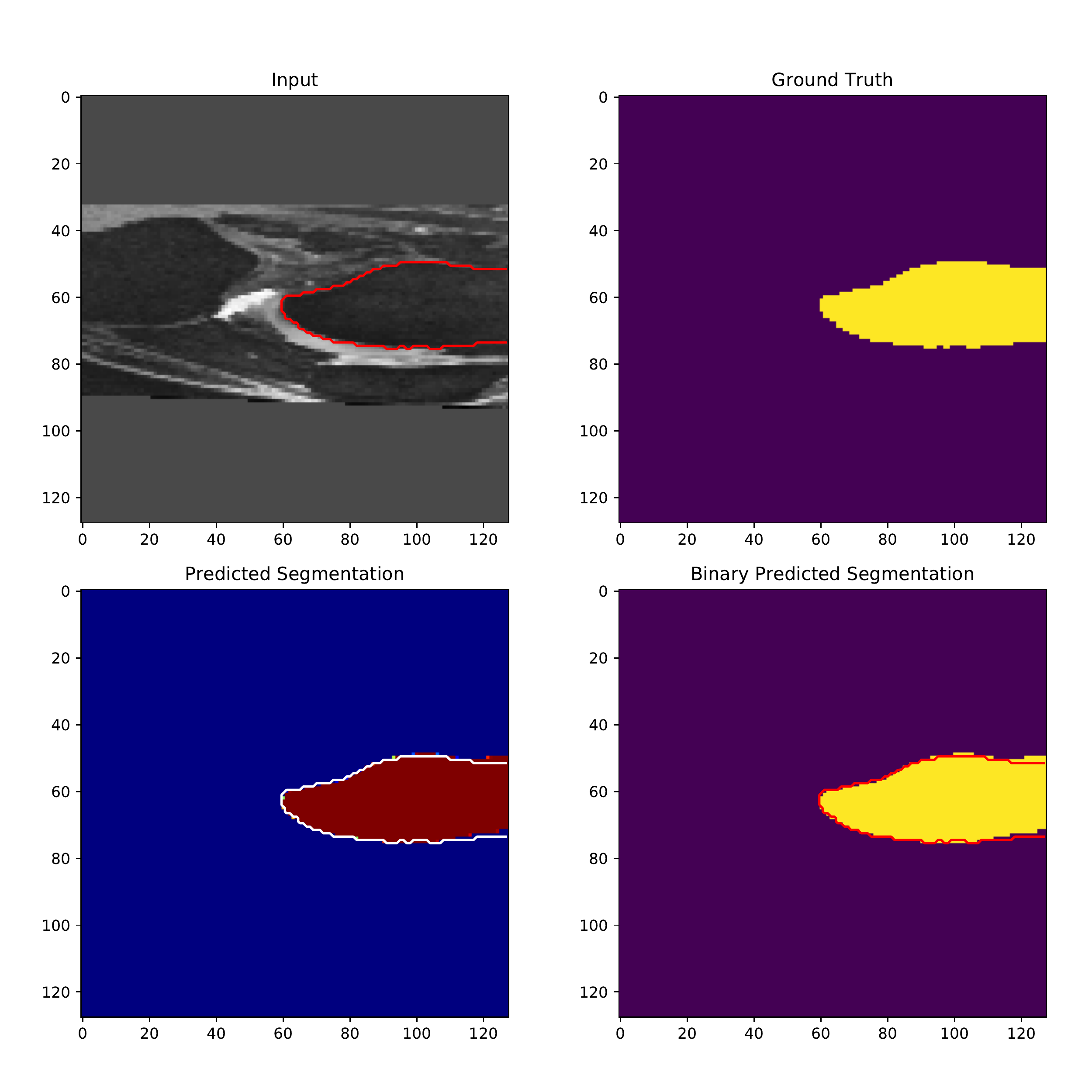}
    \caption{Segmentation of the femur bone with U-net in the sagittal plane}
\end{figure}

\begin{figure}
\centering
	\includegraphics[width=1.0\textwidth]{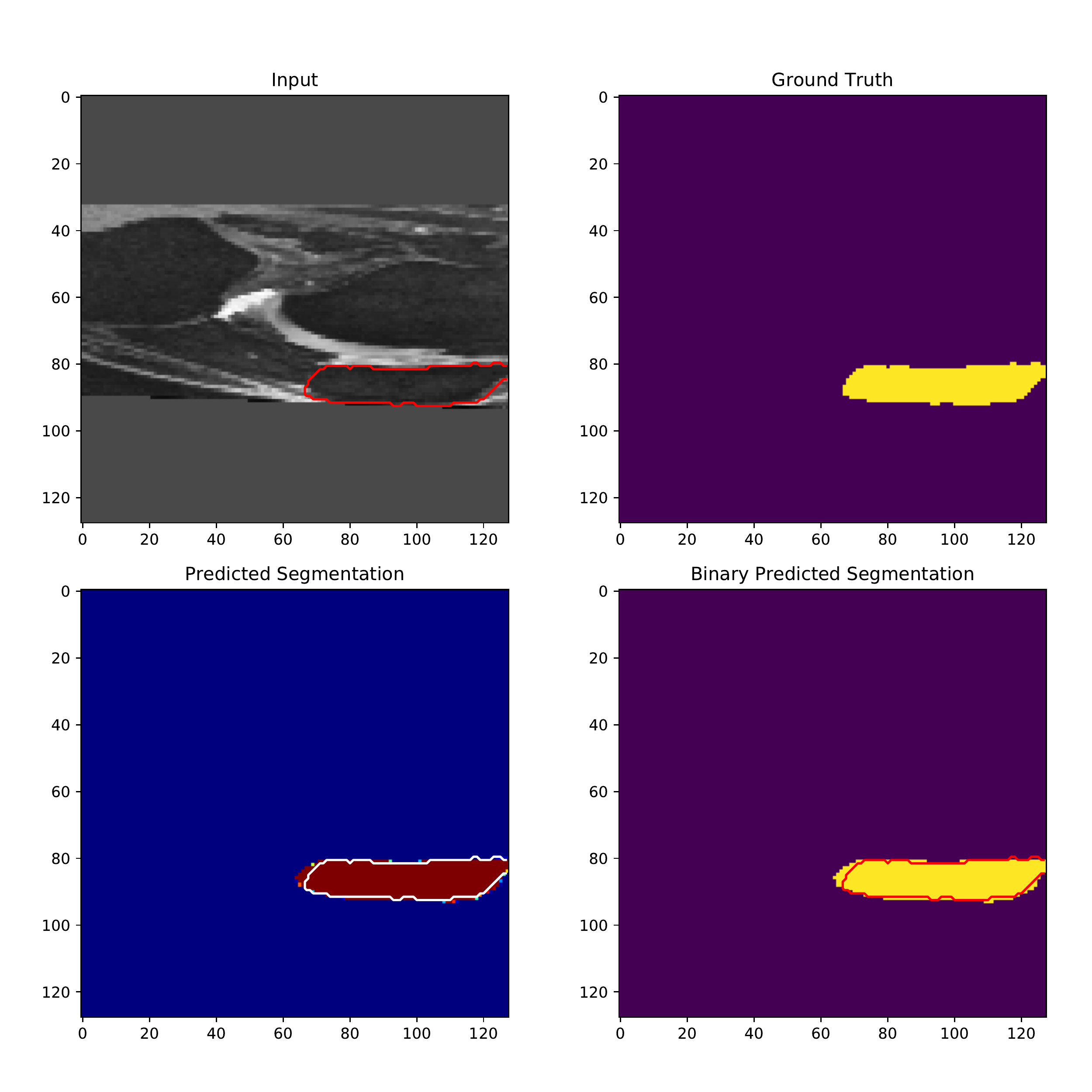}
    \caption{Segmentation of the patella bone with U-net in the sagittal plane}
\end{figure}

\begin{figure}
\centering
	\includegraphics[width=1.0\textwidth]{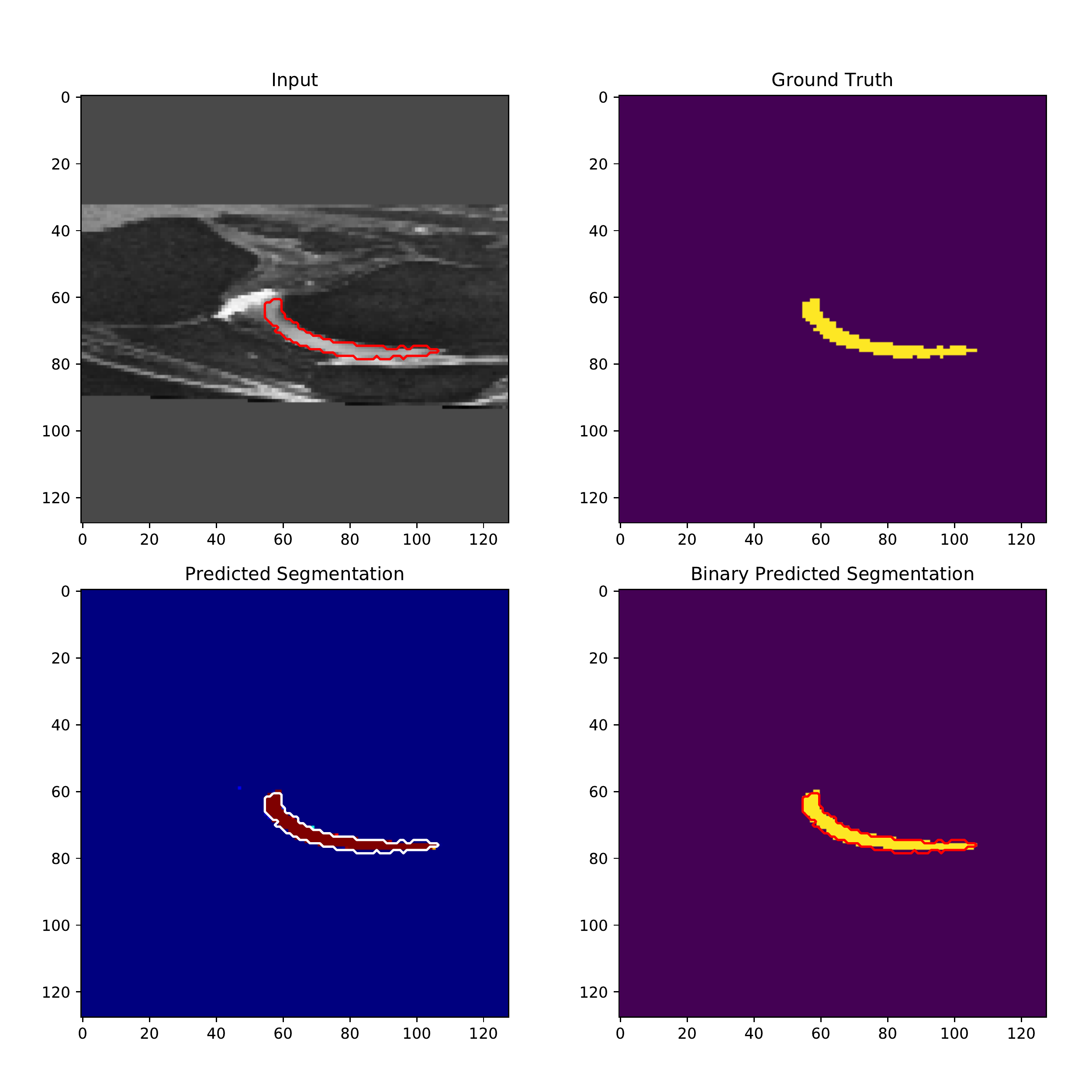}
    \caption{Segmentation of the femur cartilage with U-net in the sagittal plane}
\end{figure}

\begin{figure}
\centering
	\includegraphics[width=1.0\textwidth]{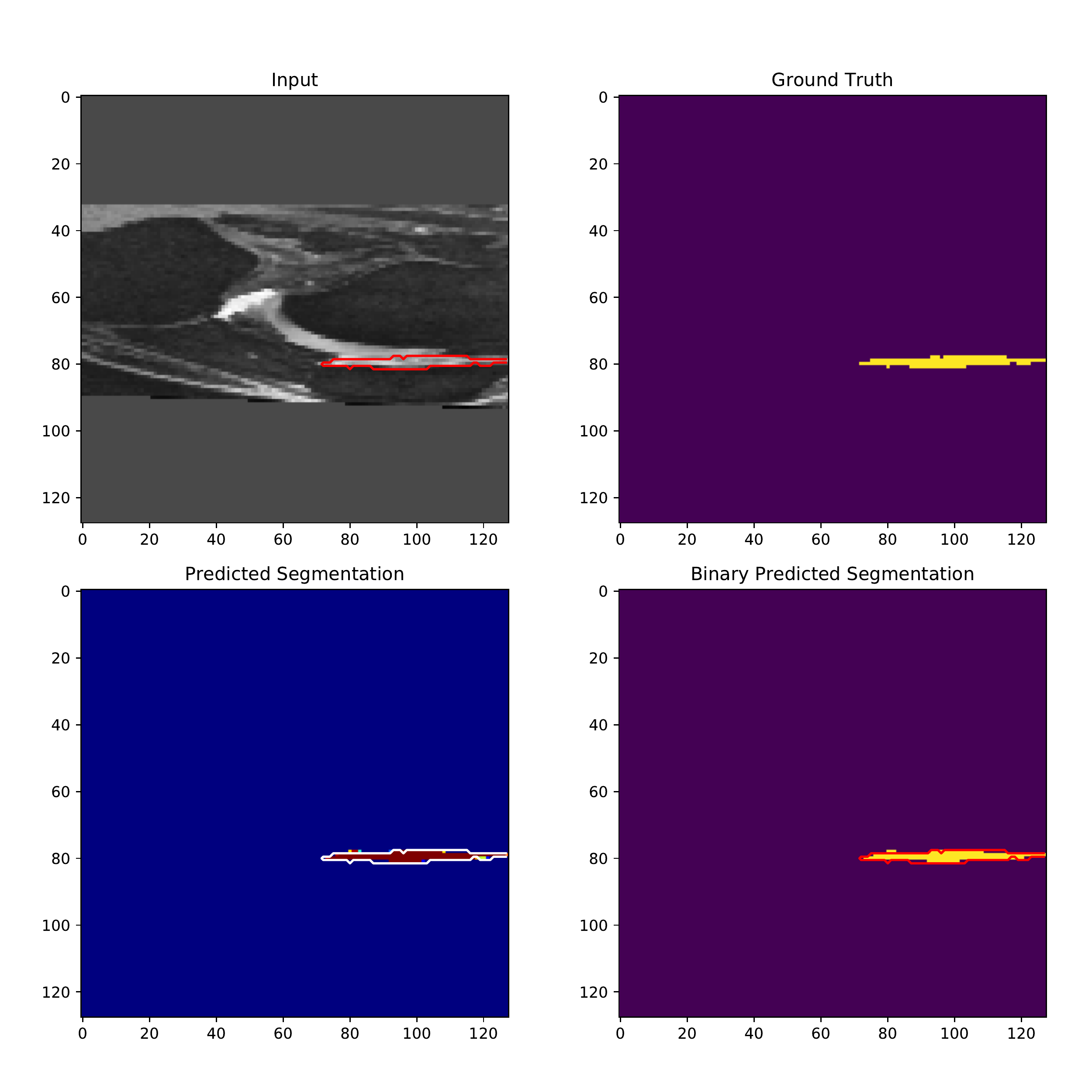}
    \caption{Segmentation of the patella cartilage with U-net in the sagittal plane}
\end{figure}

%% file: generation_results.tex
\chapter{Visual results of MRI Generation by GANs}
\label{generation_results_appendix}
\section{DCGAN 3}
\vfill
\begin{figure}[H]
\centering
	\includegraphics{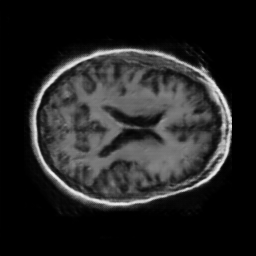}
    \caption{MRI Generation with DCGAN 3 (full resolution)}
\end{figure}
\vfill
\newpage

\begin{figure}[H]
\centering
	\includegraphics[width=0.25\textwidth]{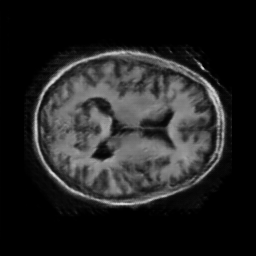}%
	\includegraphics[width=0.25\textwidth]{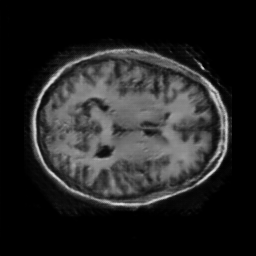}%
	\includegraphics[width=0.25\textwidth]{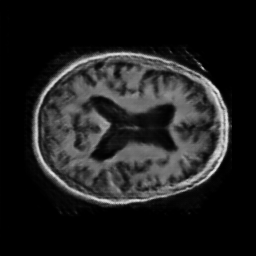}%
	\includegraphics[width=0.25\textwidth]{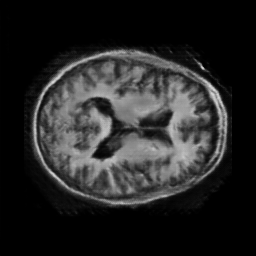}\\
	\vspace{-1pt}
    \includegraphics[width=0.25\textwidth]{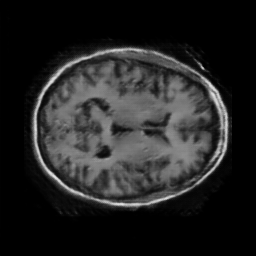}%
	\includegraphics[width=0.25\textwidth]{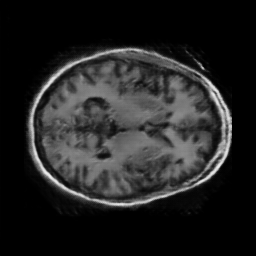}%
	\includegraphics[width=0.25\textwidth]{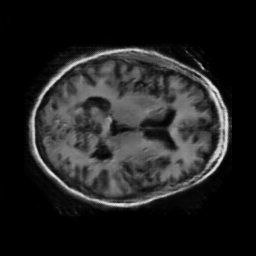}%
	\includegraphics[width=0.25\textwidth]{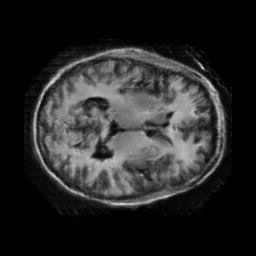}\\
	\vspace{-1pt}
    \includegraphics[width=0.25\textwidth]{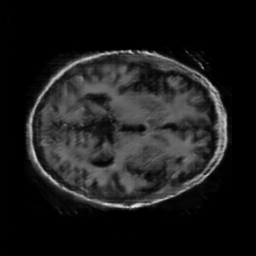}%
	\includegraphics[width=0.25\textwidth]{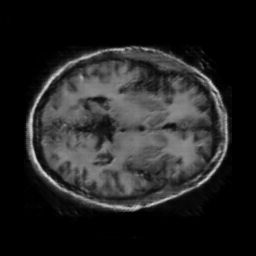}%
	\includegraphics[width=0.25\textwidth]{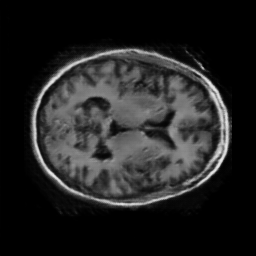}%
	\includegraphics[width=0.25\textwidth]{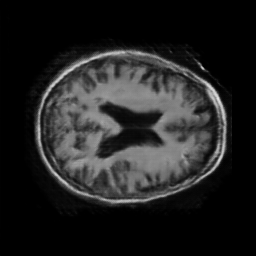}\\
	\vspace{-1pt}
    \includegraphics[width=0.25\textwidth]{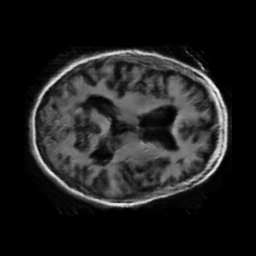}%
	\includegraphics[width=0.25\textwidth]{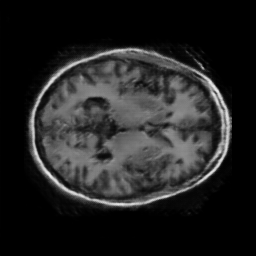}%
	\includegraphics[width=0.25\textwidth]{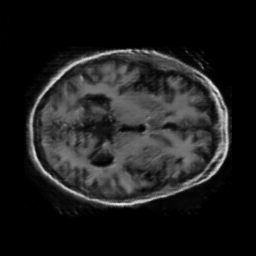}%
	\includegraphics[width=0.25\textwidth]{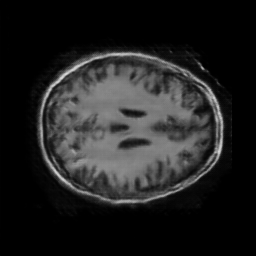}\\
	\vspace{-1pt}
    \includegraphics[width=0.25\textwidth]{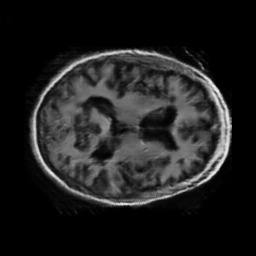}%
	\includegraphics[width=0.25\textwidth]{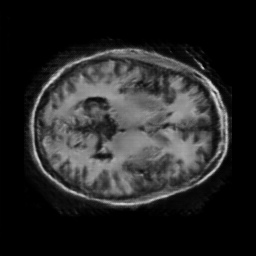}%
	\includegraphics[width=0.25\textwidth]{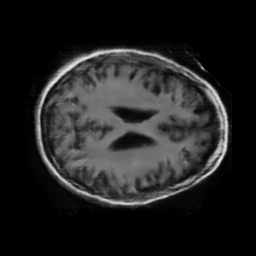}%
	\includegraphics[width=0.25\textwidth]{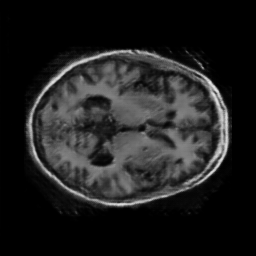}\\
	\vspace{-1pt}
    \includegraphics[width=0.25\textwidth]{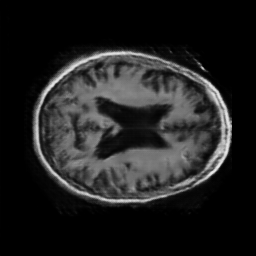}%
	\includegraphics[width=0.25\textwidth]{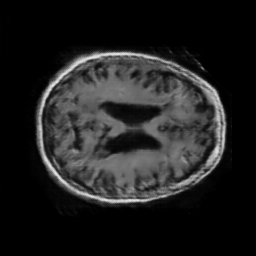}%
	\includegraphics[width=0.25\textwidth]{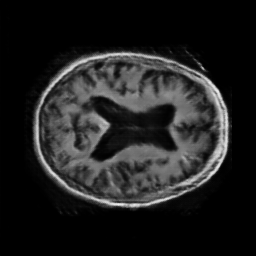}%
	\includegraphics[width=0.25\textwidth]{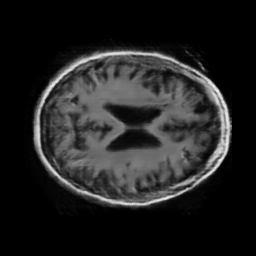}
    \caption{MRIs Generated with DCGAN 3}
\end{figure}

\newpage
\vspace*{2cm}
\section{DCGAN 3 long}
\vfill
\begin{figure}[H]
\centering
	\includegraphics{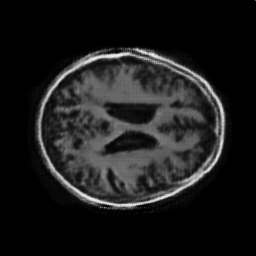}
    \caption{MRI Generation with DCGAN 3 long (full resolution)}
\end{figure}
\vfill
\newpage

\begin{figure}[H]
\centering
	\includegraphics[width=0.25\textwidth]{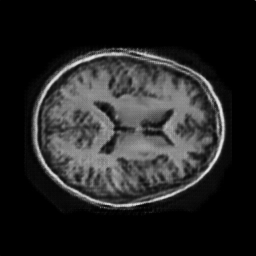}%
	\includegraphics[width=0.25\textwidth]{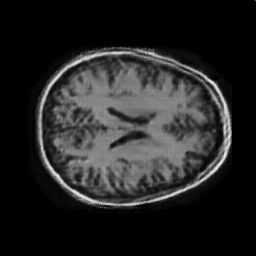}%
	\includegraphics[width=0.25\textwidth]{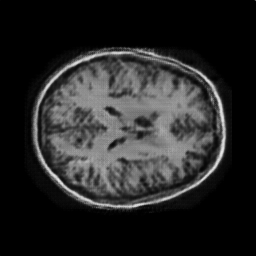}%
	\includegraphics[width=0.25\textwidth]{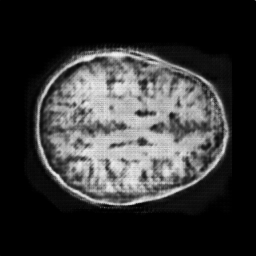}\\
	\vspace{-1pt}
    \includegraphics[width=0.25\textwidth]{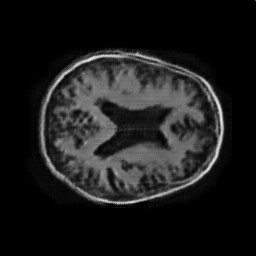}%
	\includegraphics[width=0.25\textwidth]{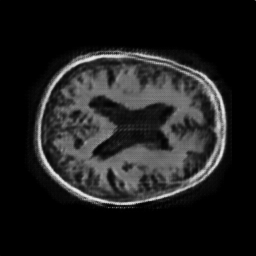}%
	\includegraphics[width=0.25\textwidth]{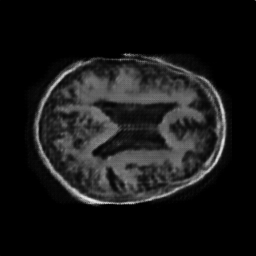}%
	\includegraphics[width=0.25\textwidth]{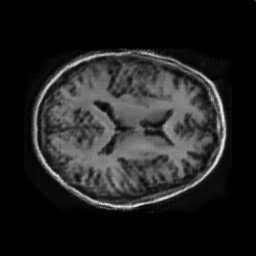}\\
	\vspace{-1pt}
    \includegraphics[width=0.25\textwidth]{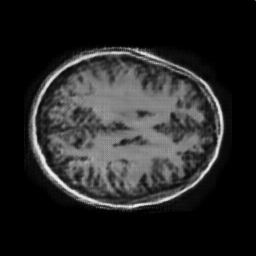}%
	\includegraphics[width=0.25\textwidth]{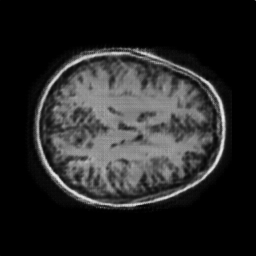}%
	\includegraphics[width=0.25\textwidth]{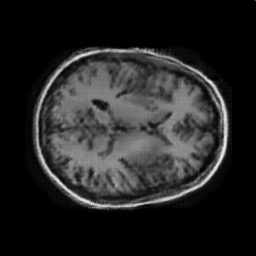}%
	\includegraphics[width=0.25\textwidth]{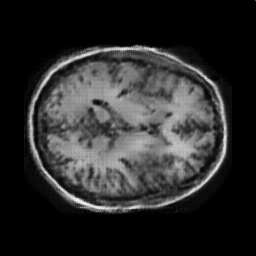}\\
	\vspace{-1pt}
    \includegraphics[width=0.25\textwidth]{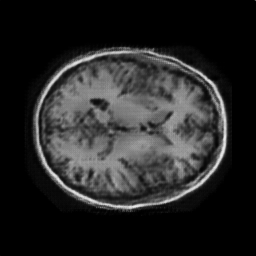}%
	\includegraphics[width=0.25\textwidth]{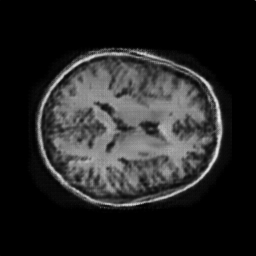}%
	\includegraphics[width=0.25\textwidth]{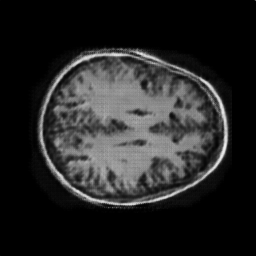}%
	\includegraphics[width=0.25\textwidth]{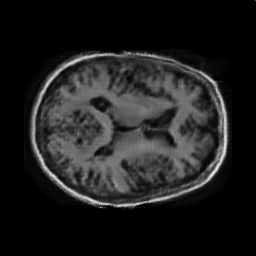}\\
	\vspace{-1pt}
    \includegraphics[width=0.25\textwidth]{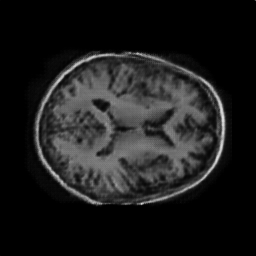}%
	\includegraphics[width=0.25\textwidth]{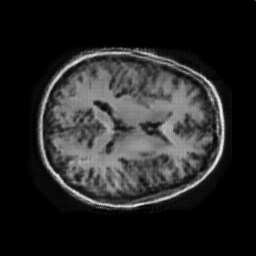}%
	\includegraphics[width=0.25\textwidth]{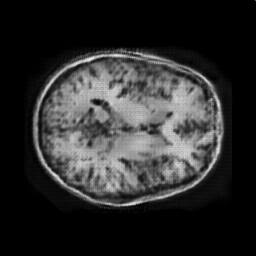}%
	\includegraphics[width=0.25\textwidth]{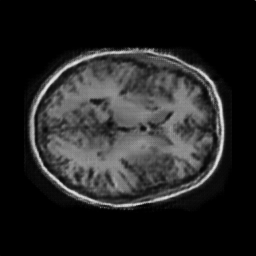}\\
	\vspace{-1pt}
    \includegraphics[width=0.25\textwidth]{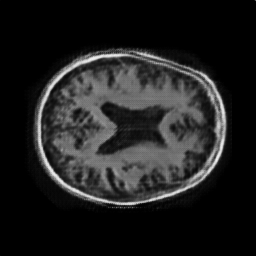}%
	\includegraphics[width=0.25\textwidth]{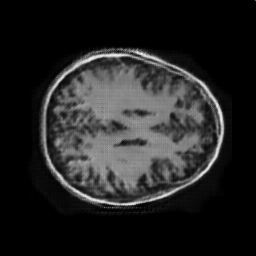}%
	\includegraphics[width=0.25\textwidth]{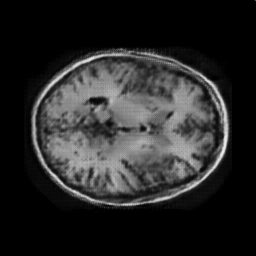}%
	\includegraphics[width=0.25\textwidth]{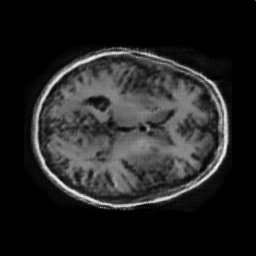}
    \caption{MRIs Generated with DCGAN 3 long}
\end{figure}

\newpage
\vspace*{2cm}
\section{SRResGAN 1}
\vfill
\begin{figure}[H]
\centering
	\includegraphics{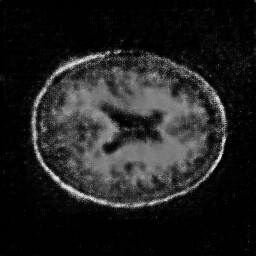}
    \caption{MRI Generation with SRResGAN 1 (full resolution)}
\end{figure}
\vfill
\newpage

\begin{figure}[H]
\centering
	\includegraphics[width=0.25\textwidth]{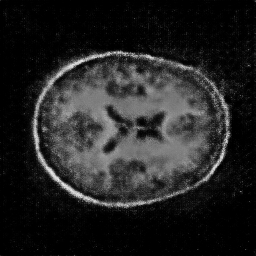}%
	\includegraphics[width=0.25\textwidth]{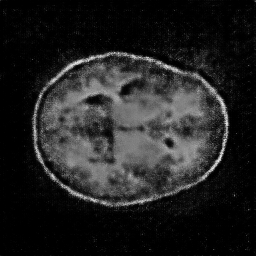}%
	\includegraphics[width=0.25\textwidth]{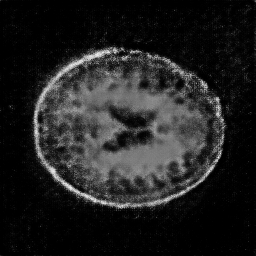}%
	\includegraphics[width=0.25\textwidth]{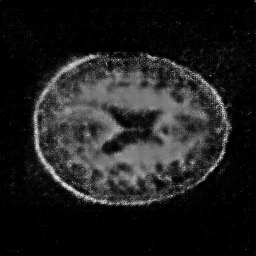}\\
	\vspace{-1pt}
    \includegraphics[width=0.25\textwidth]{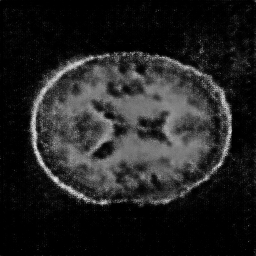}%
	\includegraphics[width=0.25\textwidth]{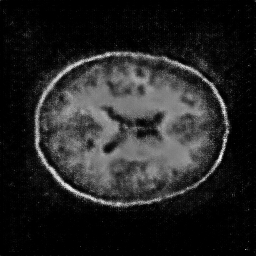}%
	\includegraphics[width=0.25\textwidth]{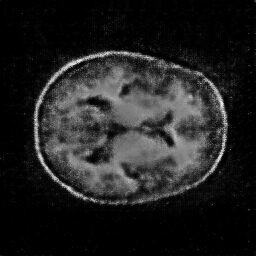}%
	\includegraphics[width=0.25\textwidth]{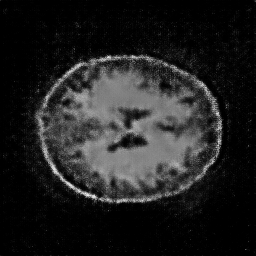}\\
	\vspace{-1pt}
    \includegraphics[width=0.25\textwidth]{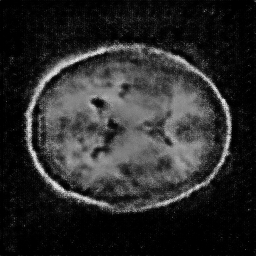}%
	\includegraphics[width=0.25\textwidth]{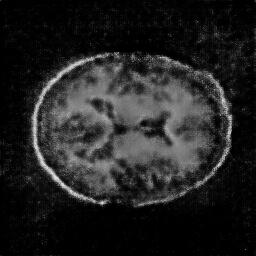}%
	\includegraphics[width=0.25\textwidth]{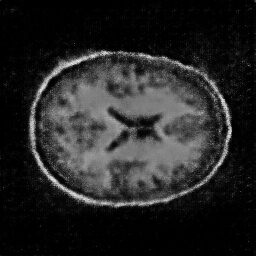}%
	\includegraphics[width=0.25\textwidth]{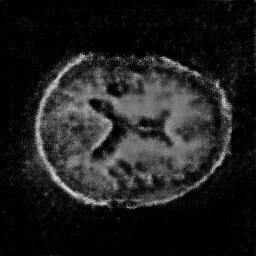}\\
	\vspace{-1pt}
    \includegraphics[width=0.25\textwidth]{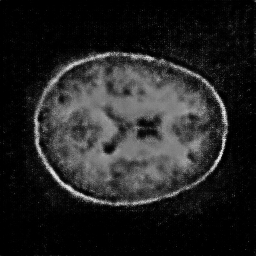}%
	\includegraphics[width=0.25\textwidth]{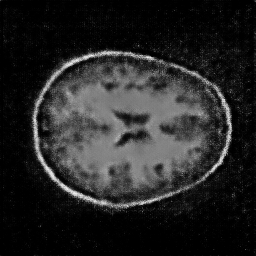}%
	\includegraphics[width=0.25\textwidth]{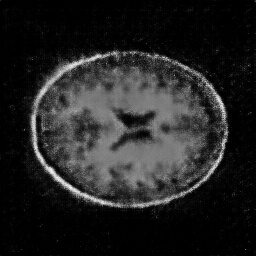}%
	\includegraphics[width=0.25\textwidth]{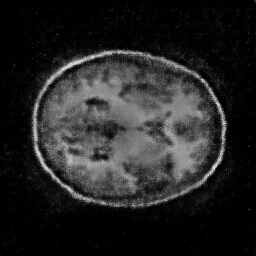}\\
	\vspace{-1pt}
    \includegraphics[width=0.25\textwidth]{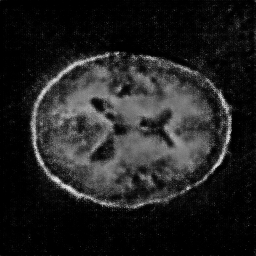}%
	\includegraphics[width=0.25\textwidth]{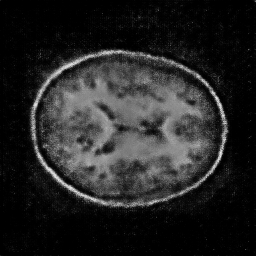}%
	\includegraphics[width=0.25\textwidth]{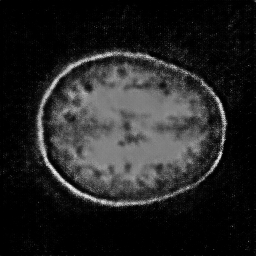}%
	\includegraphics[width=0.25\textwidth]{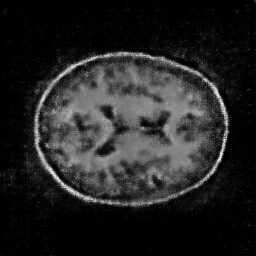}\\
	\vspace{-1pt}
    \includegraphics[width=0.25\textwidth]{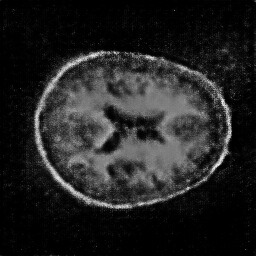}%
	\includegraphics[width=0.25\textwidth]{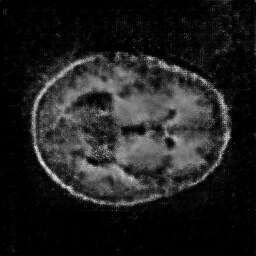}%
	\includegraphics[width=0.25\textwidth]{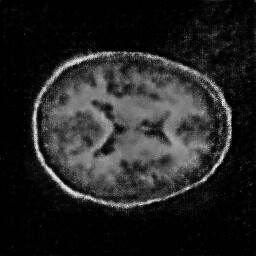}%
	\includegraphics[width=0.25\textwidth]{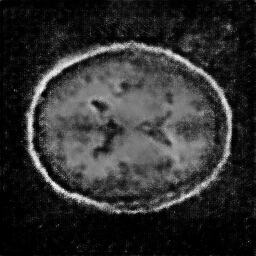}
    \caption{MRIs Generated with SRResGAN 1}
\end{figure}

\newpage
\vspace*{2cm}
\section{ProGAN 5}
\vfill
\begin{figure}[H]
\centering
	\includegraphics{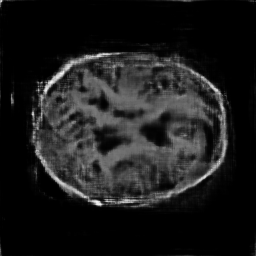}
    \caption{MRI Generation with ProGAN 5 (full resolution)}
\end{figure}
\vfill
\newpage

\begin{figure}[H]
\centering
	\includegraphics[width=0.25\textwidth]{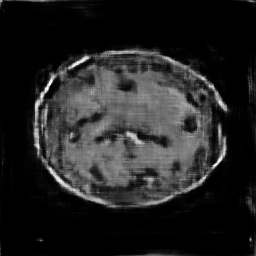}%
	\includegraphics[width=0.25\textwidth]{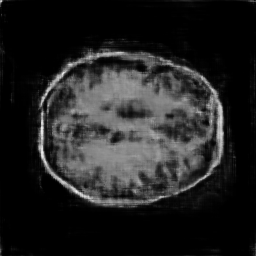}%
	\includegraphics[width=0.25\textwidth]{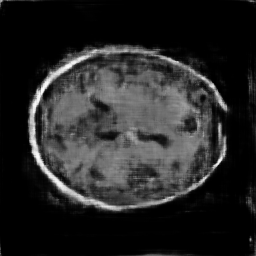}%
	\includegraphics[width=0.25\textwidth]{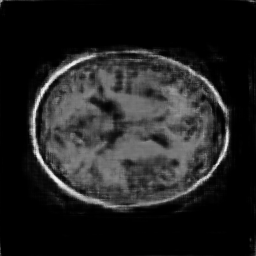}\\
	\vspace{-1pt}
    \includegraphics[width=0.25\textwidth]{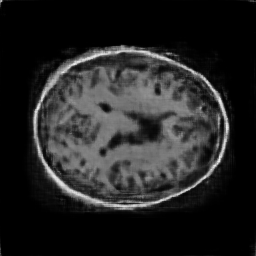}%
	\includegraphics[width=0.25\textwidth]{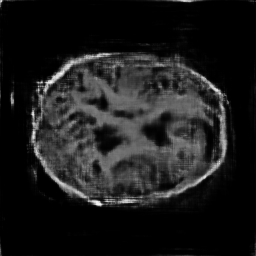}%
	\includegraphics[width=0.25\textwidth]{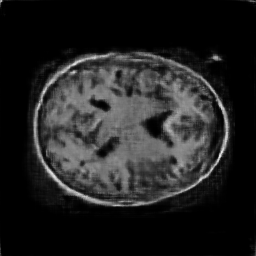}%
	\includegraphics[width=0.25\textwidth]{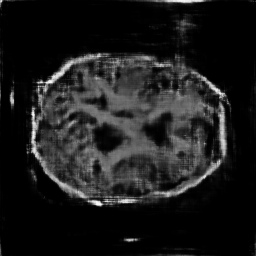}\\
	\vspace{-1pt}
    \includegraphics[width=0.25\textwidth]{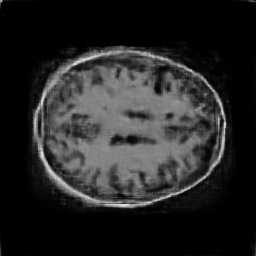}%
	\includegraphics[width=0.25\textwidth]{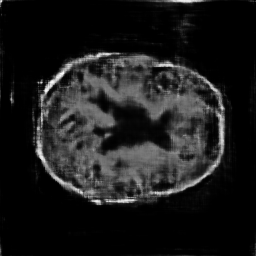}%
	\includegraphics[width=0.25\textwidth]{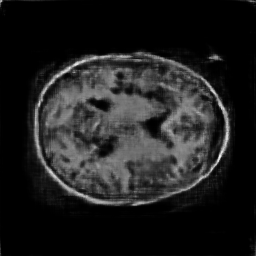}%
	\includegraphics[width=0.25\textwidth]{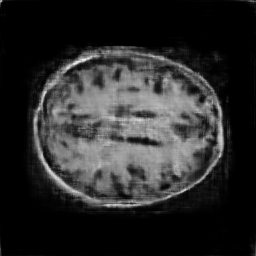}\\
	\vspace{-1pt}
    \includegraphics[width=0.25\textwidth]{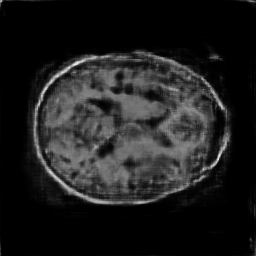}%
	\includegraphics[width=0.25\textwidth]{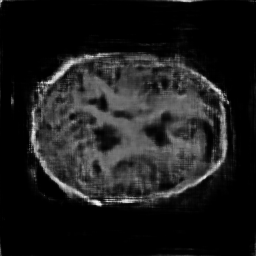}%
	\includegraphics[width=0.25\textwidth]{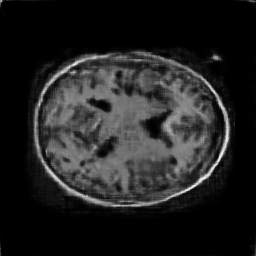}%
	\includegraphics[width=0.25\textwidth]{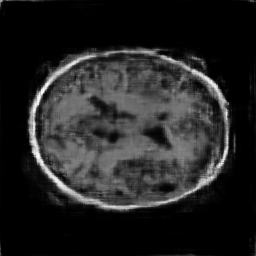}\\
	\vspace{-1pt}
    \includegraphics[width=0.25\textwidth]{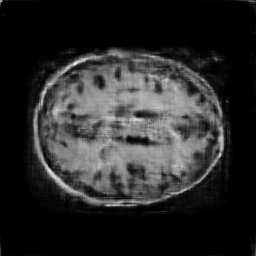}%
	\includegraphics[width=0.25\textwidth]{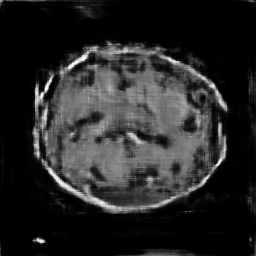}%
	\includegraphics[width=0.25\textwidth]{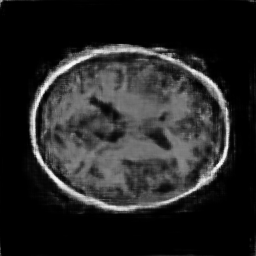}%
	\includegraphics[width=0.25\textwidth]{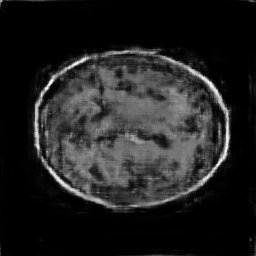}\\
	\vspace{-1pt}
    \includegraphics[width=0.25\textwidth]{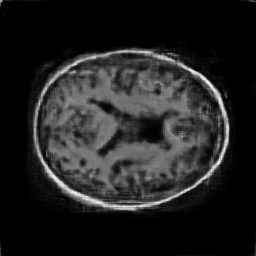}%
	\includegraphics[width=0.25\textwidth]{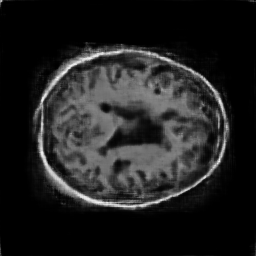}%
	\includegraphics[width=0.25\textwidth]{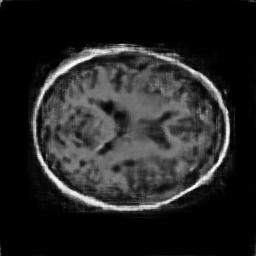}%
	\includegraphics[width=0.25\textwidth]{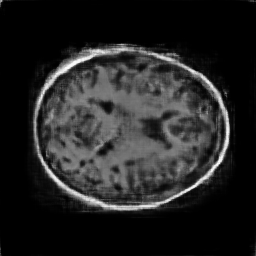}
    \caption{MRIs Generated with ProGAN 5}
\end{figure}